%% file: clustering_trajectory.tex
\documentclass[10pt]{article}
 \usepackage[utf8]{inputenc}
\usepackage{graphicx}
\usepackage{amssymb,amsmath,amsthm,bbm}
\usepackage{color}
\usepackage{url}

\usepackage{multirow,array,multicol}
\usepackage{caption, subcaption}

\usepackage[square,numbers]{natbib} 

\usepackage{setspace} 
\usepackage{lineno} 

\usepackage[dvipsnames]{xcolor}
\usepackage{cancel}
\usepackage{listings}

\title{Unsupervised detection and fitness estimation of emerging SARS-CoV-2 variants. \\Application to wastewater samples
(ANRS0160).}

\author{
Alexandra Lefebvre$^{1,2}$$^{*}$, Vincent Maréchal$^{3}$,
 Arnaud Gloaguen$^{4}$, \\ \vspace{1em}
 Obépine Consortium, Amaury Lambert$^{2,5}$$^{\P}$, Yvon Maday$^{1}$$^{\P}$ \\ 
        \small $^{1}$Sorbonne Université,  Université Paris Cité, CNRS, INRIA, \\ 
        \small Laboratoire Jacques-Louis Lions (LJLL), UMR 7598 CNRS, 75005 Paris, France\\
        \small $^{2}$Stochastic Models for the Inference of Life Evolution (SMILE), \\ \small Center for Interdisciplinary Research in Biology (CIRB), \\ 
        \small Coll\`ege de France, CNRS UMR 7241, INSERM U1050\\ \small PSL University, 75005 Paris, France\\
        \small $^{3}$INSERM, Centre de Recherche Saint-Antoine (CRSA), UMR\_S 938, \\
        \small  Sorbonne Université, 75012 Paris, France \\
        \small $^{4}$Centre National de Recherche en Génomique Humaine, \\ 
        \small Institut de Biologie François Jacob, CEA, \\
        \small Université Paris-Saclay, 91057 Évry, France \\
        \small $^{5}$Institut de Biologie de l'ENS (IBENS), \'Ecole Normale Sup\'erieure, \\
        \small CNRS UMR 8197, INSERM U1024\\ \small PSL University, 75005 Paris, France\\
        \small $^{*}$Corresponding author: Alexandra Lefebvre. E.mail: \tt{alexandra.lefebvre@math.cnrs.fr}\\
        \small $^{\P}$ These authors contributed equally to this work.
}

\date{}  

\begin{document}

\maketitle


\section*{Abstract}

Repeated waves of emerging variants during the SARS-CoV-2 pandemics have highlighted the urge of collecting longitudinal genomic data and developing statistical methods based on time series analyses for detecting new threatening lineages and estimating their fitness early in time.
Most models study the evolution of the prevalence of particular lineages over time and require a prior classification of sequences into lineages. Such process is prone to induce delays and bias. More recently, few authors studied the evolution of the prevalence of mutations over time with alternative clustering approaches, avoiding specific lineage classification. Most of the aforementioned methods are however either non parametric or unsuited to pooled data characterizing, for instance, wastewater samples. 
The analysis of wastewater samples has recently been pointed out as a valuable complementary approach to clinical sample analysis, however the pooled nature of the data involves specific statistical challenges. 
In this context, 
we propose an alternative unsupervised method for clustering mutations according to their frequency trajectory over time and estimating group fitness from time series of pooled mutation prevalence data. Our model is a mixture of observed count data and latent group assignment and we use the expectation-maximization algorithm for model selection and parameter estimation. 
The application of our method to time series of SARS-CoV-2 sequencing data collected from wastewater treatment plants in France from October 2020 to April 2021 shows its ability to agnostically group mutations according to their probability of belonging to B.1.160, Alpha, Beta, B.1.177 variants with selection coefficient estimates per group in coherence with the viral dynamics in France reported by Nextstrain. Moreover, our method detected the Alpha variant as threatening as early as supervised methods (which track specific mutations over time) with the noticeable difference that, since unsupervised, it does not require any prior information on the set of mutations.

\paragraph{keywords: } 
Time series analysis, 
mixture model, 
EM algorithm, 
Clustering trajectories,
Wastewater surveillance, 
Variant fitness.

\section{Introduction}

The Covid-19 pandemic has been characterized by the successive emergence of new SARS-CoV-2 variants
leading to several temporal waves and prompting the World Health Organization to classify certain variants as Variants Of Concern (VOC), variants of interest or variants under monitoring. Detecting variants of potential threat early in time is of great importance for an appropriate and rapid adaptation of public health responses to viral evolution.

Since SARS-CoV-2 genome can be found in feces \citep{rios2021monitoring} and other biological fluids of infected individuals, symptomatic or not, WasteWater (WW) samples give a view of SARS-CoV-2 circulation at a population level, as all infected 
individuals contribute to the sampling. 
The interest of WW surveillance has been 
highlighted in numerous previous studies \citep{wurtzer2020evaluation, martin2020tracking, anand2021review, wu2022sars, jahn2022early} as a complementary approach to clinical sample analysis.  
The statistical analysis of WW samples is challenged by their pooled nature as they contain a mixture of fragmented sequences each associated with potentially several lineages and secreted by multiple infected individuals. Moreover, most WW sequences are incomplete. On the opposite, clinical samples often contain a dominant lineage represented by almost complete sequences with a limited variability, except during persistent infections \citep{monoclonal}. The first statistical methods applied to WW samples were supervised in the sense that they track known variants using either digital or RT-qPCR \citep{ahmed2020first, medema2020presence, la2021rapid, caduff2022inferring, cluzel2022nationwide, beerenwinkel2023estimated} or high-throughput sequencing data \citep{izquierdo2021monitoring, crits2021genome, fontenele2021high, valieris2022mixture, beerenwinkel2023estimated}. They are therefore not suited for detecting newly emerging (also called cryptic) variants. Some alternative approaches based on the amplification of small and specific regions of SARS-CoV-2 genomes extracted from WW samples revealed linked polymorphisms although they seem to be inefficient in detecting threatening lineages \citep{gregory2021monitoring, smyth2022tracking}.

The amount of longitudinal genomic data available has favored the development of statistical methods for
time series analysis applied to the detection of emerging variants and the estimation of their selective advantage. There currently exists two main types of approaches. 
The first type relies on analyzing the prevalence through time of particular lineages  \citep{davies2021estimated, volz2021covid, stefanelli2022co, althaus2021tale, nishiura2021relative, dorp2021estimating, van2022global, obermeyer2022analysis} and offer statistical methods to estimate the relative fitness of lineages. Such methods require a prior clustering of sequences into lineages using, for most of them, Pango lineages \citep{rambaut2020dynamic} or current phylogenetic methods which are prone to induce delays and misinterpretation in particular for newly emerging variants. Moreover such methods are unsuited to pooled samples as they require one sample to be associated with one viral sequence. 
The second type of methods relies on analyzing the prevalence through time of mutations and includes some methods both designed for clinical and pooled samples. These methods
display a variety of clustering strategies including the k-medoids partitioning \citep{bernasconi2021data}, a weighted mutation network \citep{huang2022new}, the Levenshtein distance between sequences \citep{de2022variant}, latent epidemiological variables \citep{park2023epidemiological} or latent population genetic structure \citep{donker2024estimation}.

In this work we propose an alternative unsupervised method that falls into that second category for clustering mutations according to their frequency trajectories over time and estimating cluster fitness from time series of pooled mutation count data. 
As our parameter to estimate is our clustering criteria itself, we take advantage of the statistical power gain of both clustering and estimating fitness at once.
Our model is suited for pooled data and therefore it is particularly useful for analyzing WW samples although it can be applied to an aggregation of clinical samples. In this paper we start with a presentation of our method in Section~\ref{sec:method} before applying it to a variety of simulated datasets to present its strengths and weaknesses in Section~\ref{sec:simu}. In Section~\ref{sec:real}, we assess our model over two WasteWater Treatment Plants (WWTP) datasets collected in Nantes, France, during the emergence of B.1.1.7 (Alpha) and B.1.351 (Beta), the decline of B.1 and the transition of B.1.177 VOC. We demonstrate its ability to group mutations according to their belonging to B.1.160, Alpha, Beta, B.1.177 variants, in retrospective accordance with viral dynamics in France at the time of data collection. We also show its capacity to detect the Alpha variant as threatening as early as supervised methods with the noticeable difference that, since unsupervised, it does not require any prior knowledge on mutations. We finally discuss the limits of our model and propose future perspectives in Section~\ref{sec:perspectives}.

In order to avoid any confusion between the term \textit{cluster} in statistical clustering and cluster in epidemiology, we will use the term \textit{group} in the following.

\section{Method}\label{sec:method}

\subsection{Data}\label{sec:data}

Time series of SARS-CoV-2 sequences collected daily or weekly from October 2020 until May 2021 from two wastewater treatment plants in Nantes (France) were downloaded in fatsq format at \url{https://data-dataref.ifremer.fr/bioinfo/ifremer/obepine/lsem/data/dna-sequence-raw/}. Sample collection, data preparation and base calling are detailed in \citep{barbe2022sars}. The authors kindly provided us with files in .bam format (mapping procedure also detailed in \citep{barbe2022sars}). 
We obtained 12 samples spaced 5 to 25 days apart collected from 2020-10-20 until 2021-04-06 for the first WWTP (WWTP1) and 16 samples spaced 1 to 15 days apart collected from 2020-11-04 until 2021-04-20 for the second WWTP (WWTP2). This time period is characterized by the emergence of Alpha and Beta (mid-November 2020 and beginning of January 2021 respectively) and the decline of B.1.160 (starting around mid-October 2020) in France. Estimated frequencies of main circulating variants in France are reported by the Nextstrain project~(\url{https://nextstrain.org}) at the link \url{https://nextstrain.org/groups/neherlab/ncov/france?d=frequencies&dhttps://nextstrain.org/groups/neherlab/ncov/france%3Fd=frequencies&f_country=France&m=div&p=full&r=division}. 
We used VaRaPS \citep{djaout2024varaps} for single nucleotide variants and Indels calling with no filter using the Wuhan-Hu-1/2019 as reference genome (GenBank:MN908947.3) and samtools for read depth per position.
Mutations sequenced strictly no more than three times or of frequency below 0.05 in all samples (i.e. for all time points) were removed. We finally obtained 3640 and 4189 mutations respectively for WWTP1 and WWTP2 datasets. Read depths quantiles at 0.25, 0.50 and 0.75 are 113, 378 and 395 (respectively 107, 372 and 394) for WWTP1 (respectively WWTP2) all base positions and time points pooled. They range from 0 to 800 for both WWTPs.

The logit of mutation frequencies trough time is graphically represented in Fig~\ref{fig:data.ifremer} for each dataset. Each line is associated with one mutation with no color code. We used a logit transformation for a better visualization. The logit of a zero frequency (respectively a frequency taking value 1) is threshold at -7 (respectively 7) and ignored in case of a zero read depth on the graph. 

\input{fig_data_ifremer}

\subsection{Mathematical modeling}\label{sec:model}

Our model is a mixture of observed count data and latent group assignment variables. We consider one group under no selection named the neutral group and $K \geq 0$ groups under strictly positive or negative selection, named non-neutral groups. Group number $k\in\{0,\ldots,K\}$ will be denoted $G_k$. Let $n$ be the number of mutations in the dataset, compared to a reference sequence, we introduce $Z = \{Z_1, \ldots, Z_n\} \in  \{0,\ldots,K\}^n$ where $Z_i$ denotes the group assignment for mutation $i\in\{1,\ldots,n\}$. The neutral group is associated to value $0$ such that $\{Z_i=0\}$ means mutation $i$ belongs to the neutral group. 

Sequencing data are collected at $m+1$ increasing time points $t_0, \ldots, t_m$ leading to time series of mutation count data and read depths. We denote by $\mathcal T = (t_0-t_0=0, t_1-t_0, \ldots, t_m-t_0=T)$ the vector of differences, usually in days, between sampling date and date of origin $t_0$. Let $X_{i,t}$ (respectively $d_{i,t}$) be the mutation number $i\in\{1,\ldots,n\}$ count at time $t\in\mathcal T$ (respectively the read depth at related genome position at time $t\in\mathcal T$) . We assume a multinomial distribution for the latent variables with parameter $\pi = \{\pi_0,\ldots,\pi_K\}$ such that, for $i \in \{1,\ldots,n\}$, for $k \in \{0,\ldots,K\}$,
$$\mathbb P(Z_i = k) = \pi_k \qquad \text{where} \qquad \sum_{k=0}^K \pi_k=1.$$ 

We assume a generalized linear model with a binomial family for modeling the distribution of mutation counts conditional on group assignment as follows. 
For all $k\neq 0$, mutation $i$ count at time $t \in \mathcal T$ conditional on $\{Z_i=k\}$ follows a binomial distribution with a logit link function such that 
$$
\{X_{i,t} \,|\, Z_i=k, k\neq 0 \}  \sim \text{Binomial} \left(d_{i,t}, \frac{e^{(\mu_{k} + s_k t)}}{1+e^{(\mu_{k} + s_k t)}}\right)
$$
where $\mu_k \in\mathbb R$ and $s_k\in\mathbb{R}^\star$ are respectively the intercept and the selection coefficient associated with group $k$, $k\neq 0$. We assume no recombination event and a constant selection coefficient over the time period $[t_0, t_m]$. That latter assumption will be discussed in Section~\ref{sec:perspectives}. Note also that a selection coefficient is a direct estimate of the slope of the trajectory conditional on group assignment with no distinction between evolutionary or epidemiological parameters.  
As most mutations are neutral (under no selection, constant frequency trajectory) and because we have a particular interest in groups under positive (or negative) selection, mutation $i$ count conditional on $\{Z_i=0\}$ follows a beta-binomial distribution in order to absorb the variability of mutation frequencies at time origin for the neutral group. Therefore for all $i\in\{1,\ldots,n\}$ and $t \in \mathcal T$, we assume that 
$$
\{X_{i,t} \,|\, Z_i=0 \}  \sim \text{Binomial} \left(d_{i,t}, u \right) \qquad \text{with} \qquad u \sim \text{Beta}(\alpha, \beta)
$$
where $\text{Beta}(\alpha, \beta)$ is the beta distribution of parameters $\alpha$ and $\beta$. 

Let $\theta = (\pi, \mu, s, \alpha, \beta)$ be the set of parameters where $\pi = (\pi_0, \ldots, \pi_K) \in [0,1]^{K+1}$, $\mu = (\mu_1, \ldots, \mu_K) \in \mathbb R^{K}$, $s = (s_1, \ldots, s_K) \in \mathbb R^{*K}$and $\alpha, \beta \in \mathbb R^{*+}$, let $Z=\{Z_1,\ldots,Z_n\}$ and $X = \{X_1,\ldots,X_n\}$ where, for $i\in\{1,\ldots,n\}$, $X_i=(X_{i,t})_{t=(0,\ldots,T)}$, the joint probability of $X$ and $Z$ writes 
$$
\mathbb P(X, Z\,|\,\theta)=\prod_{i=1}^n\mathbb P(Z_i\,|\,\pi) \prod_{t\in\mathcal T}\mathbb P(X_{i,t}\,|\,Z_i; \mu, s, \alpha, \beta).
$$

\subsection{Clustering and parameter estimation}
We use the expectation-maximization (EM) algorithm \citep{dempster1977maximum} for clustering and parameter estimation. We select the number of groups either as the one that minimizes the Bayesian Information Criteria (BIC) or Integrated Complete-data Likelihood (ICL) \citep{biernacki2000assessing} or with the elbow method applied to the ICL according to the context.  We perform maximization steps of the EM algorithm with the Lagrange multipliers for updating parameter $\pi$, maximum likelihood estimator with the \texttt{glm} function of the \texttt{stats R package} for updating $\mu$ and $s$ and the \texttt{vglm} function of the \texttt{VGAM R} package for updating $\alpha$ and $\beta$ with posterior group assignment probabilities, that is, for $k\in\{0,\ldots,K\}, \mathbb P(Z_i=k\,|\,X_i)$, as weights in generalized linear models. The expectation steps are simply given by
$$
\mathbb P(Z_i = k, k\neq0 \,|\, X_i) \propto \pi_k \prod_{t\in\mathcal T} f_{i,t,k}(x_{i,t})
$$
where $f_{i,t,k}$ is the density of the binomial distribution of parameters $(d_{i,t}, \exp(\mu_k + s_k t) / (1+\exp(\mu_k + s_k t))$ for non-neutral groups and by 
\begin{equation}\label{eq:betabinom}
\mathbb P(Z_i=0\,|\,X_i) \propto \pi_0 \prod_{t \in \mathcal T} \binom{d_{i,t}}{x_{i,t}} \frac{\prod_{a = 0}^{\sum_{t\in\mathcal T} (x_{i,t}-1)} (\alpha+a) \prod_{b = 0}^{\sum_{t\in\mathcal T} (d_{i,t} -x_{i,t}- 1)} (\beta+b) }{\prod_{c = 0}^{\sum_{t\in\mathcal T} (d_{i,t}-1)} (\alpha+\beta+c)}
\end{equation}
for the neutral group, using the fact that the beta distribution is a conjugate prior for the binomial distribution (see proof of Equation~(\ref{eq:betabinom}) in Appendix Section~\ref{sec:app.proof.eq1}).

We faced a high sensitivity of the EM algorithm to initialization, when performed on real datasets. That issue was solved when applying the following procedure. 
Initialization was performed on posterior group assignment probabilities, gathered, for the sake of clarity, in a $n\times(K+1)$ matrix denoted $\eta$. We repeated several times the following two steps. 
During the first step, we tested several random matrices $\eta$ as initial values for few iterations of an EM algorithm applied to an alternative model where each of the $K+1$ group is assumed to be non-neutral and the neutral group is removed from the model.
During the second step, the updated $\eta$ matrix associated with the highest log-likelihood is used to initialize our algorithm, for few iterations, setting the neutral group as the one associated with the lowest absolute value of the estimated selection coefficients. These two steps are repeated few times and the updated $\eta$ matrix associated to the highest log-likelihood is then used for initializing our algorithm and run it until convergence. The R code of the EM algorithm applied to our model and the EM algorithm applied to the alternative model is provided in Appendix Section~\ref{sec:app.em}. 

Confidence Intervals (CI) from Section~\Ref{sec:real} are empirically computed with the observed Fisher information matrix using \texttt{numDeriv R} package.

\section{Results}\label{sec:results}

\subsection{Simulations}\label{sec:simu}

We started the assessment of our model in the number of groups selection and parameter estimation over various datasets simulated with the same model. Read depths per position and time points are simulated with a Poisson distribution of parameter denoted $\lambda$. For clarity, quantities and parameters set in each simulation scheme are summarised in Table~\ref{tab:summary.simu}.
\input{tab_summary_simu}

\paragraph{Selection of the number of groups.}

In order to evaluate the capacity of our model in selecting the number $K$ of non-neutral groups, we simulated a collection of datasets setting $K=1$ non-neutral group (simulation scheme A) or $K=3$ non-neutral groups (simulation scheme B), $n=100$ mutations, vector of time points $\mathcal T = (0, 5, 12, 20)$, $\alpha = 10$, $\beta = 50$ and $\lambda = 40$. 
In simulation scheme A, we set $\mu = -1.5$, used various selection coefficients ($s = 0.025$, $s = 0.050$, $s = 0.075$, $s = 0.100$)  and two different vectors of group proportions $\pi = (0.8, 0.2)$ and $\pi = (0.95, 0.05)$. In simulation scheme B, we set the vector of intercepts at $\mu = (1.5, -1, -2.5)$, the vector of non-zero selection coefficients $s = (-0.1, 0.05, 0.1)$ and various vectors of group proportions ($\pi = (0.4, 0.2, 0.2, 0.2)$, $\pi = (0.6, 0.2, 0.1, 0.1)$ and $\pi=(0.8, 0.08, 0.06, 0.06)$) (see Table~\ref{tab:summary.simu}). We performed 100 replications for each simulation scheme and reported the proportion of models associated with minimal BIC and ICL criteria in Table~\ref{tab:sel}.
In this example and as expected, we can see that minimizing a Bayesian criteria for selecting the number of non-neutral groups is highly dependent on the strength of the signal ($s$ in this example) and on group proportions $\pi$.  
As expected, the higher the selection coefficient in simulation scheme A, the better the model performance in selecting the true number of non-neutral groups. Moreover as one or more group proportion(s) tend(s) towards zero (simulation scheme A and B), the model performances are declining in determining the number of groups and tends to gather groups together. In particular, in simulation scheme B (Table~\ref{tab:selK3}), group $G_2$ and group $G_3$ are the two most similar ones with a negative intercept and a positive selection coefficient. As both their proportion decrease, the model tends to fuse them into one group resulting in a rising proportion of models containing 2 non-neutral groups selected. 
 
Moreover, in each simulation scheme, the ICL tends to select a lower number of groups than the BIC which is consistent with the fact that the ICL adds a penalty according to clustering entropy. 
\input{tab_sel}

\paragraph{Estimation.}

In this paragraph, we assume that a number of groups has been estimated as illustrated in the previous paragraph and we assess the performances of our model in parameter estimation conditional on a fixed number of groups. We handled label switching using the minimum of the mean square error between true (used for simulations) and estimated parameters $\mu$ and $s$ where $s$ is multiplied by $T$ in order to obtain quantities of similar range. We excluded $\pi$ in this mean square error for it to be of very different range. 
 We used various simulated datasets setting $K=2$ non-neutral groups, $\pi = (0.6, 0.3, 0.1)$, $\mu =  (0.5,-3.0)$, $s = (-0.05, 0.10)$, $\alpha = 10$ and $\beta = 50$ for each one of them. In simulation scheme C we set $\mathcal T = (0,5,12,20)$, $\lambda = 40$ and used various number of mutations ($n=25$, $n=50$, $n=100$, $n=200$, $n=400$). In simulation scheme D we set $n=200$, $\lambda = 40$ and used various number of times points, $\mathcal T = (0, 5)$, $\mathcal T = (0, 5, 9)$, $\mathcal T = (0, 5, 9, 12)$ and $\mathcal T = (0, 5, 9, 12, 20)$. In simulation scheme E we set $n=200$, $\mathcal T = (0, 5, 12, 20)$ and used various Poisson parameter for sampling read depths ($\lambda=10$, $\lambda=25$, $\lambda=100$, $\lambda=200$, $\lambda=400$) (see Table~\ref{tab:summary.simu}). For each mutation, posterior group assignment was performed using maximum a posteriori probability, $\arg\max_{k\in\{0,\ldots,K\}} \mathbb P(Z_i=k\,|\,X_i=x_i)$, where $(x_i)$ denotes the observed vector of mutation counts. We performed 200 replications for each simulation scheme and reported parameter estimates as well as Area Under the ROC Curve (AUC) of posterior group assignment in Fig~\ref{fig:estim.n} (respectively \ref{fig:estim.T} and~\ref{fig:estim.d}) for simulation scheme C (respectively D and E). As expected, as the number of observation increases (increasing $n$, $T$ or $\lambda$), interquartile ranges of boxplots of parameter estimates shrink in expected proportions and AUCs tend toward 1. 
 We can also note that, the greater the proportion of a group is, the more narrow the interquartile ranges associated to its parameters.
The parameter estimator being the maximum likelihood estimator, its bias with a limited number of observations $n$ is corrected with an increasing number of observations and becomes negligible from $n=100$ mutations in simulation scheme C. The parameter estimator bias should be limited in the framework of a real dataset of mutation count from time series of SARS-CoV-2 genome sequencing in WWTP samples as the number of mutations sequenced in such sample is usually very high. 
We can finally notice the very accurate posterior group assignment probabilities with AUCs first quartiles above 0.95 in each simulation scheme. 
\input{fig_boxplot_n}
\input{fig_boxplot_T}
\input{fig_boxplot_d}

\subsection{Real data}\label{sec:real}

We present in this section results obtained from WWTP1 and WWTP2 datasets presented in Section~\ref{sec:data} and Fig~\ref{fig:data.ifremer}. We firstly consider the whole set of time points in Section~\Ref{sec:sel} and Section~\ref{sec:estim.aff} before restricting datasets to a selection of time points between November and December 2020 in Section~\Ref{sec:detection}. 
For the sake of clarity, analyses performed in the following are summarized in Table~\ref{tab:summary.analyses} with their related dataset, time period studied, number of non-neutral groups estimated and related Figures and/or Tables.

\input{tab_summary_analyses}

\subsubsection{Selection of the number of groups.}\label{sec:sel}
In order to select the number of groups in our datasets, we computed the BIC and ICL criteria of models composed of $K=0$ to $K=9$ non-neutral groups. Results are reported in Fig~\ref{fig:sel} along with minus two times the log-likelihood ($-2\log L$) from $K=1$ non-neutral group. For a better visualization, we omitted values associated to $K=0$ in the graphical representation for them to be very much higher and to offer limited information. 
Reminding that the entropy of the model, for a fixed number of groups $K+1$, is given by $- \sum_{i=1}^n \sum_{k=0}^K \mathbb P(Z_i=k\,|\,X_i=x_i) \log\mathbb P(Z_i=k\,|\,X_i=x_i)$, it is quite straightforward to note that the maximal entropy is given by $n\log(K+1)$ (reached for $\mathbb P(Z_i=k\,|\,X_i) = 1/(K+1)$ for all $i\in\{1,\ldots,n\}$ and all $k\in\{0,\ldots,K\}$). The ratio between clustering entropy and maximal entropy is added in Fig~\ref{fig:sel} and will be called entropy ratio.

We firstly notice that the BIC and the ICL decrease with an increasing number of groups up to $K=9$ non-neutral groups and their values are almost equal to $-2\log L$. Reminding that the BIC penalizes $-2\log L$ with a function of the number of parameters in the model and that the ICL adds a penalty to the BIC according to model entropy, this observation means that each additional group leads to such a gain in log-likelihood that penalizing the model with the number of parameters and model entropy is not sufficient to draw conclusions from Bayesian criteria.
This result  reflects a limit of our model in capturing sufficient variability in the datasets mostly explained by the high number of circulating lineages.
We remind indeed that we assume a constant selection coefficient over the studied time period and, moreover, the variability laying in binomial distributions is increased for the neutral group (beta-binomial distribution) but not for non-neutral groups (logit transformation with constant parameters). Furthermore, one group is not associated to one variant but to a group of mutations of similar frequency trajectory. Many mutations are shared by multiple lineages leading to additional subgroups with their own parameter estimates. These limits are discussed in Section~\ref{sec:perspectives}. 

In such context, we propose to use the elbow method applied to the ICL complemented with the elbow method applied to the BIC inside a range of associated low values of entropy ratio, for selecting the number of groups.
In such manner we control the selected number of groups and we ensure that groups are well separated as posterior group affectation probabilities tend toward 0 or 1 (low entropy). Moreover, selecting a model of low entropy drives our choice of maximum a posteriori probability for determining posterior group assignment in the following. 
These choices lead to analysing WWTP1 dataset conditional on $K=4$ non-neutral groups (analysis A) and WWTP2 dataset conditional on $K=3$ and $K=6$ non-neutral groups (respectively analysis B and C).

\input{fig_main_selection}

\subsubsection{Parameter estimation and posterior group assignment.}\label{sec:estim.aff}

In this paragraph we assume that the number of groups is selected as described in Section~\ref{sec:sel} and we report and analyze quantities computed conditional on a fixed number of groups. Estimated quantities will be denoted with a hat symbol and, for the sake of clarity, exponentiated with the letter of the associated analysis. For instance $\hat\pi^\text{A} = (\hat\pi_0^\text{A}, \ldots, \hat\pi_{\hat K^\text{A}}^\text{A})$ denotes the vector of parameter estimate for $\pi$ in Analysis A where $\hat K^\text{A}=4$ is the number of groups estimated as detailed in the previous section.
As we often comment our results for a selection of combined analyses in the following, groups $G_1,\ldots,G_K$ will be exempted from this choice of notation in order to facilitate the reading. Group $G_1$ may mean group number 1 estimated in Analysis A, or group number 1 estimated in Analysis B, etc. 
We computed Confidence Intervals (CI) for intercepts and selection coefficients of non-neutral groups as they are the parameters of interest. All CIs in the following are given at 95\%. We therefore can determine in particular if a selection coefficient is significantly different from zero with a threshold at 2.5\% with its associated CI. Note finally that time origins may vary from one analysis to the other. Times are expressed in days from the time origin associated with the related analysis.

In order to verify our outputs in the following, we firstly propose to compare them to supervised results reported by Barbé et al. \citep{barbe2022sars}. We will secondly verify in which extend the retrospective probability of belonging to a circulating variant stratified on posterior group assignment estimated by our model is in coherence with viral dynamics in France reported by Nextstrain. In order to perform that second verification, we will use a mutation profile matrix defined as a mutation $\times$ lineage matrix filled with the probability for each mutation to belong to each lineage. We followed the procedure provided by Virpool \citep{gafurov2022virpool} to compute this matrix using script $\texttt{src/gisaid/process\_gisaid.pl}$ with GISAID sequences collected between 2020-01-01 and 2021-05-15 and default parameter `MAX PER MONTH = 50000'; `MIN LENGTH = 25000' for a random sampling of sequences and script $\texttt{src/profile\_estimation.py}$ for the inference. Both scripts are available at \url{https://github.com/fmfi-compbio/virpool}. We extracted lineages B.1.1.7 (Alpha), B.1, B.1.160, B.1.177, B.1.351 (Beta), B.1.367 and B.1.1 from VirPool's outputs, for them to be the main circulating lineages at the time of the analysis, and we computed the frequency of each mutation observed at least once for at least one of these lineages.

Parameter estimates, their graphical representation with resulting group frequency trajectories as well as mutation profiles stratified on maximum a posteriori probability of group assignment are reported in Fig~\ref{fig:estim.w1k4}, \ref{fig:estim.w2k3} and~\ref{fig:estim.w2k6} for Analysis A, B and C respectively. In the following, the estimated frequency at time origin of the neutral group is computed from estimates of $\alpha$ and $\beta$ such that $\hat f_0(0)=\hat\alpha/(\hat\alpha+\hat\beta)$ and the associated intercept $\hat\mu_0$ is given by the logit transformation of $\hat f_0(0)$. Moreover, estimated frequencies at time origin for non-neutral groups are computed from estimated intercepts such that, for $k\in\{1,\ldots,K\}$, $\hat f_k(0) = \exp(\hat\mu_k)/(1+\exp(\hat\mu_k))$.
\input{fig_main_WWTP1_K4_estimation}
\input{fig_main_WWTP2_K3_estimation}
\input{fig_main_WWTP2_K6_estimation}

Similar results reported in Fig~\Ref{fig:estim.w1k4}, \ref{fig:estim.w2k3} and~\ref{fig:estim.w2k6} are consistent with the fact that both sites, WWTP1 and WWTP2, are closely located in Nantes, France. The estimated proportion of the neutral group is as high as $\hat\pi^\text{A}_0 = 0.968$, $\hat\pi^\text{B}_0 = 0.976$ and $\hat\pi^\text{C}_0 = 0.954$ for Analysis A, B and C respectively, which is consistent with the fact that most mutations are under no selection. All selection coefficients are significantly different from 0 with a threshold below 2.5\% as no 95\%CI contain value 0.  

We distinguish for each dataset one rising trajectory associated with group $G_1$ of very low estimated frequency at time origin $\hat f^\text{A}_1 (0) = \exp(\hat\mu^\text{A}_1) / (1+\exp(\hat\mu^\text{A}_1)) = 0.03$, $\hat f^\text{B}_1 (0) = 0.02$ and $\hat f^\text{C}_1 (0) = 0.02$ and high positive selection coefficient estimate $\hat s^\text{A}_1=2.73 \,[2.71; 2.74] \times 10^{-2}$, $\hat s^\text{B}_1=3.42 \,[3.41; 3.44] \times 10^{-2}$ and $\hat s^\text{C}_1=3.40 \,[3.38; 3.41] \times 10^{-2}$ respectively for analyses A, B and C. We also distinguish a declining trajectory (group $G_2$) starting with the highest estimated frequency at time origin among all groups with $\hat f^\text{A}_2 (0) = 0.66$, $\hat f^\text{B}_2 (0) = 0.61$ and $\hat f^\text{C}_2 (0) = 0.62$ and a negative selection coefficient estimate $\hat s^\text{A}_2 = -1.61 \,[-1.63; -1.59] \times 10^{-2}$, $\hat s^\text{B}_2 = -2.05 \,[-2.07; -2.03] \times 10^{-2}$ and $\hat s^\text{C}_2 = -2.06 \,[-2.08; -2.04] \times 10^{-2}$ respectively for analyses A, B and C. 
For each analysis all mutations assigned to group $G_1$ (respectively group $G_2$) with maximum a posteriori probability are associated with a probability of belonging to Alpha (respectively B.1.160) greater than 0.995 (respectively 0.973) (Fig~\ref{fig:signature.w1k4}, \ref{fig:signature.w2k3} and~\ref{fig:signature.w2k6}) except one to two mutations per group. 

These results are consistent with the viral clades dynamics in France during the time period considered reported by Nextstrain with B.1.160 VOC dominating at the beginning of the time period and starting to decline while Alpha was emerging and replacing it. Let us also point at the fact that one group is not associated with one variant stricto sensu but it is a set of mutations with similar frequency tendency during the time period studied. In particular, mutations may be shared by several lineages, which may also explain why the global tendency is captured while an estimated group frequency trajectory is not fully related to a particular variant. 

For a better visualization of the estimated frequency trajectories of group $G_1$ and group $G_2$ represented in Fig~\ref{fig:traj.w1k4}, \ref{fig:traj.w2k3} and~\ref{fig:traj.w2k6}, their computed values at a selection of time points are given in Table~\ref{tab:freq} and compared to estimated frequencies reported by Nextrain. The selection of time points is simply motivated by a subset of those chosen by Nextstrain. 
We can see that our model captures the global tendency of Alpha but tends to overestimate its frequency until mid-November 2020, underestimate it between mid-November 2020 and the end of January 2021 and overestimate it from February 2021 until the end of March. These alternations between overestimation and underestimation are explained by the fact that we assume a constant selection coefficient over the time period studied and therefore, a global estimate, with no consideration of time-varying changes influenced by public health strategies, vaccinations, the late emergence of a new variant, etc. 
In order to overcome this limitation, as discussed in Section~\ref{sec:perspectives}, we plan on including piecewise constant selection coefficients and perform break point detection in the future. 
On the contrary and for similar reasons, B.1.160 VOC frequencies are underestimated by our model until the end of November or December 2020, depending on the analysis, and overestimated afterwards. 

\input{tab_freq}

Conditioning the analysis of WWTP2 dataset on 6 non-neutral groups (Fig~\ref{fig:estim.w2k6}, Analysis C) allows for a better characterization of B.1.177 mutations with declining group $G_3$ associated with a negative selection coefficient $\hat s^\text{C}_3 = -1.30\,[-1.34; -1.27] \times 10^{-2}$ and a frequency at time origin $\hat f^\text{C}_3  (0)= \exp(\hat \mu^\text{C}_3 / (1+ \exp(\hat \mu^\text{C}_3)) = 0.20$. This group gather 19 mutations among which 6 (respectively one) are associated with a probability above 0.984 (respectively 0.665) to belong to B.1.177. 
The frequency trajectory of that group partly reflects the overall B.1.177 frequency evolution reported by Nextstrain around 7\% at the end of October 2020, 1\% by mid-March 2021 and nearly 0\% by the beginning of April 2021. As previously noted, because of the assumption of constant selection coefficients, our model misses the full trajectory of B.1.177 VOC with its increased frequency between November 2020 and February 2021 but it captures its overall tendency throughout the entire time period studied.

\subsubsection{Reduced datasets.}\label{sec:reduced.dataset}

The estimated proportion of neutral mutations in previous analyses ranges between 95.4\% and 97.6\%. Such high values may have an impact on statistical power for clustering groups under positive and negative selection. In order to avoid an accumulation of neutral mutations, mutations under purifying selection and sequencing error data, we performed similar analyses on datasets restricted to mutations associated with a probability above 0.005 to belong to at least one of the main circulating lineages at the time of the analysis. Choosing B.1.1.7 (Alpha), B.1, B.1.1, B.1.160, B.1.177, B.1.351 (Beta), B.1.367 as these lineages, WWTP1 and WWTP2 datasets are reduced to 281 and 309 mutations respectively. BIC and ICL criteria associated to models composed of 1 to 7 non-neutral groups are graphically represented in Fig~\ref{fig:reduced.sel} and lead, as in previous analyses, to an inconclusive number of groups. We suggest to apply the elbow method for the ICL or the elbow method for the BIC combined with low values of entropy ratio leading to a choice of $\hat K^\text{D} = \hat K^\text{E} = 2$ non-neutral groups for both reduced WWTP1 and WWTP2 datasets (respectively Analysis D and E) and $\hat K^\text{F} = 6$ non-neutral groups for reduced WWTP1 (Analysis F). Parameter estimates, frequency trajectories and mutation signatures stratified on posterior group assignment are reported in Fig~\ref{fig:reduced.estim.w1k2}, \ref{fig:reduced.estim.w2k2} and~\ref{fig:reduced.estim.w1k6} respectively for analysis D, E and F. We firstly can note a decreased proportion of neutral mutations when compared to previous analyses, although that group still contain a vast majority of mutations with $\hat \pi_0^\text{D}=86.5\%$, $\hat \pi_0^\text{E}=86.1\%$ and $\hat \pi_0^\text{F}=76.4\%$ despite our data reduction. 

\input{fig_reduced_data_selection}

\input{fig_reduced_data_WWTP_K2_estimation}
\input{fig_reduced_data_WWTP1_K6_estimation}

Similar results for Analyses D and E (Fig~\ref{fig:reduced.estim.w.k2}) are expected due to the close localization of both treatment plants in France, as also noted in the previous section. Both analysis D and E reveal one emerging group (group $G_1$) with  $\hat\mu^\text{D}_1 = -3.43 \,[-3.45; -3.41]$, $\hat s^\text{D}_1 = 2.71 \,[2.70; 2.73] \times 10^{-2}$ and $\hat\mu^\text{E}_1 = -3.91 \,[-3.93; -3.89]$, $\hat s^\text{E}_1 = 3.42 \,[3.41; 3.44] \times 10^{-2}$ and one declining group (group $G_2$) with $\hat\mu^\text{D}_2 = 0.68 \,[0.66; 0.70]$, $\hat s^\text{D}_2 = -1.62 \,[-1.64; -1.60] \times 10^{-2}$ and $\hat\mu^\text{E}_2 = 0.46 \,[0.45; 0.48]$, $\hat s^\text{E}_2 = -2.05 \,[-2.07; -2.03] \times 10^{-2}$. The emerging (respectively declining) group gather 24 (respectively 14) mutations, all of them are associated with a probability above 0.995 (respectively 0.973) to belong to Alpha (respectively B.1.160). Frequency trajectories of emerging and declining groups are similar to those associated with Alpha and B.1.160 mutations previously computed in Analyses A, B and C (Fig~\ref{fig:estim.w1k4}, \ref{fig:estim.w2k3} and~\ref{fig:estim.w2k6}) as well as the tendency of Alpha and B.1.160 variants in France at the time period studied (Table~\ref{tab:freq}). 

Conditioning the analysis of the reduced WWTP1 dataset on 6 instead of 2 non-neutral groups (Analysis F versus D) reveals several subgroups. We firstly note that group $G_1$ in Analysis D is roughly split into two positively selected subgroups with estimated proportions, intercepts and slopes $\hat \pi_1^\text{F}=0.050$, $\hat \pi_5^\text{F}=0.055$, $\hat \mu_1^\text{F}=-2.91\,[-2.94; -2.89]$, $\hat \mu_5^\text{F}=-5.65\,[-5.68; -5.61]$ and $\hat s_1^\text{F}=2.45\,[2.43; 2.47] \times 10^{-2}$, $\hat s_5^\text{F}=4.08\,[4.05; 4.11] \times 10^{-2}$. This distinction was not possible over the whole WWTP1 dataset due to the lack of power induced by the proportion of neutral mutations. The set of mutations assigned to each of these groups is listed in Table~\ref{tab:signature.w1} and contains only mutations with probability above 0.995 to belong to Alpha. Note also that group $G_1$ estimated intercept is much higher than the one associated to group $G_5$, suggesting that mutations assigned to that group appeared in Nantes before those of group $G_5$. This result is consistent with Fig 5 in \citep{barbe2022sars} which represents the frequency of each mutation throughout time. All mutations assigned to group $G_1$ (Table~\ref{tab:signature.red}) or to group $G_5$ (Table~\ref{tab:signature.magenta}) were firstly detected by \citet{barbe2022sars} between 2020-11-17 and 2020-12-25 or between 2020-12-25 and 2021-02-10 respectively, except C15279T which was firstly detected on the $4^\text{th}$ of December 2020 by the authors. However we can see that among all mutations studied in Fig 5 in \citep{barbe2022sars}, C15279T is one of the most rapid increase in frequency which could explain its assignment to group $G_5$ that is the one of highest selection coefficient estimate.

\input{tab_signature_WWTP1_rising}

The third rising group revealed in Analysis F (group $G_4$, cyan in Fig~\ref{fig:reduced.estim.w1k6}) is associated to the lowest positive selection coefficient with $\hat s^\text{F}_E= 1.80 \,[1.77; 1.84] \times 10^{-2}$. That group is composed of 14 mutations including three mutations associated with probabilities ranging from 0.329 to 0.861 to belong to Alpha and three mutations associated with probabilities ranging from 0.168 to 0.967 to belong to Beta variants. One of such mutations is shared by both variants. 
Reminding that we assume a constant selection coefficient, that group partly reflects the viral dynamics of Beta with low frequency at time origin with $\hat f^\text{F}_4 (0) = \exp(\hat\mu^\text{F}_4 )/(1+\exp(\hat\mu^\text{F}_4)) = 0.007$ and an overall increased frequency by April 2021 resulting in a group of positive but limited selection coefficient.  As shown in Nextstrain reports with Beta variant frequency estimates at 0\% on
2020-10-12, 6\% on 2021-01-08, 21\% on 2021-02-24, 22\% on 2021-03-18 and 27\% on 2021-04-09, Beta was almost absent in France in December 2020 and started
to emerge at the beginning of January 2021.  All other Beta VOC mutations were assigned by our model to the neutral group except G25563T and A23063T. The beta-binomial distribution associated to the neutral group indeed tends to absorb part of the variability in the data. Moreover G25563T and A23063T are respectively assigned to declining group $G_2$ and emerging group $G_1$ which can be explained by the fact that G25563T is also signature of B.1.160 (declining variant at the time of the analysis) and A23063T is also signature of Alpha (emerging variant at the time of the analysis).

Group $G_3$ of negative selection coefficient with $\hat s^\text{F}_3=-1.52\,[-1.58; -1.46]$ gather 4 on 10 mutations with probability above 0.984 to belong to B.1.177. With similar estimated parameters and composition than those associated to group $G_3$ in Analysis C, that group partly reflects the global tendency of B.1.177 variant over the entire studied period of time and under the assumption of a constant selection coefficient. 

Let us finally mention group $G_6$ (yellow group in Fig~\ref{fig:reduced.estim.w1k6}) with very high frequency throughout time and very low selection coefficient absolute value, although still significantly different from zero with a threshold at 2.5\%. It is associated with estimated parameters $\hat\mu_6^\text{F}=2.53\,[2.45;2.60]$ leading to an estimated frequency $\hat f^\text{F}_6 = 0.926$ and $\hat s^\text{F}_6=-0.11\,[-0.19; -0.04]$. That group is composed of C241T, C3037T and A23403G which are all associated to a probability above 0.996 to belong to all main circulating variants through the time period (B.1.1.7, B.1.1, B.1.160, B.1.177, B.1.351)  which is also in coherence with expected frequency trajectory, one dominant variant being replaced by another one leading to a high and near constant frequency of mutations shared by all main lineages. 

\subsubsection{Fitness detection.}\label{sec:detection}

In order to assess the performances of our model in detecting new variants of increased fitness early in time, we performed several analyses over WWTP1 and WWTP2 datasets restricted to the time period of Alpha emergence, between the end of October and mid-December 2020. 
In Analysis G and Analysis H, WWTP1 dataset is restricted respectively to the first two time points (2020-10-20 and 2020-11-04) and the second and third time points (2020-11-04 and  2020-11-17). In analysis I (respectively Analysis J) WWTP2 is restricted to time points 2020-11-09 and 2020-12-04 (respectively 2020-12-04 and 2020-12-18). The choice of time origin for WWTP2 is motivated by Fig 6 in \citep{barbe2022sars} where SARS-CoV-2 Alpha VOC mutations were detected above 5\% from 2020-12-18 in WWTP2 and a month earlier (2020-11-17) in WWTP1. As described in Section~\ref{sec:data}, mutations sequenced strictly no more than three times or of frequency below 0.05 in all samples (i.e. for all selected time points) were removed. We finally obtained 1242, 1177, 1458 and 1611 mutations respectively for Analyses G, H, I and J. 
BIC and ICL criteria along with entropy ratio associated to models composed of increased number of groups are graphically represented in Fig~\ref{fig:detect.sel} and lead to a choice of 3 non-neutral groups for Analyses G, H and I and 2 non-neutral groups for Analysis J when applying the elbow method on the ICL or the elbow method on the BIC associated to low values of entropy ratio. Another choice of 4 non-neutral groups for Analysis G could be an option and led to similar results (not shown).
Group frequency trajectories drawn from parameter estimates as well as Alpha and B.1.160 VOC mutation signature stratified on maximum a posteriori group assignment probability are displayed in Fig~\ref{fig:detect.estim}.  

\input{fig_detection_selection}

\input{fig_detection_estimation}

No rising group gathering Alpha VOC mutations was detected before 2020-11-04 in WWTP1 (Analysis G, Fig~\ref{fig:detect.estim1}) nor before 2020-12-04 in WWTP2 (Analysis I, Fig~\ref{fig:detect.estim2}), so did not \citet{barbe2022sars} as shown in their Fig 6 with a threshold below 5\%. 

Both Analyses G and I reveal a rising group $G_3$ with moderate to high intercept $\hat f^\text{G}_3(0)= \exp(\hat\mu^\text{G}_3) / (1+\exp(\hat\mu^\text{G}_3)) =  0.34$ and $\hat f^\text{I}_3 = 0.34$ 
and high positive selection coefficient with $\hat s^\text{G}_3=0.124\,[0.118; 0.130]$ and $\hat s^\text{I}_3=0.113\,[0.109; 0.117]$. That group gather 9 (respectively 10) mutations all of them (respectively 6 of them) associated with probability above 0.973 to belong to B.1.160. Analysis I also reveals a declining group $G_2$ of much lower proportion ($\hat\pi^\text{I}_2=00.1 < \hat\pi^\text{I}_3=0.007$) with high intercept ($\hat f^\text{I}_2=0.71$) and negative selection coefficient ($\hat s^\text{I}_2= -0.170 \,[-0.186; -0.154]$). That group is composed of 2 mutations, C11497T and G25563T, associated with probability 0.994 to belong to B.1.160. These results are consistent the fact that B.1.160 is the dominant variant before mid-November 2020 (analysis G) and starting to be replaced by mid-December 2020 (Analysis I) in accordance with B.1 VOC frequencies in France reported by Nextstrain at 64\%, 68\% 64\% respectively at time points 2020-10-23, 2020-11-25, 2020-12-17. 

Note that the second rising group in Analysis G (group $G_1$) with estimated selection coefficient $s^\text{G}_1 = 0.191\,[0.185; 0.197]$ is composed of 25 mutations, one of them (G28883C) associated with probability above 0.997 to belong the Alpha and B.1.1 (a fluctuating variant of relatively low frequency at the time of the analysis). This result partly reveals a strength in the model in detecting threatening variant very early in time and weakness as most mutations assigned to that group are actually no signature of any VOC at the time of the analysis suggesting that our model may be prone to data variability or noise in particular when choosing solely two time points. This weakness could be avoided with more time points and break point detection and will be discussed in Section~\ref{sec:perspectives}. 

Analyses H and J both reveal emerging group $G_1$ of low intercept and declining group $G_2$ of high intercept with $\hat f^\text{H}_1(0) = 0.02$ , $\hat f^\text{H}_2(0) = 0.80$, $\hat s^\text{H}_1=0.217 \,[0.212; 0.222]$, $\hat s^\text{H}_2=-0.211 \,[-0.219; -0.203]$ for Analysis H and $\hat f^\text{J}_1(0) = 0.02$, $\hat f^\text{J}_2(0) = 0.85$, $\hat s^\text{J}_1=0.293 \,[0.288; 0.299]$, $\hat s^\text{J}_2=-0.298 \,[-0.306; -0.289]$ for Analysis J. 
Group $G_1$ in Analysis H gather 32 mutations among which 6 (respectively 2 and 3) are associated with probability above 0.995 (respectively 0.994 and 0.994) to belong to B.1.1.7, Alpha (respectively B.1.177 and B.1.367). One of them (C3267T) is shared by Alpha and B.1.367.
The low proportion of Alpha mutations in that group may be explained by the fact that the Alpha variant was at the beginning of its emergence and led to all first three quartiles near zero in the associated boxplot
(middle column in Fig~\ref{fig:detect.estim}, Analysis H). 
Group 1 in Analysis J gather 17 mutations, 10 of them are associated with probability above 0.995 to belong to Alpha. 
Group 2 in Analysis H (respectively Analysis J) gather 7 (respectively 13) mutations among which 7 (respectively 8) are associated to B.1.160 with probability above 0.989 (respectively 0.973). These results are consistent with the viral dynamics in France with Alpha starting to replace B.1.160 at the time period considered. Moreover B.1.177 was still a rising variant during the time period of Analysis H with frequencies 23\% on 2020-11-04 and 25\% on 2020-11-18 according to Nextstrain estimates. Estimates and group assignment reported in Fig~\ref{fig:detect.estim} show the ability of our model to detect a potentially threatening group of mutations associated to the emerging Alpha variant as early as \citet{barbe2022sars} as shown in their Fig 6, that is 2020-11-17 for WWTP1 (Analysis F) and 2020-12-18 for WWTP2 (Analysis H)  with the major difference that our method is unsupervised, that is, we do not use any prior information about mutations present in our dataset.

Let us analyze the set of mutations assigned to the emerging group $G_1$ in Analysis H along with their profile listed in Table~\ref{tab:signature.detection}. 
This group is flagged as threatening by our model with $\hat s^\text{H}_1= 0.217 \,[0.212; 0.222] >> 0$ with data collected earlier than 2020-11-17. As previously mentioned, this group gather 32 mutations among which six Alpha mutations, two B.1.177 mutations and two B.1.367 mutations for a total of 31\% of mutations with probabilities above 0.994 to belong to a threatening emerging variant. Interestingly all Alpha VOC mutations reported in Table~\ref{tab:signature.detection} were assigned to group $G_1$ in Analysis F over the whole time period until April 2021 (see Table~\ref{tab:signature.red}). This remark echoes a previous one when comparing intercept estimates of group $G_1$ (red) and group $G_5$ (magenta) in Analysis F (Fig~\ref{fig:reduced.estim.w1k6}), $\hat\mu^\text{F}_1 >> \hat \mu^\text{F}_5$, hence a higher group 1 frequency at time origin. We therefore expect mutations assigned to that group to be detected earlier than those of group $G_5$, that is confirmed by Analysis H.
\input{tab_signature_detection}

We finally notice that absolute values of estimated selection coefficients from datasets restricted to time intervals included in November - December 2020 (Analyses G, H, I, J) are about ten times higher than those over the whole datasets, until April 2021 (Analyses A, B, C, D, E, F in Sections~\ref{sec:estim.aff} and~\ref{sec:reduced.dataset}) as they are computed from data collected before and during the second lockdown in France. 

\section{Discussion and perspectives}\label{sec:perspectives}

We presented an unsupervised method for clustering mutation frequency trajectories and estimating group fitness from time series of SARS-CoV-2 genome sequences. Our method takes time series of SARS-CoV-2 sequencing data as input and returns an estimated number of non-neutral groups, group proportions, frequency at time origin and selection coefficient estimates associated to each group. Our method is suited for fragmented and pooled genomes of multiple lineage origin, typically found in WW samples. Although only tested on the difficult case of wastewater sample analysis, it could also be applied to an aggregation of clinical samples. 
We applied our method to publicly available WWTP datasets presented in \citep{barbe2022sars} and collected between October 2020 and April 2021. We demonstrated that our method highlights groups of mutations who's frequency trajectory estimates and in particular frequency at time origin and selection coefficient estimates are consistent with the observations of the authors regarding Alpha as well as VOC dynamics presented in Nextstrain reports for  B.1.1.7 (Alpha), B.1.160 and B.1.177. Moreover, restricting the analysis to a period of time shortly before and at the beginning of the emergence of Alpha (until 2020-11-17 for WWTP1 and 2020-12-18 for WWTP2) leads to the detection of a group of high positive selection coefficient. This group is mostly composed of Alpha VOC mutations detected by Barbé et al. \citep{barbe2022sars} at the same date. 

In summary, our results are consistent with those of \citet{barbe2022sars} and estimated variant frequencies in France reported by Nextstrain over the time period considered with the noticeable difference that our method is unsupervised, that is, it does not require any prior knowledge on the set of mutations contained in the dataset. It is therefore suited for detecting newly emerging variants. It is adapted to pooled fragmented genomes, hence particularly useful for WW samples although it can be applied to any pooled dataset. 

We also applied our algorithm over WWTP1 dataset covering the whole time period (until April 2021) restricted to mutations associated with a probability above 0.005\% to belong to at least one lineage among B.1.1.7 (Alpha), B.1, B.1.1, B.1.160, B.1.177, B.1.351 (Beta), B.1.367 in order to explore its performances in a context of limited proportion of mutations under no selection.
We showed its capacity to group  Alpha VOC mutations into two distinct groups according to their temporal emergence consistently with the time of their detection by \citet{barbe2022sars}, that is before versus after 2020-12-25. We also obtained groups of mutations which are consistent with B.1.1.7 (Alpha), B.1.160, B.1.177 and B.1.351(Beta) dynamics reported in Nextstrain over the time period in France. 

Our method however presents some limitations. Its main weakness is the lack of robust criteria, such as Bayesian criteria, for determining the number of groups to select. This task was however well performed over simulated datasets. Such limitation reflects its poor capacity in capturing residual variability to which SARS-CoV-2 variants dynamics and WWTP data are particularly prone. Along with the variance inherent to multinomial and binomial distributions, the only additional variability lays in the beta-binomial distribution of neutral mutations counts. We indeed assumed a generalized linear model with a binomial family with fixed effect (constant parameters) for modeling the distribution of mutation counts conditional on a non-neutral group assignment. We would like to relax such assumption with a mixed effect model composed of a random intercept and/or random selection coefficient. 

We also assumed constant selection coefficients restricting our method to limited periods of time, which can be sequentially repeated. That limitation leads to ignoring break points in frequency trajectories which characterizes the emergence of a mutation conferring a selective advantage. We therefore consider to further develop the model with piecewise constant selection coefficients and break point detection.

This development leads to a natural extension of our work to a model based on a random walk.
As previously noted \citep{malaspinas2012estimating, ferrer2016approximate, paris2019inference}, a common way for modeling an evolutionary process is a hidden Markov model with an underlying Wright-Fisher diffusion process. These methods are however computationally intensive.  
We ignore, in the present work, part of the temporal structure of time series data and we would like to extend our model with an underlying hidden Markov model composed of Gaussian latent variables denoted $X_{i,t}$ and observed mutation counts denoted $Y_{i,t}$ for $i \in \{1,\ldots,n\}$ and for $t\in \mathcal T$  such that, for $i\in\{1,\ldots,n\}$, $Z_i$ follows a multinomial distribution of parameter $\pi$, $X_{i,0}$ follows a probability distribution taking its values in $\mathbb R$, for $t>0$, $\{X_{i,t} - X_{i,t-1} \,|\,Z_i=k \}  \sim \mathcal N(s_k \Delta t, \sigma \Delta t)$ where $\mathcal N(\mu,\sigma)$ is the Gaussian distribution of mean $\mu$ and variance $\sigma$ and $\Delta t$ is the difference between $t$ and $t-1$ and $\{Y_{i,t} \,|\, X_{i,t} = x \}  \sim \mathcal B \left(d_{i,t}, \frac{e^x}{1+e^x}\right)$. 
Note finally that WWTP datasets are highly fragmented but still partly contain haplotype information that would be a valuable input to be taken into account in future developments.

\section*{Authors contributions}

\noindent $\bullet$ Alexandra Lefebvre: Conceptualization, Methodology (designed mathematical model, discussed and revised mathematical model, selected statistical methods, discussed and revised statistical methods), Algorithm (designed and implemented algorithm), Data preparation (prepared data provided by Obepine), Analysis (performed simulations, analyzed simulated and real data), Writing (wrote original draft), Review (discussed and commented manuscript).

\noindent $\bullet$ Vincent Maréchal: Conceptualization, Funding acquisition, Project administration, Supervision, Review (discussed and commented manuscript).

\noindent $\bullet$ Arnaud Gloaguen: Data (helped in data collection, produced mutation profile matrix used for analyses), Review (discussed and commented manuscript). 

\noindent $\bullet$ Amaury Lambert: Conceptualization, Methodology (designed mathematical model, discussed and revised mathematical model), Supervision, Review (commented and reviewed manuscript).

\noindent $\bullet$ Yvon Maday: Conceptualization, Methodology (discussed and revised mathematical model, discussed and revised statistical methods), Funding acquisition, Project administration, Supervision, Review (commented and reviewed manuscript).

\section*{Acknowledgements}

This study was supported in part by the ANRS-MIE (EmerEaUde project ANRS0160), together with the french state aid managed by the Agence nationale de la recherche under the France 2030 programme, reference ANR-24-MIEM-0004 and has benefited from the environment of the multidisciplinary scientific group of interest GIS-OBEPINE (Isabelle Bertrand, Mickaël Boni, Christophe Gantzer, Soizick Le Guyader, Yvon Maday, Vincent Maréchal, Jean-Marie Mouchel, Laurent Moulin).

We would like to thank Laure Barbé and Marion Desdouit for their help in data collection, preprocess and for providing us aligned files. We warmly thank Marie Courbariaux, El  Hacene  Djaout and Nicolas Cluzel for their help in data preparation and their guidance in the use of VaRaPS package. We deeply thank Stéphane Robin and Grégory Nuel for their generous advises and for sharing their knowledge in computational statistics. We also would like to thank Sébastien Wurtzer and Philippe Lopez for their help and explanations regarding data preparation and variant calling as well as Samuel Alizon for his precious advises and discussions on viral evolution and epidemiology. 

\section*{Competing interests}

The authors declare no competing interests. 

\bibliographystyle{unsrtnat}    
\bibliography{library}

\appendix

\section{Proof of Equation (1)}\label{sec:app.proof.eq1}

We have 
\begin{align*}
\mathbb P(Z_i=0 | X_i, \theta) & \propto \pi_0 \int_{0}^1 \prod_{t=0}^T \mathbb P(X_{i,t}|u, Z_i=0) \mathbb P(u) \ du \\
& \propto \pi_0 \int_{0}^1 \prod_{t=0}^T {d_{i,t} \choose x_{i,t}} u^{x_{i,t}} (1-u)^{d_{i,t}-x_{i,t}} \mathbb P(u) \ du
\end{align*}
where we recall that $\mathbb P(Z_i=0)=\pi_0$ and $u\sim\text{Beta}(\alpha,\beta)$, $\alpha, \beta > 0$, where $\text{Beta}$ is the beta distribution. Returning to the definition of the beta distribution, we have
$$
\mathbb P(u) = C_{\alpha,\beta} \ u^{\alpha-1} (1-u)^{\beta-1} \quad \text{where} \quad C_{\alpha,\beta} = \left(\int_0^1 u^{\alpha-1} (1-u)^{\beta-1} \ du\right)^{-1} = \left(\text{B}(\alpha,\beta)\right)^{-1}
$$
where $\text{B}$ is the beta function. 

Let $A = \int_{0}^1 \prod_{t=0}^T \mathbb P(X_{i,t}|u, Z_i=0) \mathbb P(u) \ du$, we therefore have 

\begin{align*}
A & = C_{\alpha,\beta} \prod_{t=0}^T {d_{i,t} \choose x_{i,t}} \int_0^1 u^{\sum_{t=0}^T x_{i,t} + \alpha - 1} (1-u)^{\sum_{t=0}^T d_{i,t} - x_{i,t} + \beta - 1} \ du \\
A & = C_{\alpha,\beta} \prod_{t=0}^T {d_{i,t} \choose x_{i,t}} \text{B}\left(\sum_{t=0}^T x_{i,t} + \alpha, \sum_{t=0}^T d_{i,t}-x_{i,t}+\beta\right).
\end{align*}

Recalling that $\text B(\alpha,\beta) = \frac{\Gamma(\alpha)\Gamma(\beta)}{\Gamma(\alpha+\beta)}$, we obtain 
$$
A = \prod_{t=0}^T {d_{i,t} \choose x{i,t}}
 \frac{\Gamma(\sum_{t=0}^T x_{i,t}+\alpha)\Gamma(\sum_{t=0}^T d_{i,t} - x_{i,t}+\beta)}{\Gamma(\sum_{t=0}^T d_{i,t} +\alpha+\beta)}
  \frac{\Gamma(\alpha+\beta)}{\Gamma(\alpha)\Gamma(\beta)}.
$$
We notice that 
\begin{align*}
\frac{\Gamma(\sum_{t=0}^T x_{i,t}+\alpha)}{\Gamma(\alpha)} & = \frac{1}{\cancel{\Gamma(\alpha)}}{\left(\sum_{t=0}^T x_{i,t}+\alpha-1\right) \left(\sum_{t=0}^T x_{i,t}+\alpha-2\right) \ldots \left(\alpha+1\right)\alpha \cancel{\Gamma(\alpha)}} \\
 & = \prod_{a=0}^{\sum_{t=0}^T x_{i,t}-1} (\alpha+a).
\end{align*}
Similarly we have 
$$
\frac{\Gamma(\sum_{t=0}^T d_{i,t} - x_{i,t}+\beta)}{\Gamma(\beta)}  = \prod_{b=0}^{\sum_{t=0}^T d_{i,t} - x_{i,t}-1} (\beta + b)
$$
and 
$$
\frac{\Gamma(\sum_{t=0}^T d_{i,t}+\alpha+\beta)}{\Gamma(\alpha+\beta)}  = \prod_{c=0}^{\sum_{t=0}^T d_{i,t} - 1} (\alpha + \beta + c) 
$$
which concludes the proof.

\section{R code for the EM algorithm}\label{sec:app.em}

\subsection{Packages and functions}

\begin{lstlisting}[language=R, backgroundcolor=\color{lightgray!22}, breaklines=true]
# required libraries 
library(VGAM)

# logsumexp avoids computational underflow 
logsumexp = function(x){
  i = which.max(x); 
  res = x[i] + log1p(sum(exp(x[-i] - x[i]))); 
  if (is.nan(res)) res = -Inf; 
  return(res)
}
\end{lstlisting}

\subsection{EM algorithm}

 \begin{lstlisting}[language=R, backgroundcolor=\color{lightgray!22}, breaklines=true]
em = function(x, d, K = NULL, eta = NULL, niter = 2000, tol = 1e-3) {

  # n: number of mutations 
  # m: number of sampling times (including the first one)
  # K: total number of groups minus one
  
  # time: vector of size m, differences (usually in days) between sampling  date and time origin. First vector entry is 0.
  
  # x: n.m matrix of mutation counts per time point
  # d: n.m matrix of read depth at related positions
  
  # eta: n.(K+1) matrix of posterior group assignments

  # pi: vector of size K+1, group proportions	
  # alpha and beta: single values, parameters of the beta distribution associated to the neutral group
  # mu: vector of size K, intercepts of non-neutral groups 
  # s: vector of size K, selection coefficients of non-neutral groups 
    
  # niter: number of iterations
  # tol: convergence tolerance for algorithm break
  
  if (K==0) { # no group under selection
    # alpha and beta estimates
    fit = vglm(cbind(apply(x,1,sum), apply(d-x, 1, sum)) ~ 1,family = betabinomialff())
    alpha = Coef(fit)[1] # Coef and not coef function
    beta = Coef(fit)[2] # Coef and not coef function
    
    # compute loglik
    loglik = 0
    for (i in 1:n) {
      res = sum(lchoose(d[i,],x[i,]))
      for (a in 0:(sum(x[i,])-1)) res = res + log(alpha + a)
      for (a in 0:(sum(d[i,]-x[i,])-1)) res = res + log(beta + a)
      for (a in 0:(sum(d[i,])-1)) res = res - log(alpha + beta + a)
      loglik = loglik + res
    }
    
    return(list(loglik = loglik, alpha = alpha, beta = beta))
    
  } else { # at least one group under selection
    
    # theta.tol introduced to check convergence
    theta.tol = rep(0, 2*K+2)
    
    # initialization
    if(is.null(eta)) { 
      # if no initial eta is provided
      # random initialization 
      eta = matrix(NA,n,K+1)
      for (i in 1:n) {tmp = runif(K+1); eta[i,] = tmp / (sum(tmp))}
    } else { 
      # otherwise retrieve K 
      K = ncol(eta)-1 
    }
    
    for (iter in 1:niter) {
      
      # M step
      ## update pi
      pi = apply(eta,2,sum) / n
      
      ## update alpha and beta
      weights = eta[,1];
      weights[weights==0]=1e-300 # avoid null weights
      fit = vglm(cbind(apply(x,1,sum), apply(d-x, 1, sum)) ~ 1,family = betabinomialff(), weights = weights)
      alpha = Coef(fit)[1] 
      beta = Coef(fit)[2] 
  
      ## update mu and s
      df = data.frame(x = rep(c(t(x)), K), d = rep(c(t(d)), K), time = rep(time, n*K), z = as.factor(rep(1:K, each = n*m)))
      weights = NULL
      for (k in 1:K) weights = c(weights, rep(eta[,k+1], each = m))
      if(K>1) {
        fit = glm(cbind(x, d - x) ~ 0 + z + time:z, "binomial", df, weights)
      } else {
        fit = glm(cbind(x, d - x) ~ 1 + time, "binomial", df, weights)
      }
      mu = coefficients(fit)[1:K]
      s = coefficients(fit)[-(1:K)]

      # E step
      ## eta matrix and log-likelihood
      tmp = matrix(NA,n,K+1)
      for (i in 1:n) {
        # neutral group
        res = sum(lchoose(d[i,],x[i,]))
        for (a in 0:(sum(x[i,])-1)) res = res + log(alpha+a)
        for (a in 0:(sum(d[i,]-x[i,])-1)) res = res + log(beta+a)
        for (a in 0:(sum(d[i,])-1)) res = res - log(alpha + beta + a)
        tmp[i,1] = res + log(pi[1])

        # non-neutral groups
        for (k in 1:K) {
          prob = exp(mu[k]+s[k]*time)
          prob = prob/(1+prob)
          tmp[i,k+1] = sum(dbinom(x[i,], d[i,], prob, log=TRUE)) + log(pi[k+1])
        }
      }
      aux = apply(tmp, 1, logsumexp)
      ## log-likelihood
      loglik = sum(aux) 
      ## eta
      eta = exp(tmp - aux)
  
      #check convergence
      if(max(abs(theta.tol - c(mu, s, alpha, beta)) / abs(c(mu, s, alpha, beta))) < tol) break;
      theta.tol = c(mu,s, alpha, beta) 
    }
    
    return(list(loglik = loglik, pi = pi, mu = mu, s = s, alpha = alpha, beta = beta, eta = eta)) 
  }
  
\end{lstlisting}

\subsection{EM algorithm applied on the alternative model for initialization}
In the alternative model we assume that all groups are non-neutral (positively of negatively selecte) such that each group is associated with an intercept and a selection coefficient.  

\begin{lstlisting}[language=R, backgroundcolor=\color{lightgray!22}, breaklines=true]
em.init = function(x, d, K = NULL, eta = NULL, niter = 2000, tol = 1e-3) {
  
  # n: number of mutations 
  # m: number of sampling times 
  # K: number of groups minus one 
  
  # time: vector of size m, differences (usually in days) between sampling  date and time origin. First vector entry is 0.
  
  # x: n.m matrix of mutation counts per time point
  # d: n.m matrix of read depth at related positions

  # eta: n.(K+1) matrix of posterior group assignments
  
  # pi: vector of size K+1, group proportions
  # mu: vector of size K, intercepts 
  # s: vector of size K, selection coefficients
  
  # niter: number of iterations
  # tol: convergence tolerance for algorithm break

 # initialization
    if(is.null(eta)) { 
      # if no initial eta is provided
      # random initialization 
      eta = matrix(NA,n,K+1)
      for (i in 1:n) {tmp = runif(K+1); eta[i,] = tmp / (sum(tmp))}
    } else { 
      # otherwise retrieve K 
      K = ncol(eta)-1 
    }
        
  # theta.tol introduced to check convergence 
  theta.tol = rep(0, 3*(K+1)) 
  
  for (iter in 1:niter) {
    
    # M step
    ## update pi
    pi = apply(eta,2,sum) / n

    ## update mu and s
    df = data.frame(
      x = rep(c(t(x)), K+1), 
      d = rep(c(t(d)), K+1), 
      time = rep(time, n*K+1), 
      z = as.factor(rep(0:K, each = n*m)) # group affectation
    )
    weights = NULL
    for (k in 0:K) 
      weights = c(weights, rep(eta[,k+1], each = m))
    if(K > 0) {
      fit = glm(cbind(x, d - x) ~ 0 + z + time:z, "binomial", df, weights)
    } else {
      fit = glm(cbind(x, d - x) ~ 1 + time, "binomial", df, weights)
    }
    mu = coefficients(fit)[1:(K+1)]
    s = coefficients(fit)[-(1:(K+1))]
  
    # E step
    ## eta matrix and log-likelihood
    tmp = matrix(NA,n,K+1)
    for (i in 1:n) {
      for (k in 0:K) {
        prob = exp(mu[k+1]+s[k+1]*time)
        prob = prob/(1+prob)
        tmp[i,k+1] = sum(dbinom(x[i,], d[i,], prob, log = TRUE)) + log(pi[k+1])
      }
    }
    aux = apply(tmp, 1, logsumexp)
    ## log-likelihood
    loglik = sum(aux) 
    ## eta
    eta = exp(tmp - aux)
  
  #check convergence
   if(max(abs(theta.tol - c(pi,mu, s)) / abs(c(pi,mu, s))) < tol) break;
   theta.tol = c(pi,mu, s) 

  }
  return(list(loglik = loglik, pi = pi, mu = mu, s = s, alpha = alpha, beta = beta, eta = eta))
}
\end{lstlisting}

\end{document}

%% file: fig_data_ifremer.tex
\begin{figure}
\centering 
\begin{subfigure}{0.50\textwidth}
\includegraphics[width=\linewidth]{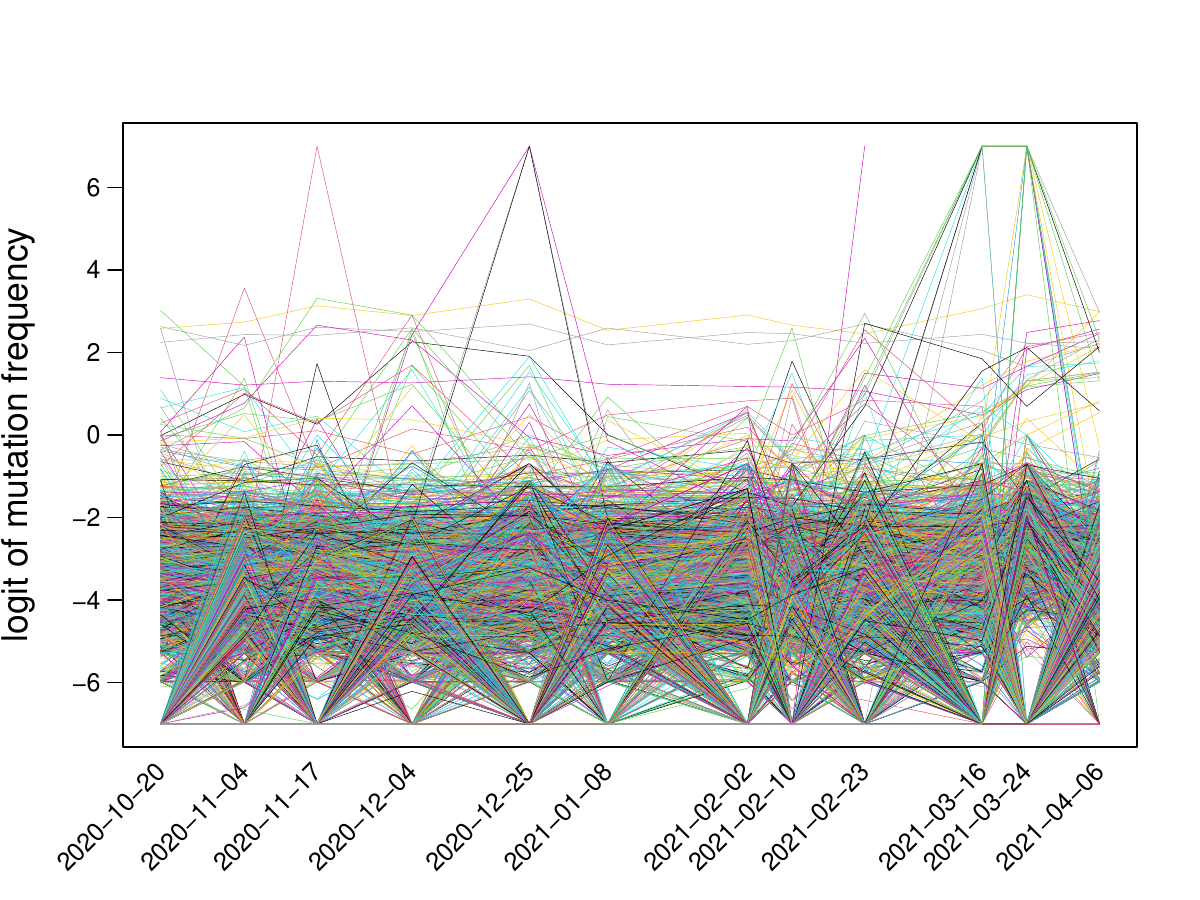}
\caption{WWTP1}
\end{subfigure}\hfil 
\begin{subfigure}{0.50\textwidth}
\includegraphics[width=\linewidth]{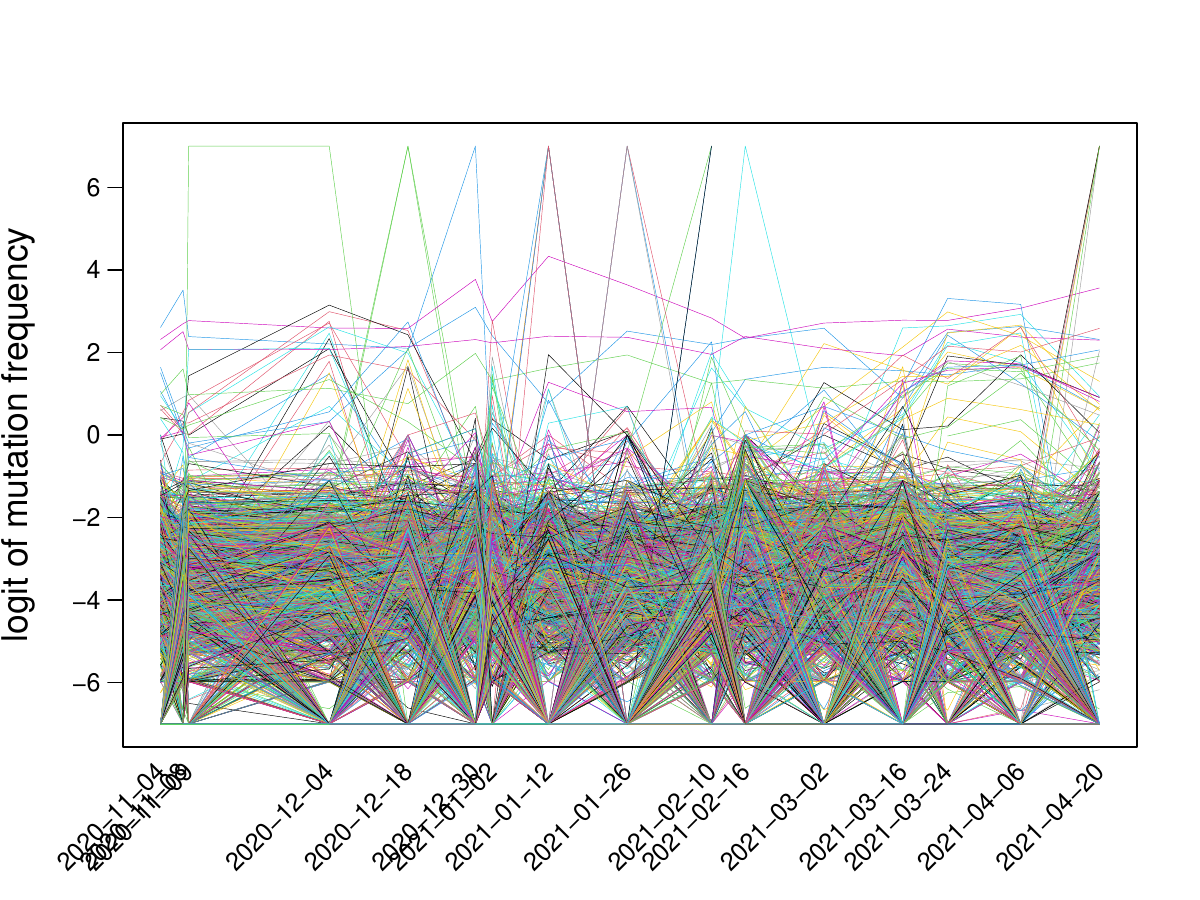}
\caption{WWTP2}
\end{subfigure}\hfil 
\caption{Logit of mutation frequency trajectories from times series of sequencing data collected from WWTP1 (left) and WWTP2 (right).
}
\label{fig:data.ifremer}
\end{figure}

%% file: tab_summary_simu.tex
\begin{table}
\centering
\setlength{\tabcolsep}{3pt}
\begin{tabular}{c|ccccccc}
Scheme&n&K&$\mathcal T$&$\pi$&$\mu$&$s$&$\lambda$\\
\hline
A&$100$&$1$&$(0, 5, 12, 20)$&vary&$-1.5$&vary&$40$\\
B&100&3&$(0, 5, 12, 20)$&vary&(1.5, -1, -2.5)&(-0.1, 0.05, 0.1)&40\\
C&vary&2&$(0, 5, 12, 20)$&(0.6, 0.3, 0.1)&(0.5, -3.0)&(-0.05, 0.10)&40\\
D&200&2&vary&(0.6, 0.3, 0.1)&(0.5, -3.0)&(-0.05, 0.10)&40\\
E&200&2&$(0, 5, 12, 20)$&(0.6, 0.3, 0.1)&(0.5, -3.0)&(-0.05, 0.10)&vary\\
\hline
\end{tabular}
\caption{Summary of quantities and parameters set for simulation schemes named in the first column. In all simulation schemes, parameters $\alpha$ and $\beta$ are set to 10 and 50 respectively.}
\label{tab:summary.simu}
\end{table}

%% file: tab_sel.tex
\begin{table}

\begin{subtable}{1\textwidth}
\centering
\begin{tabular}{r|cccc c|c cccc}
\multicolumn{1}{c}{} & \multicolumn{4}{c}{BIC}    &&& \multicolumn{4}{c}{ICL} \\
Number of groups & 0 & \textbf 1 & 2 & 3 &&& 0 & \textbf 1 & 2 & 3 \\
\hline
$s = 0.025$ & 29 & \bf 71 & 0 & 0 &&& 98 & \bf 2 & 0 & 0\\
$s = 0.050$ & 0 & \bf 100 & 0 & 0 &&& 16 & \bf 84 & 0 & 0\\
$s = 0.075$ & 0 & \bf 100 & 0 & 0 &&& 0 & \bf 100 & 0 & 0\\
$s = 0.100$ & 0 & \bf 99 & 1 & 0 &&& 0 & \bf 100 & 0 & 0\\
$s = 0.500$ & 0 & \bf 100 & 0 & 0 &&& 0 & \bf 100 & 0 & 0\\
\hline
\end{tabular}
\caption*{$\pi=(0.80,0.20)$}
\end{subtable}
    
\vspace{1em}

\begin{subtable}{1\textwidth}
\centering
\begin{tabular}{r|cccc c|c cccc}
\multicolumn{1}{c}{} & \multicolumn{4}{c}{BIC}    &&& \multicolumn{4}{c}{ICL} \\
Number of groups & 0 & \textbf 1 & 2 & 3 &&& 0 & \textbf 1 & 2 & 3 \\
\hline
\hline
$s = 0.025$ & 100 & \bf 0 & 0 & 0 &&& 100 & \bf 0 & 0 & 0\\
$s = 0.050$ & 75 & \bf 24 & 1 & 0 &&& 91 & \bf 9 & 0 & 0\\
$s = 0.075$ & 30 & \bf 67 & 2 & 1 &&& 38 & \bf 62 & 0 & 0\\
$s = 0.100$ & 6 & \bf 69 & 25 & 0 &&& 15 & \bf 68 & 17 & 0\\
$s = 0.500$ & 0 & \bf 95 & 4 & 1 &&& 1 & \bf 95 & 4 & 0\\
\hline
\end{tabular}
\caption*{$\pi=(0.95,0.05)$}
\caption{Simulation scheme A with $K=1$, $n=100$, $\mathcal T = (0, 5, 12, 20)$, $\mu=-1.5$, $\alpha = 10$, $\beta = 50$, $\lambda = 40$ and varying $s$ (in rows) and $\pi$ (one per table).}
\label{tab:selK1}
\end{subtable}

\hspace{2em}

\begin{subtable}{1\textwidth}
\centering
\setlength{\tabcolsep}{5pt}
\begin{tabular}{r|cccccc | cccccc}
\multicolumn{1}{c}{} & \multicolumn{6}{c}{BIC} & \multicolumn{6}{c}{ICL} \\
Number of groups & 0  & 1 & 2 & \bf 3 & 4 & 5 & 0 & 1 &2 & \bf 3 & 4 & 5\\
\hline
$\pi = (0.80, 0.08, 0.06, 0.06)$ & 0 & 5 & 60 & \bf 32 & 3 & 0 & 0 & 12 & 73 & \bf 15 & 0 & 0\\
$\pi = (0.60, 0.20, 0.10, 0.10)$  & 0 & 1 & 20 & \bf 79 & 0 & 0 &  0 & 1 & 52 & \bf 47 & 0 & 0\\
$\pi = (0.40, 0.20, 0.20, 0.20)$   & 0 & 0 & 0 & \bf 95 & 4 & 1 & 0 & 0 &  2 & \bf 94 & 3 & 1\\
\hline
\end{tabular}
\caption{Simulation scheme B with $K=3$, $n=100$, $\mathcal T = (0, 5, 12, 20)$, $\mu = (1.5, -1, -2.5)$, $s = (-0.1, 0.05, 0.1)$, $\alpha = 10$, $\beta = 50$, $\lambda = 40$ and varying $\pi$ (in rows).}
\label{tab:selK3}
\end{subtable}
\caption{Proportion of the number of non-neutral groups associated with the likelihood that minimizes the BIC (left) and ICL (right) criteria in simulation scheme A (Table~\ref{tab:selK1}) and simulation scheme B (Table~\ref{tab:selK3}) using various selection coefficients $s$ and/or group proportions $\pi$. The columns associated with the true number of groups (used for simulations) are in bold.}
\label{tab:sel}
\end{table}

%% file: fig_boxplot_n.tex
\begin{figure}
\centering 
\begin{subfigure}{0.50\textwidth}
\includegraphics[width=\linewidth]{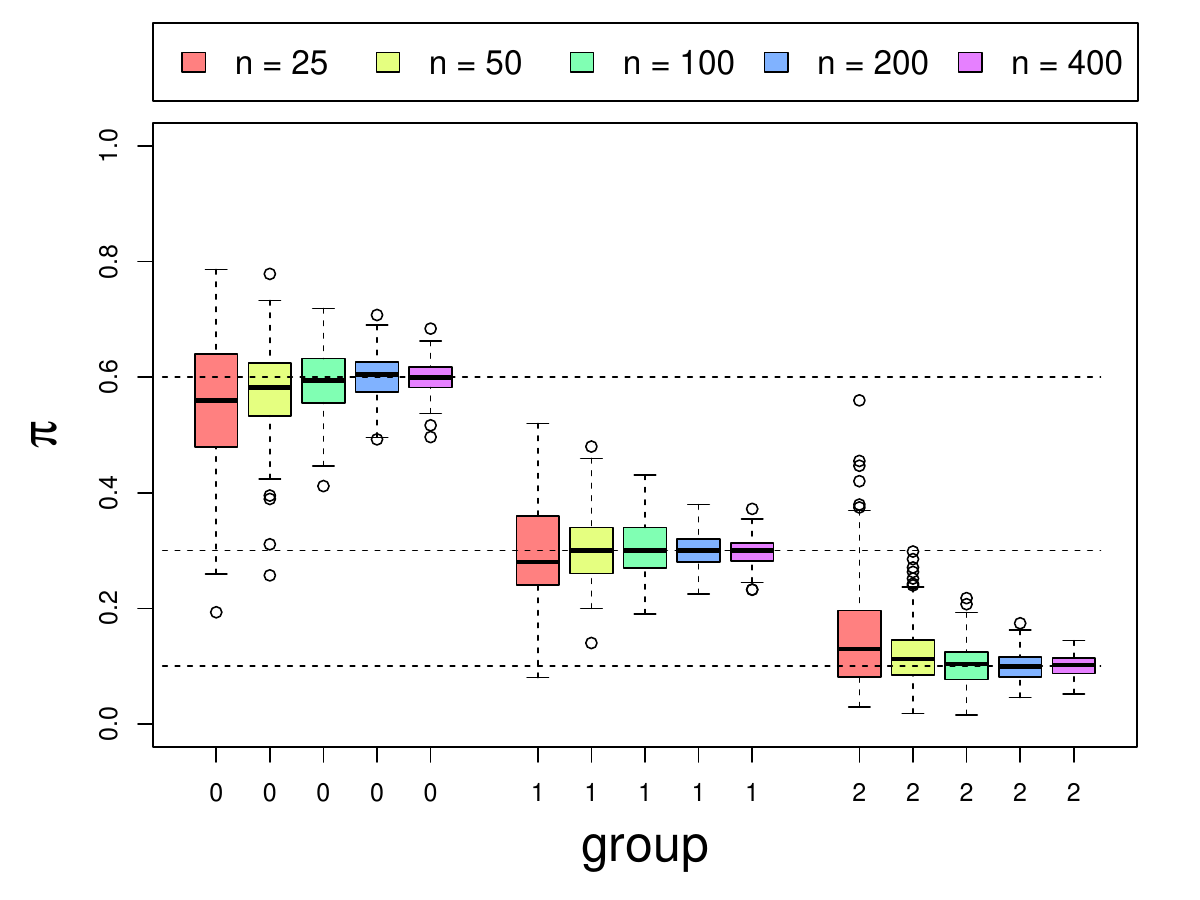}
\end{subfigure}\hfil 
\begin{subfigure}{0.50\textwidth}
\includegraphics[width=\linewidth]{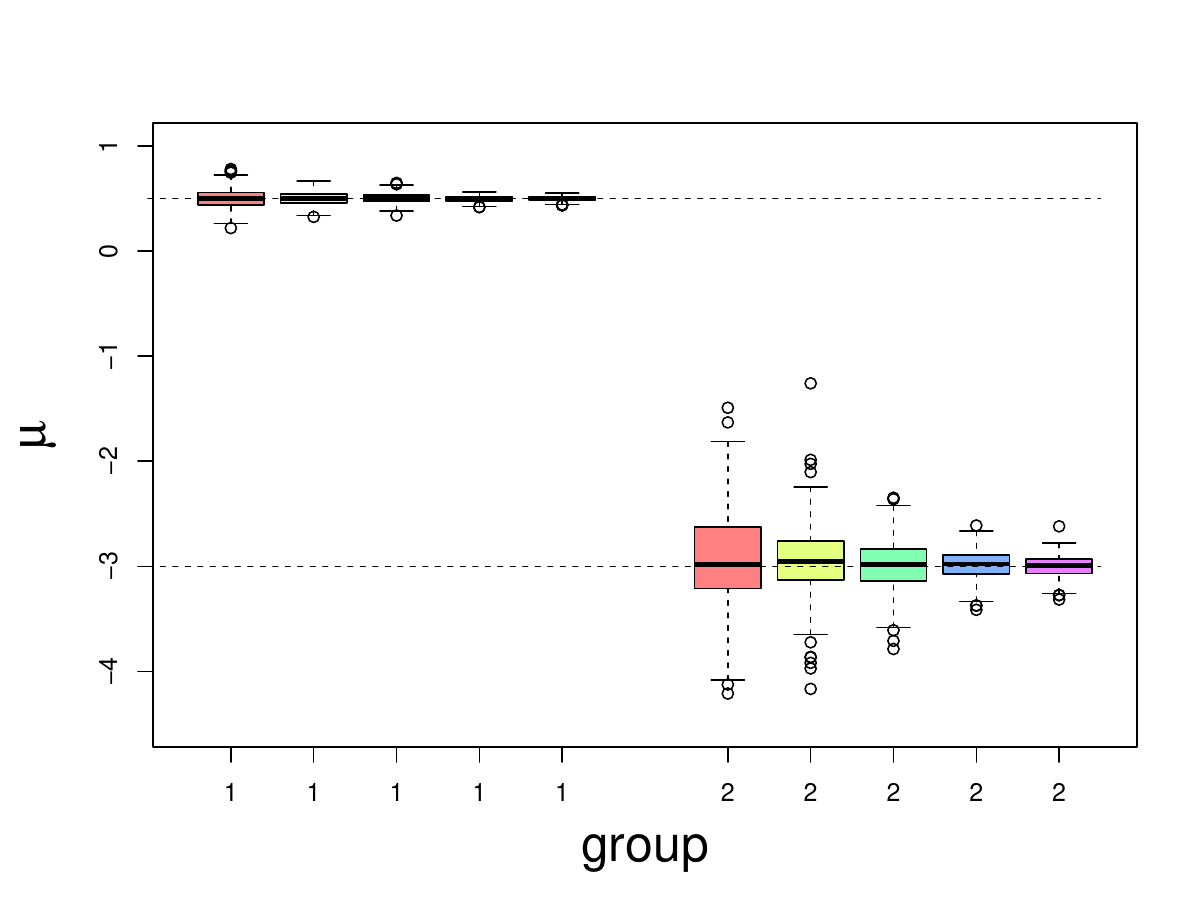}
\end{subfigure}\hfil 
\begin{subfigure}{0.50\textwidth}
\includegraphics[width=\linewidth]{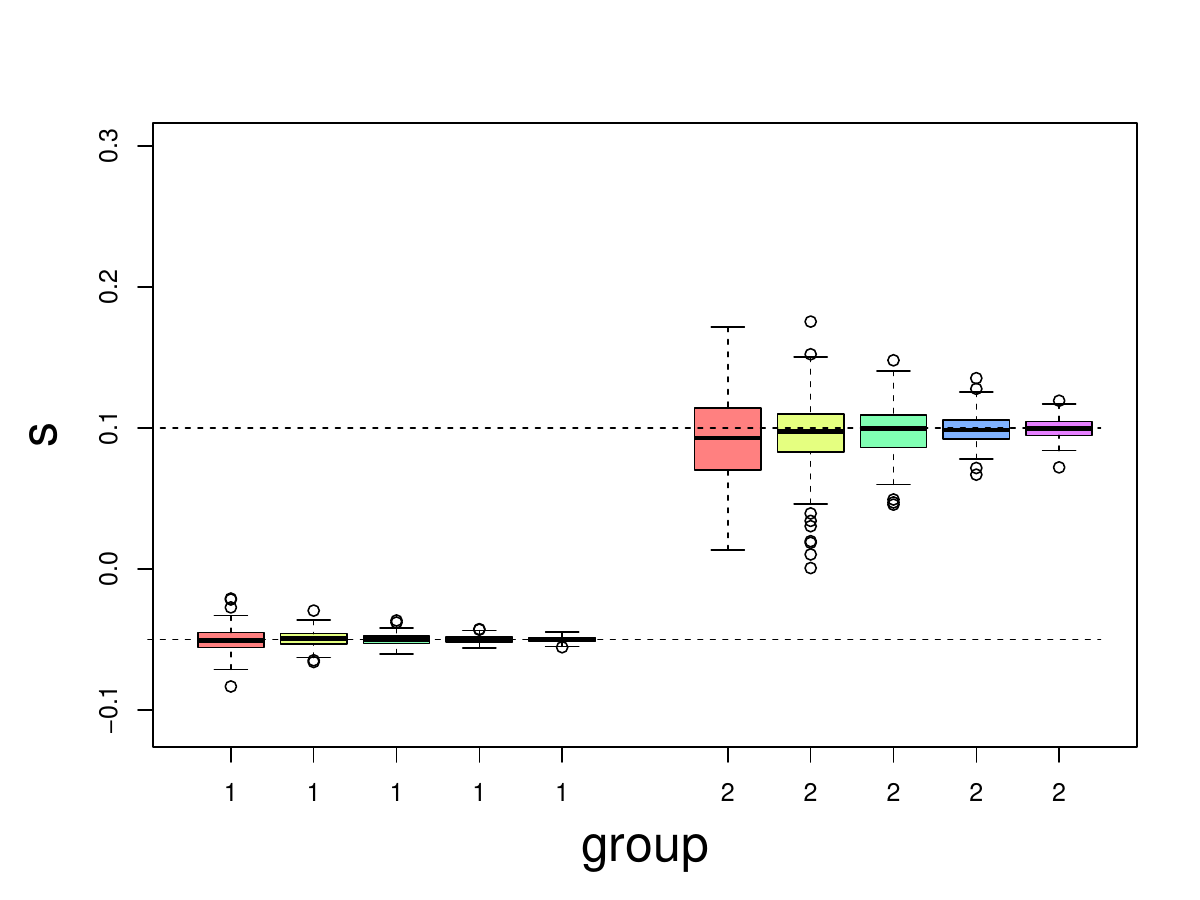}
\end{subfigure}\hfil 
\begin{subfigure}{0.50\textwidth}
\includegraphics[width=\linewidth]{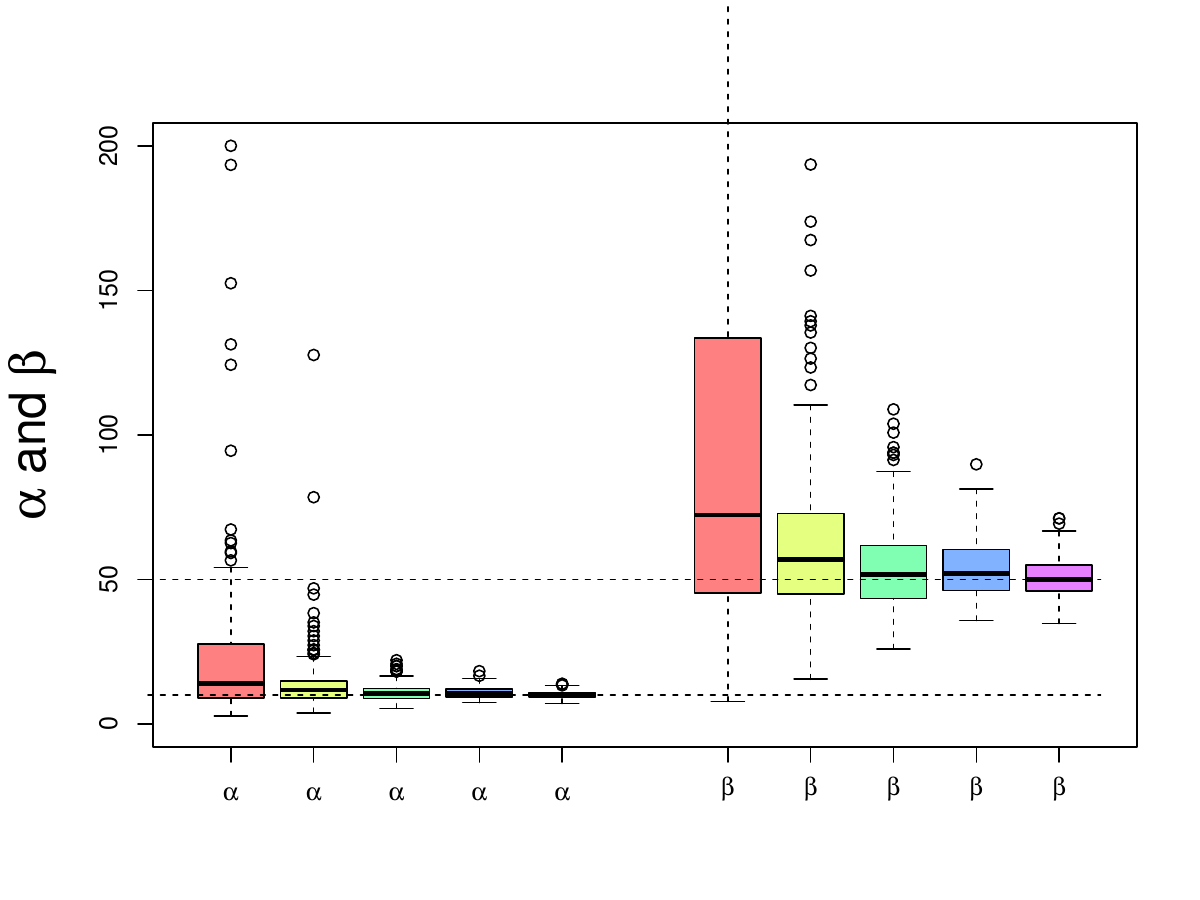}
\end{subfigure}\hfil 
\begin{subfigure}{0.50\textwidth}
\includegraphics[width=\linewidth]{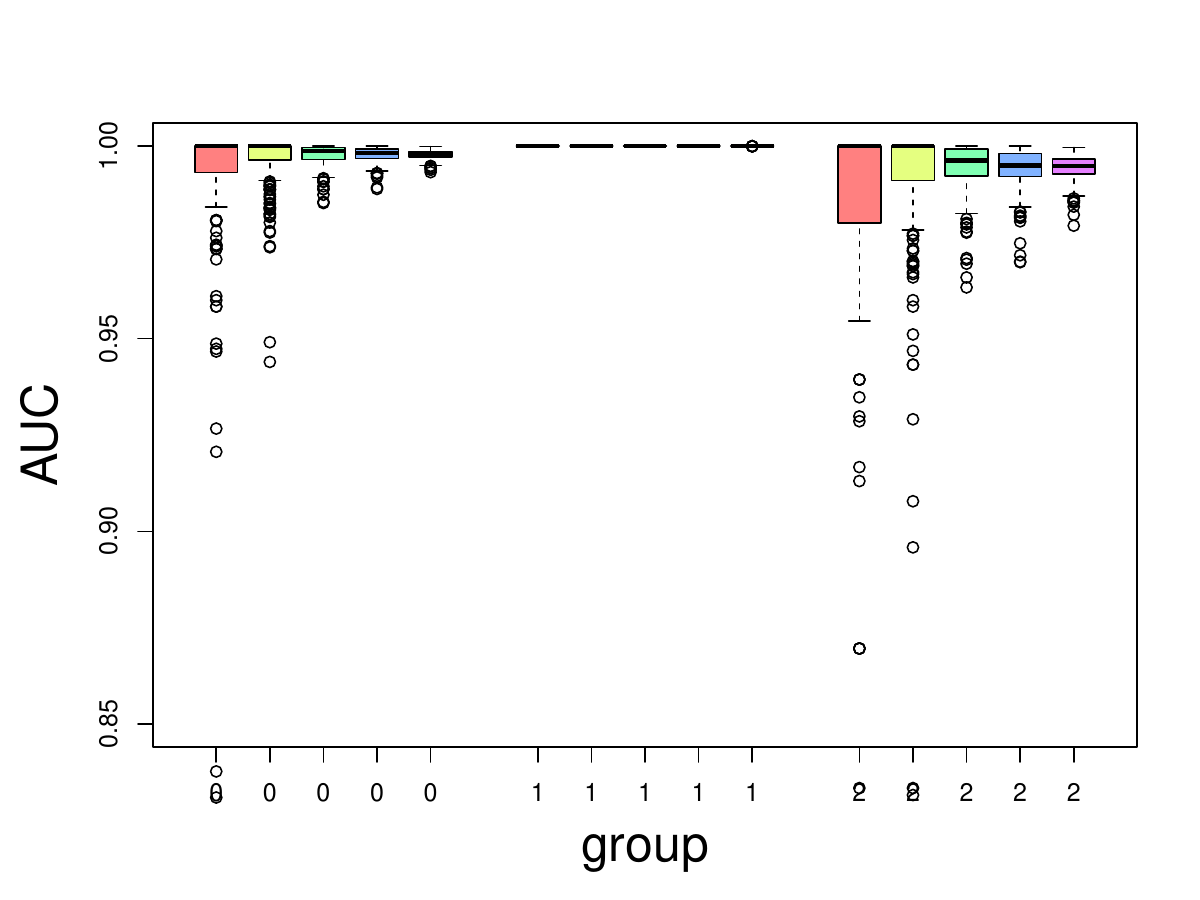}
\end{subfigure}\hfil 
\caption{Parameter estimates stratified on the number of mutations $n$ with 200 replications of simulation scheme C where $K=2$ non-neutral groups, $\mathcal T = (0, 5, 12, 20)$, $\pi = (0.6, 0.3, 0.1)$, $\alpha=10$, $\beta = 50$ and $\lambda = 40$. True parameters (used for simulations) are highlighted with horizontal dashed lines.}
\label{fig:estim.n}
\end{figure}

%% file: fig_boxplot_T.tex
\begin{figure}
\centering 
\begin{subfigure}{0.50\textwidth}
\includegraphics[width=\linewidth]{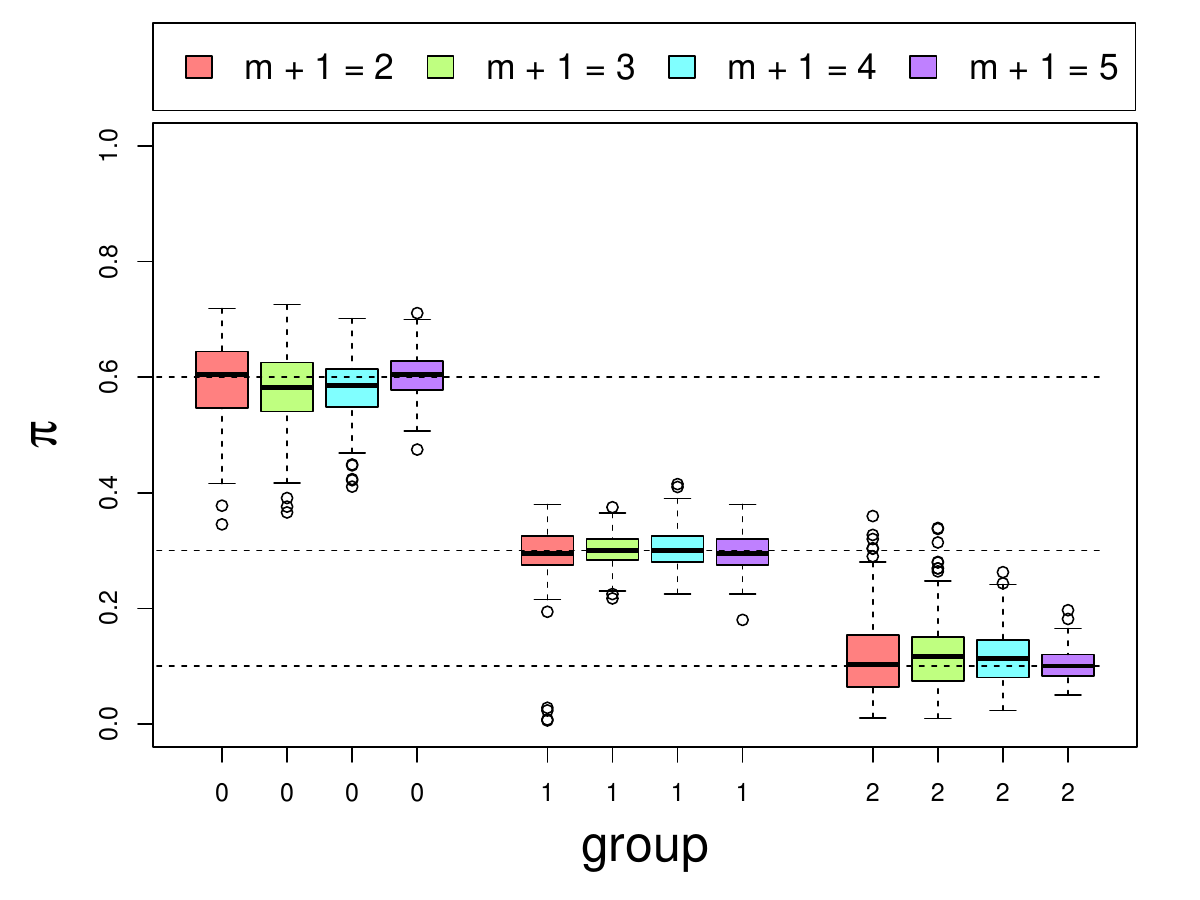}
\end{subfigure}\hfil 
\begin{subfigure}{0.50\textwidth}
\includegraphics[width=\linewidth]{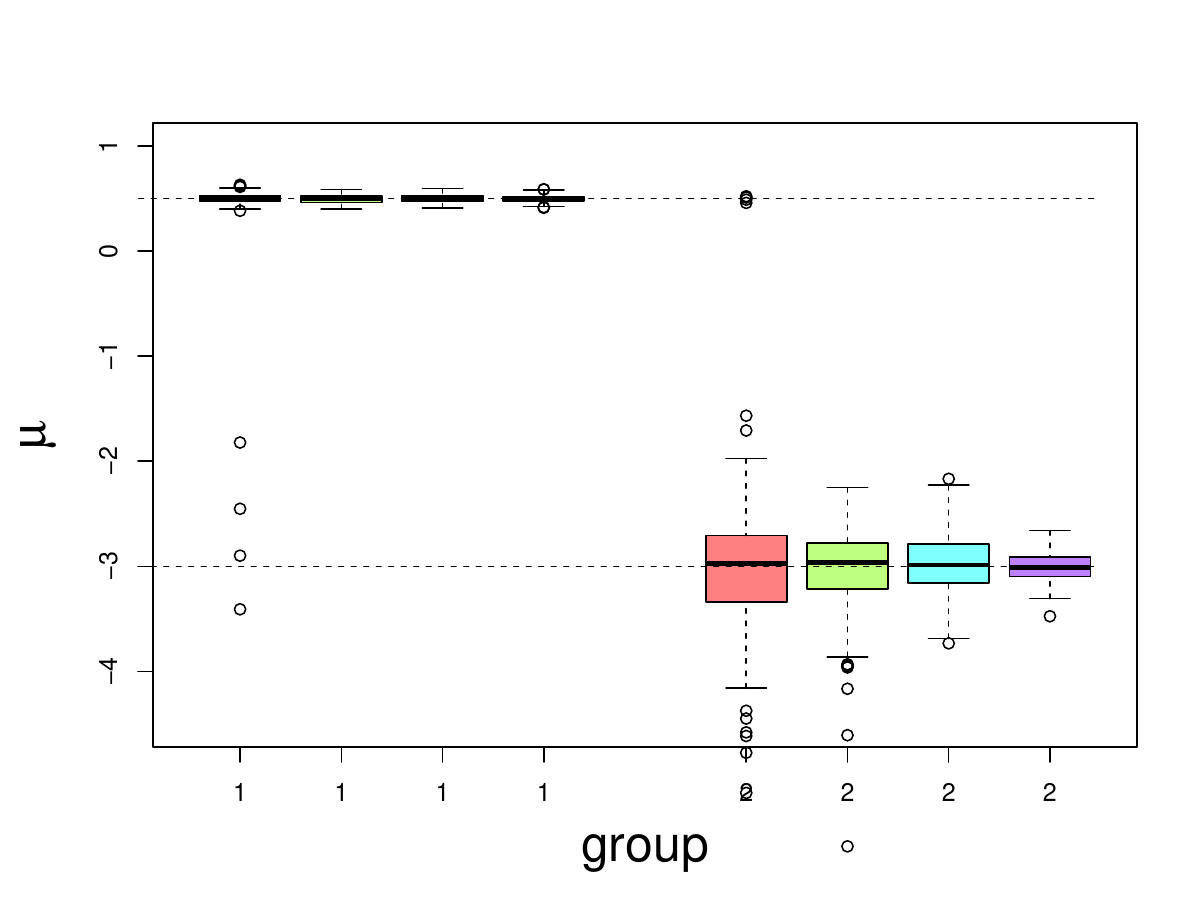}
\end{subfigure}\hfil 
\begin{subfigure}{0.50\textwidth}
\includegraphics[width=\linewidth]{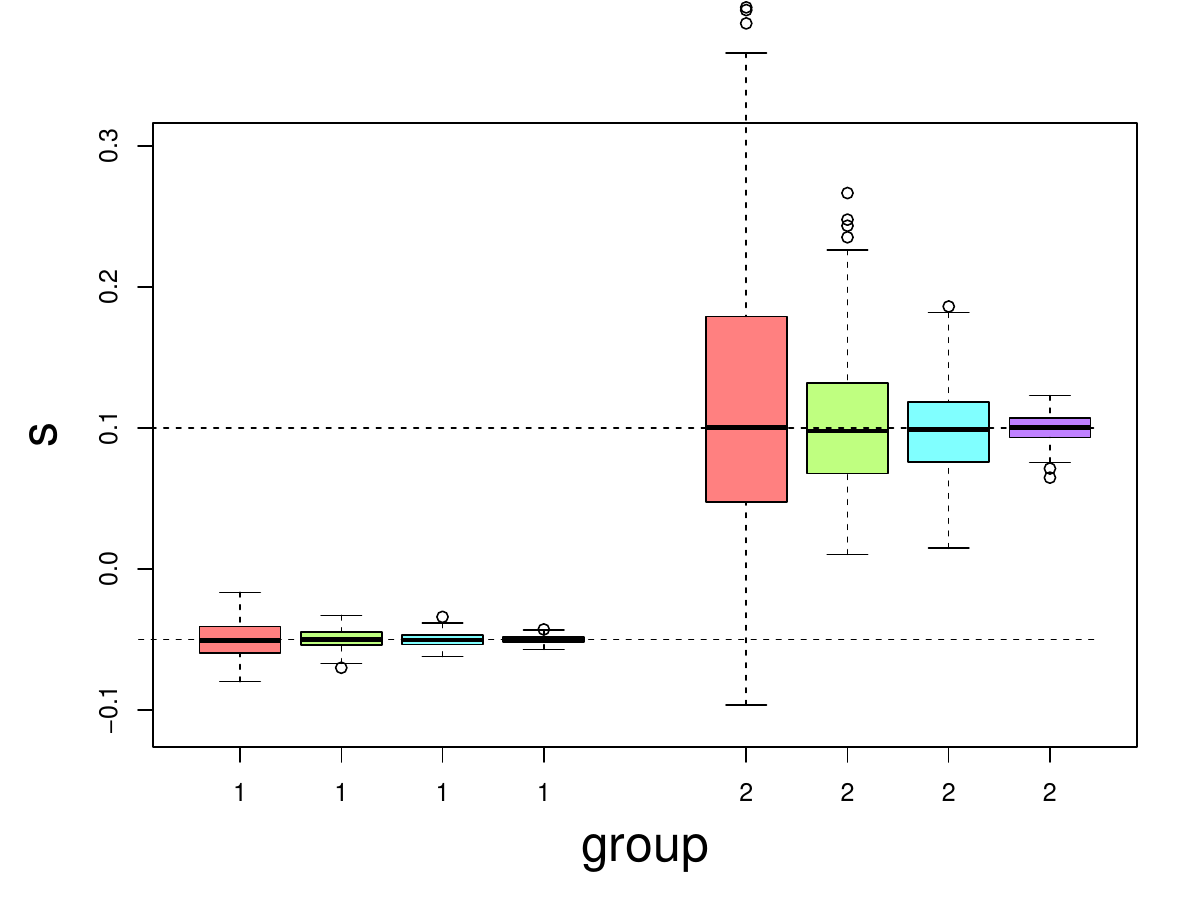}
\end{subfigure}\hfil 
\begin{subfigure}{0.50\textwidth}
\includegraphics[width=\linewidth]{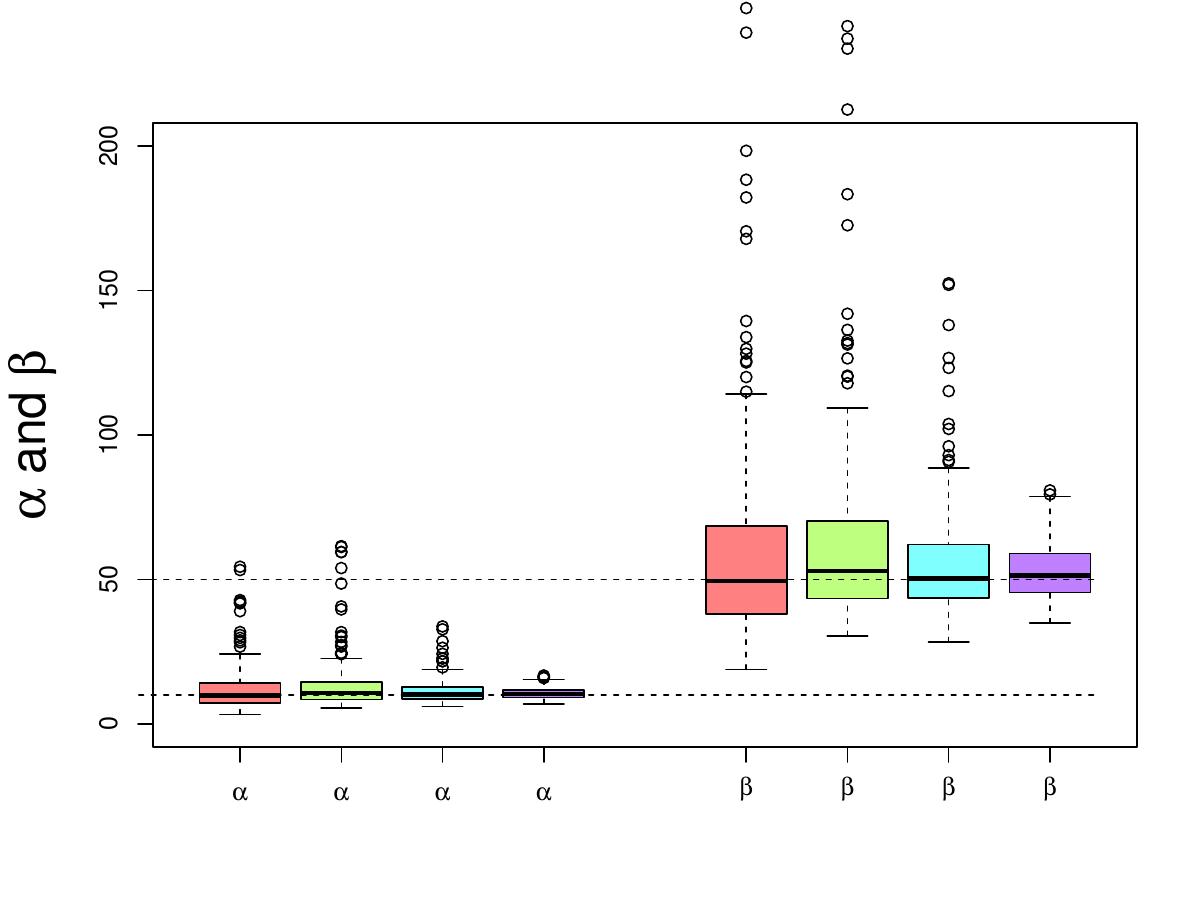}
\end{subfigure}\hfil
\begin{subfigure}{0.50\textwidth}
\includegraphics[width=\linewidth]{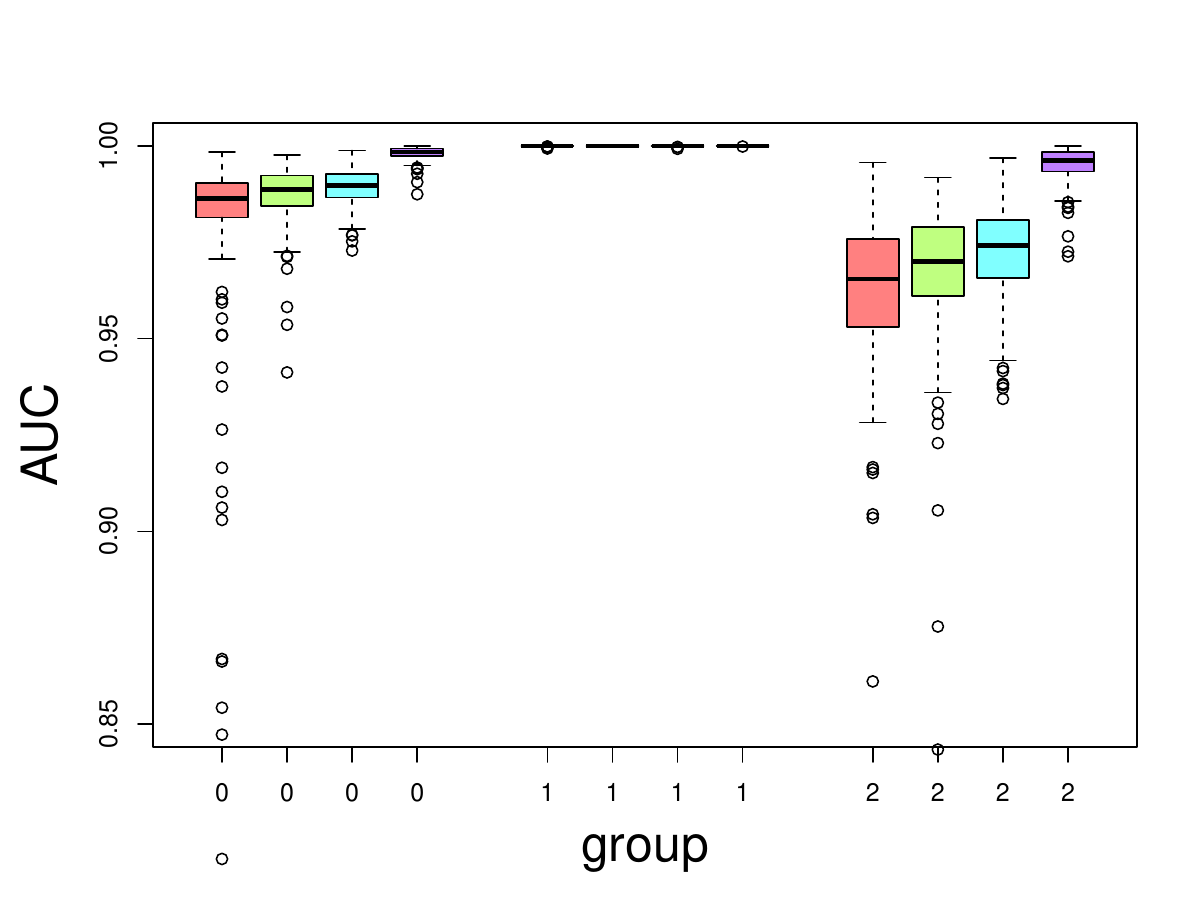}
\end{subfigure}\hfil
\caption{Parameter estimates stratified on the number of time points picked from the first to the $(m+1)^\text{th}$ value of $\mathcal T = (0, 5,9, 12, 20)$ with 200 replications of simulation scheme D where $K=2$ non-neutral groups, $n=200$ mutations, $\pi = (0.6, 0.3, 0.1)$, $\alpha=10$, $\beta = 50$ and $\lambda = 40$. True parameters (used for simulations) are highlighted with horizontal dashed lines.}
\label{fig:estim.T}
\end{figure}

%% file: fig_boxplot_d.tex
\begin{figure}
\centering 
\begin{subfigure}{0.50\textwidth}
\includegraphics[width=\linewidth]{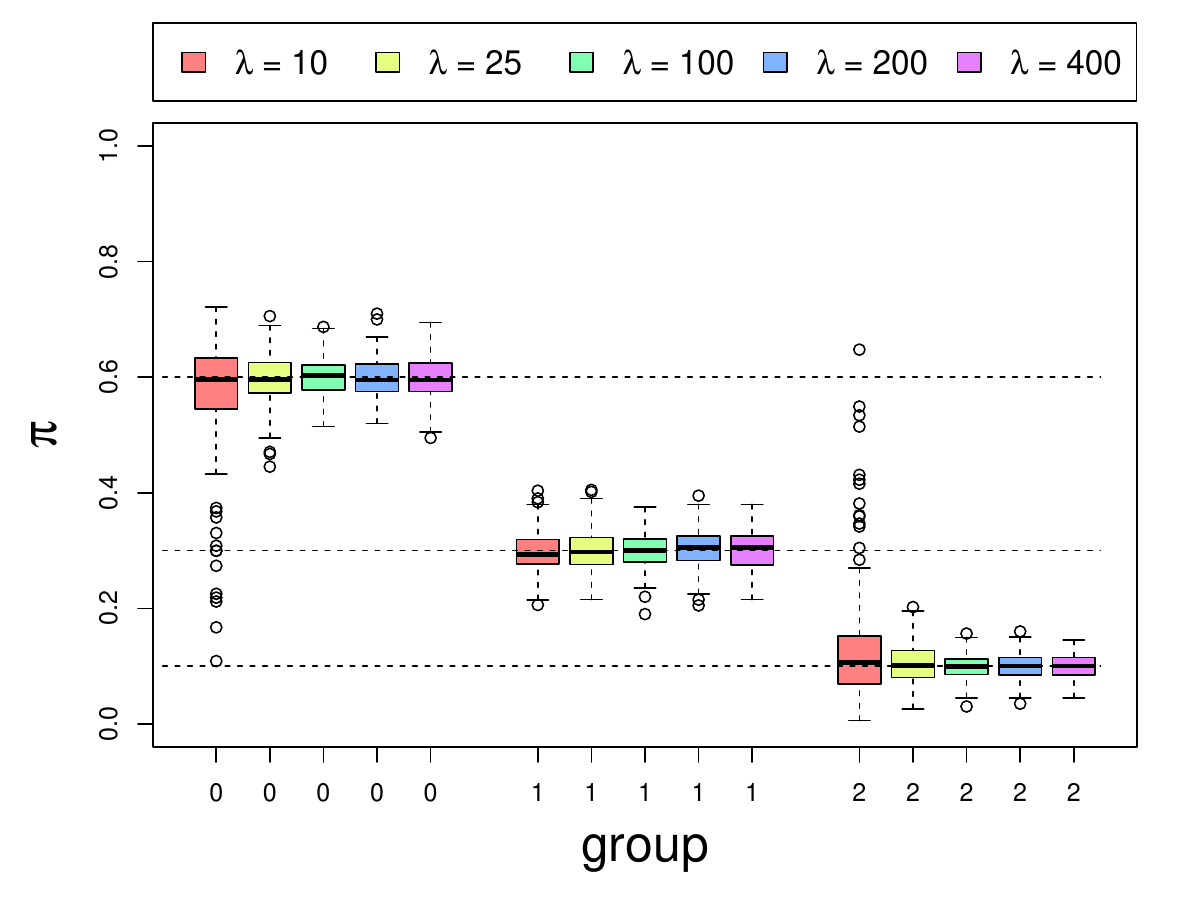}
\end{subfigure}\hfil 
\begin{subfigure}{0.50\textwidth}
\includegraphics[width=\linewidth]{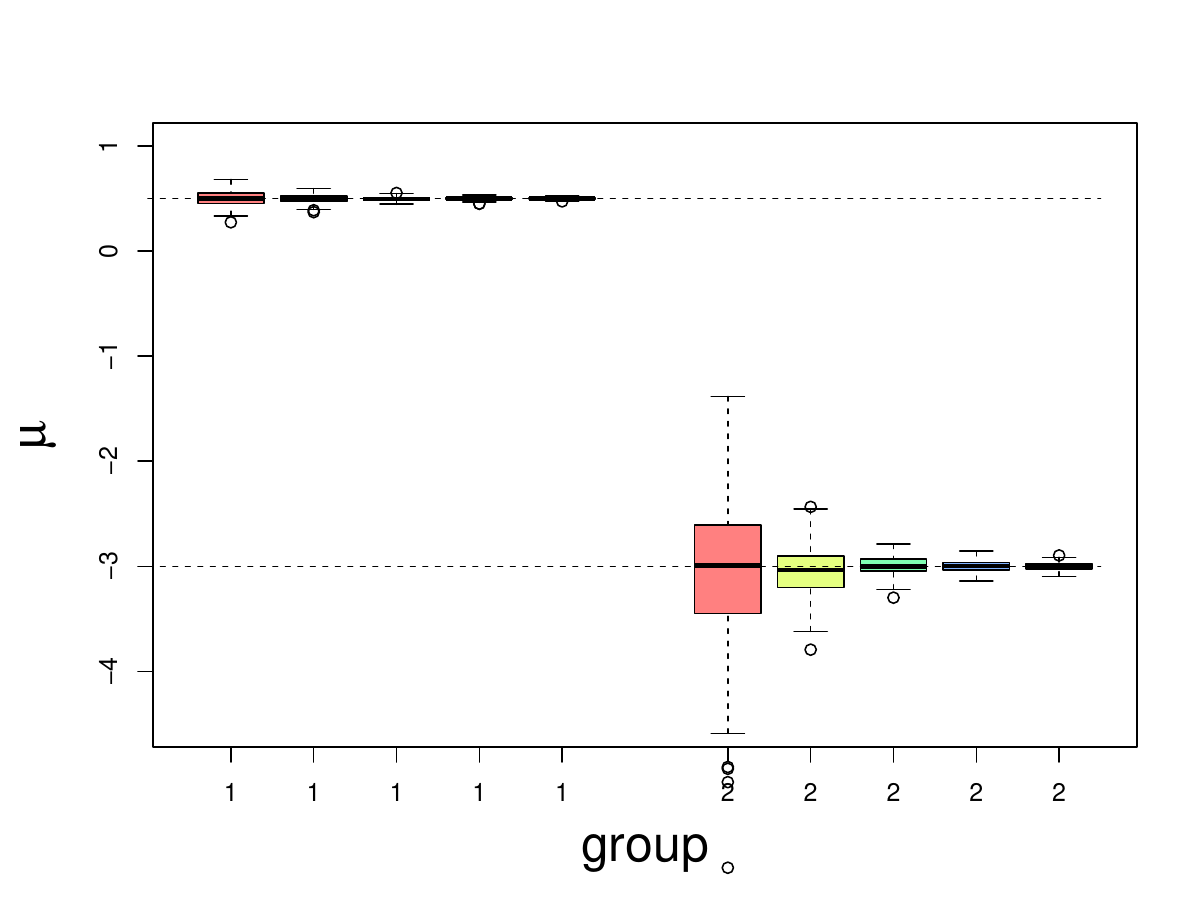}
\end{subfigure}\hfil 
\begin{subfigure}{0.50\textwidth}
\includegraphics[width=\linewidth]{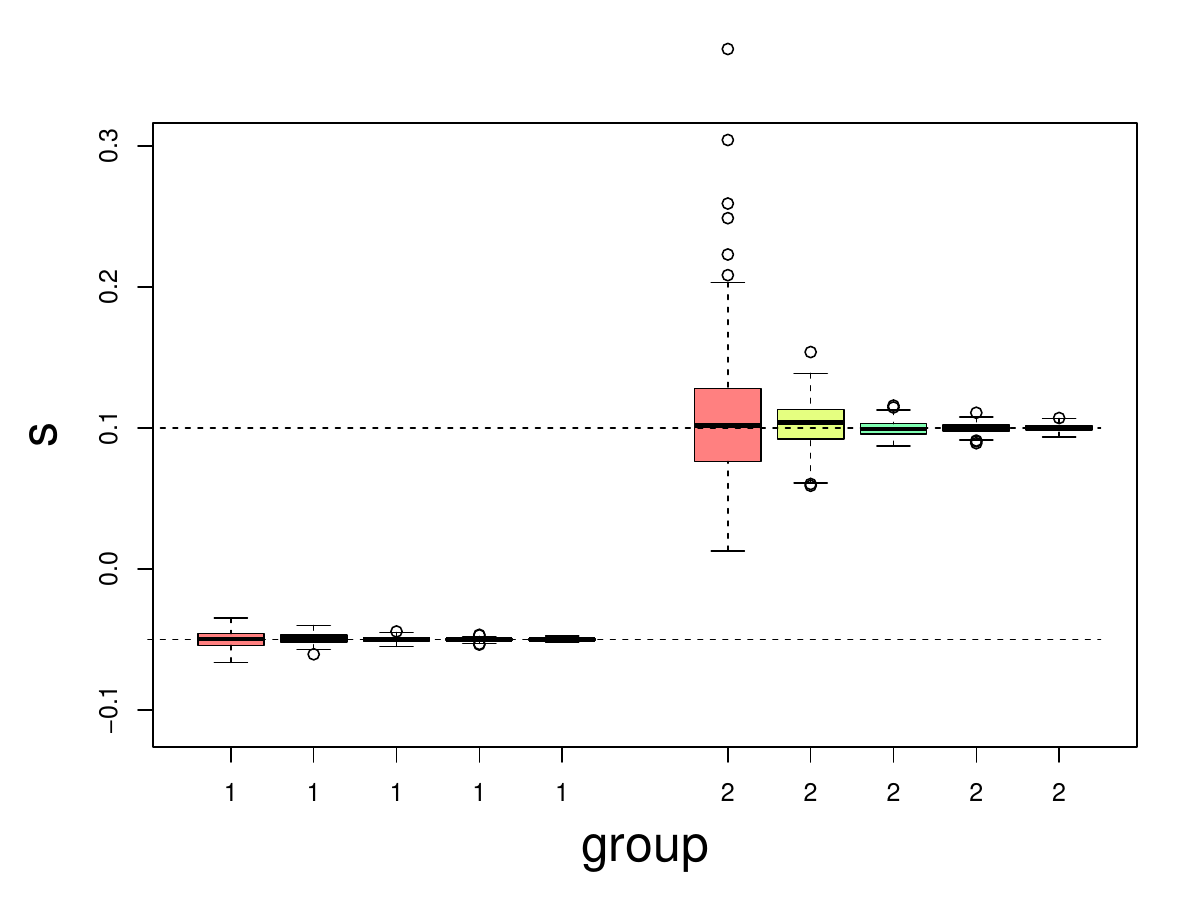}
\end{subfigure}\hfil 
\begin{subfigure}{0.50\textwidth}
\includegraphics[width=\linewidth]{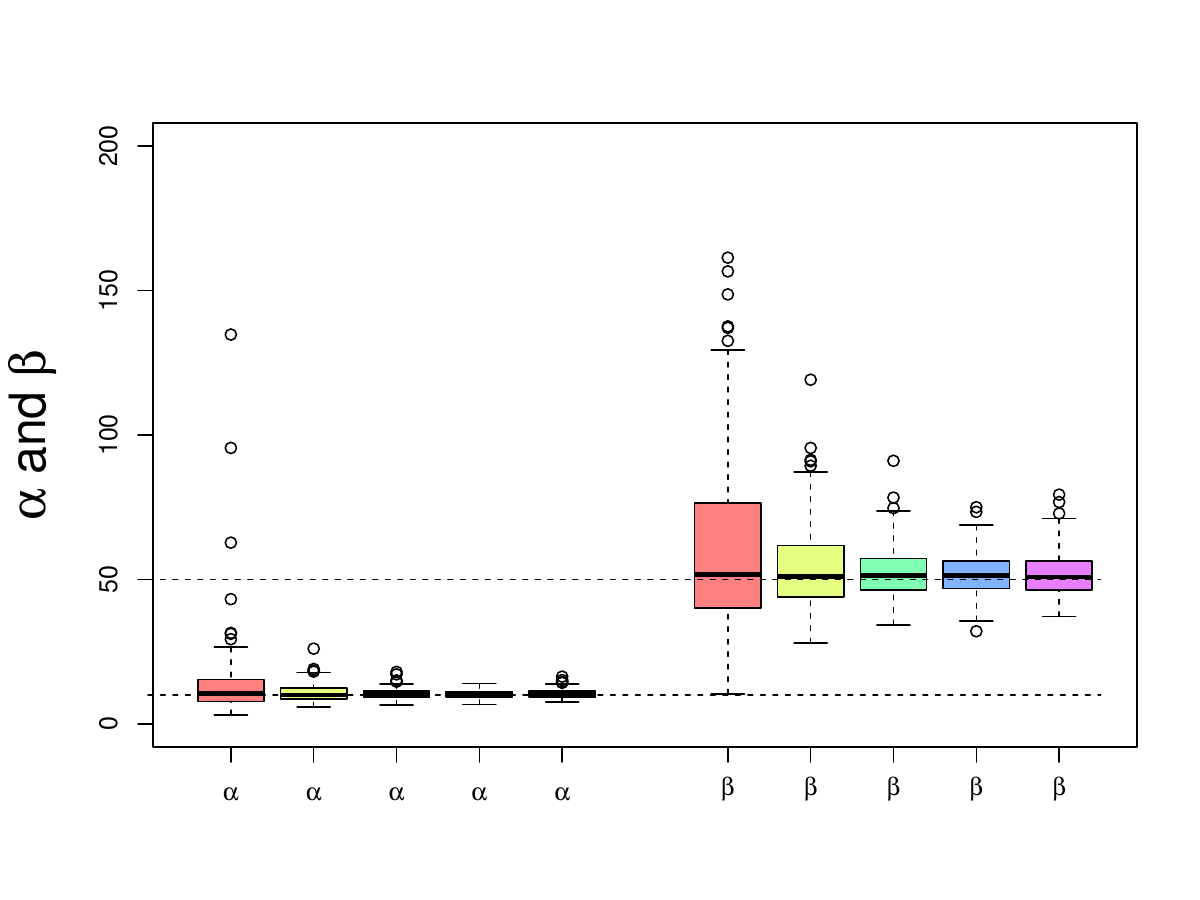}
\end{subfigure}\hfil 
\begin{subfigure}{0.50\textwidth}
\includegraphics[width=\linewidth]{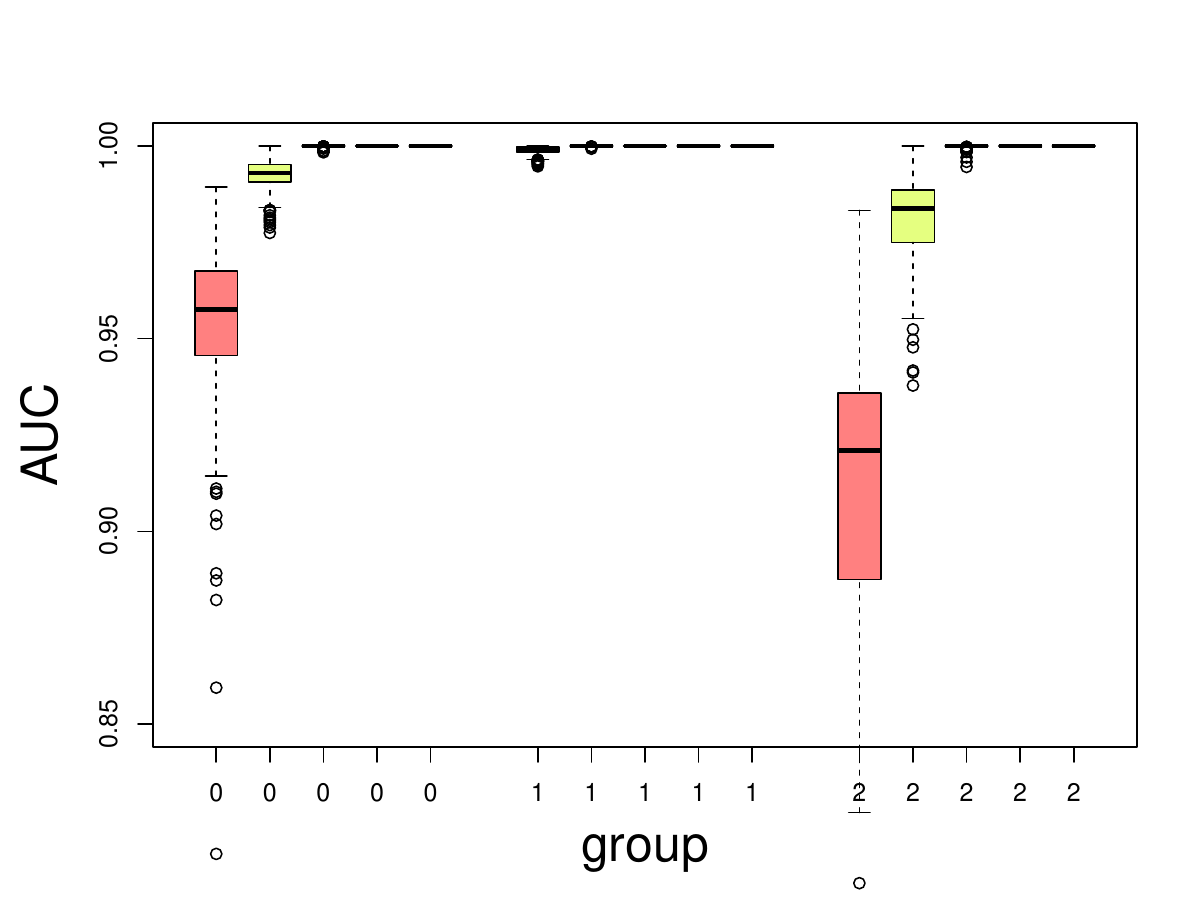}
\end{subfigure}\hfil 
\caption{Parameter estimates stratified on the value of the Poisson parameter $\lambda$ with 200 replications of simulation scheme E where $K=2$ non-neutral groups, $n=200$ mutations, $\mathcal T = (0, 5, 12, 20)$, $\pi = (0.6, 0.3, 0.1)$, $\alpha=10$ and $\beta = 50$. True parameters (used for simulations) are highlighted with horizontal dashed lines.}
\label{fig:estim.d}
\end{figure}

%% file: tab_summary_analyses.tex
\begin{table}
\centering
\setlength{\tabcolsep}{5.5pt}
\begin{tabular}{|c|c|c|c|c|}
\hline
Analysis & Dataset & time period & \# of groups & Figures \& Tables\\
\hline
A & WWTP1 & 2020-10-20& 4 & Figure~\ref{fig:sel}\\
&&2021-04-06&&Figure~\ref{fig:estim.w1k4}\\
&&&&Table~\ref{tab:freq}\\
&&&&\\
B & WWTP2 & 2020-11-04&3&Figure~\ref{fig:sel}\\
&&2021-04-20&&Figure~\ref{fig:estim.w2k3}\\
&&&&Table~\ref{tab:freq}\\
&&&&\\
C & WWTP2 & 2020-11-04&6& Figure~\ref{fig:sel}\\
&& 2021-04-20&&Figure~\ref{fig:estim.w2k6}\\
&&&&Table~\ref{tab:freq}\\
&&&&\\
D & reduced WWTP1 & 2020-10-20& 2 &Figure~\ref{fig:reduced.sel}\\
&&2021-04-06&&Figure~\ref{fig:reduced.estim.w1k2}\\
&&&&\\
E & reduced WWTP2 & 2020-11-04 & 2 &Figure~\ref{fig:reduced.sel}\\
&&2021-04-20&&Figure~\ref{fig:reduced.estim.w2k2}\\
&&&&\\
F & reduced WWTP1 & 2020-10-20 & 6 &Figure~\ref{fig:reduced.sel}\\
&&2021-04-06&&Figure~\ref{fig:reduced.estim.w1k6}\\
&&&&Table~\ref{tab:signature.w1}\\
&&&&\\
G & WWTP1 & 2020-10-20 & 3 &Figure~\ref{fig:detect.sel}\\
&&2020-11-04&&Figure~\ref{fig:detect.estim1}\\
&&&&\\
&&&&\\
H & WWTP1 &2020-11-04& 3 &Figure~\ref{fig:detect.sel}\\
&&2020-11-17&&Figure~\ref{fig:detect.estim1}\\
&&&&Table~\ref{tab:signature.detection}\\
&&&&\\
I & WWTP2 & 2020-11-09 & 3 &Figure~\ref{fig:detect.sel}\\
&&2020-12-04&&Figure~\ref{fig:detect.estim2}\\
&&&&\\
J & WWTP2 &2020-12-04& 2 &Figure~\ref{fig:detect.sel}\\
&&2020-12-18&&Figure~\ref{fig:detect.estim2}\\
&&&&\\
\hline
\end{tabular}
\caption{Summary of the analyses performed with their associated dataset, time period, estimated number of non-neutral groups and related Figures and/or Tables. A dataset is said to be reduced if restricted to mutations associated to a probability above 0.005 to belong to at least one of the main circulating lineages at the time of the study.}
\label{tab:summary.analyses}
\end{table}

%% file: fig_main_selection.tex
\begin{figure}
\centering 
\begin{subfigure}{0.45\textwidth}
\includegraphics[width=\linewidth]{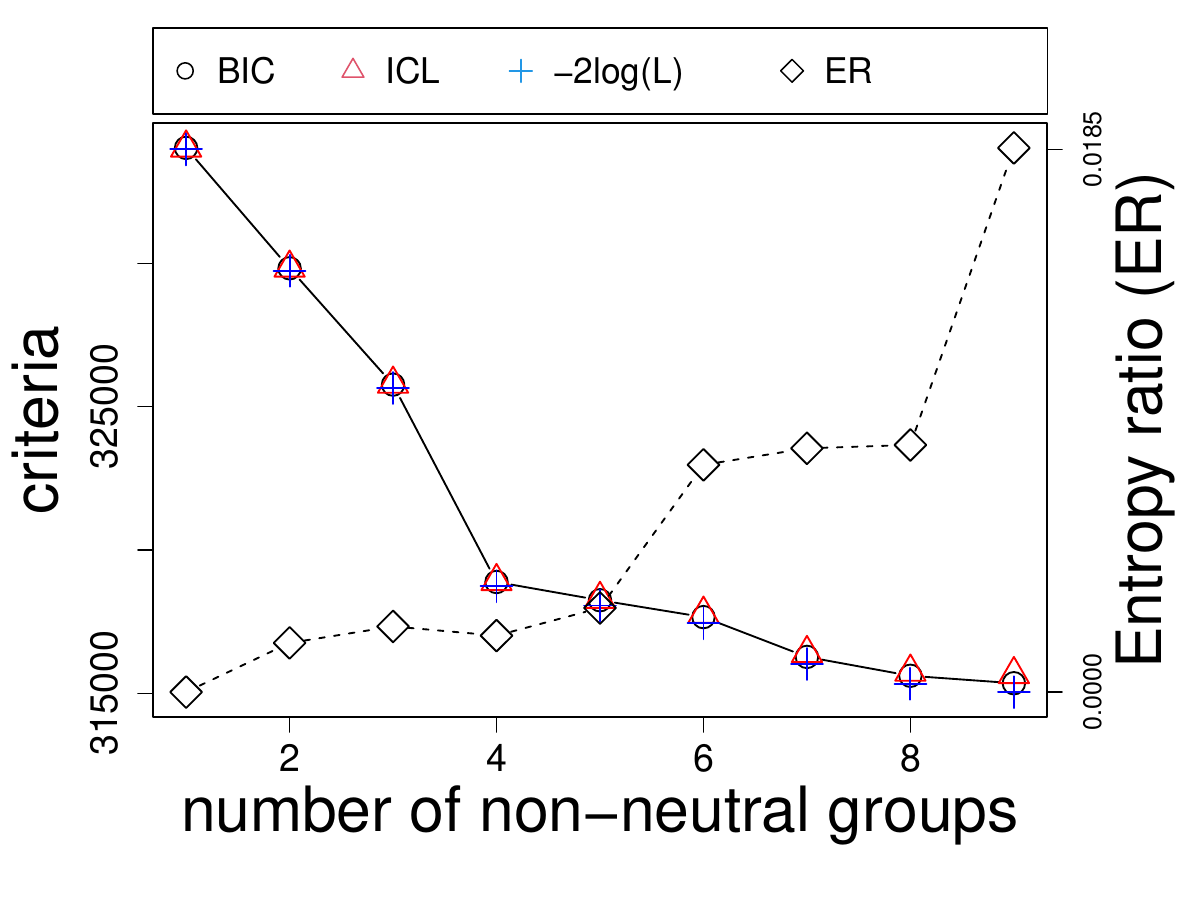}
\caption*{WWTP1 dataset}
\end{subfigure}
\hfil\hfil 
\begin{subfigure}{0.45\textwidth}
\includegraphics[width=\linewidth]{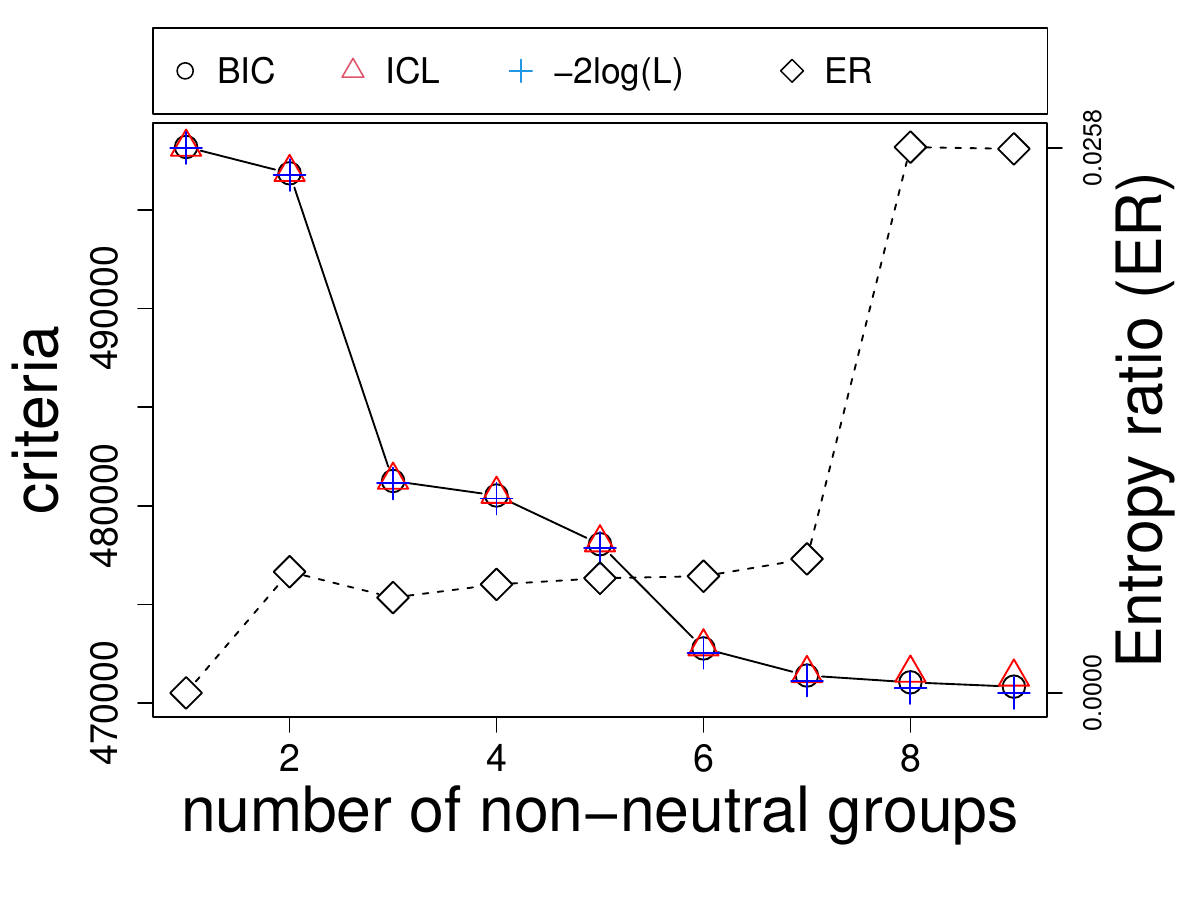}
\caption*{WWTP2 dataset}

\end{subfigure}\hfil 

\caption{BIC, ICL along with minus two times the log-likelihood ($-2\log L$) of models composed of 1 to 9 non-neutral groups with WWTP1 dataset (left figure, left axis) or WWTP2 dataset (right figure, left axis). The ratio between clustering entropy and maximal entropy denoted entropy ratio (ER), added on both figures, is associated with the right axis.}
\label{fig:sel}
\end{figure}

%% file: fig_main_WWTP1_K4_estimation.tex
\begin{figure*}
\vspace{-6em}
\begin{subfigure}{1 \linewidth}
\centering
\begin{tabular}{|r|c|c|c|c|c|}
\cline{2-6}
\multicolumn{1}{c|}{} & \multicolumn{5}{c|}{Group} \\
\cline{2-6}
\multicolumn{1}{c|}{} & 0 & 1 & 2 & 3 & 4\\
\hline
$\hat\pi^\text{A}$ &  0.968 & 0.007 & 0.004 & 0.009 & 0.012\\
$\hat\mu^\text{A}$ & -3.14 & -3.46 & 0.68 & -1.70 & -4.78\\
$[95\% \text{CI}]$& - & [-3.48; -3.44] & [0.66; 0.70] & [-1.73; -1.67] & [-4.81; -4.75]\\
$\hat s^\text{A} \times 100$&-&  2.73 & -1.61 & -2.26 & 1.75\\
$[95\% \text{CI}]$&- & [2.71; 2.74] & [-1.63; -1.59] &  [-2.32; -2.20] & [1.73; 1.77]\\
\hline
\end{tabular}
\caption{Parameter estimates. Selection coefficients and their 95\%CI boundaries are multiplied by 100. The estimate $\hat\mu^\text{A}_0$ is computed from estimates $\hat\alpha^\text{A} = 1.68$ and $\hat\beta^\text{A} = 38.8$.}
\label{fig:par.w1k4}
\end{subfigure}
 
\begin{subfigure}{1\textwidth}
\begin{multicols}{2}
    \includegraphics[width=1\linewidth]{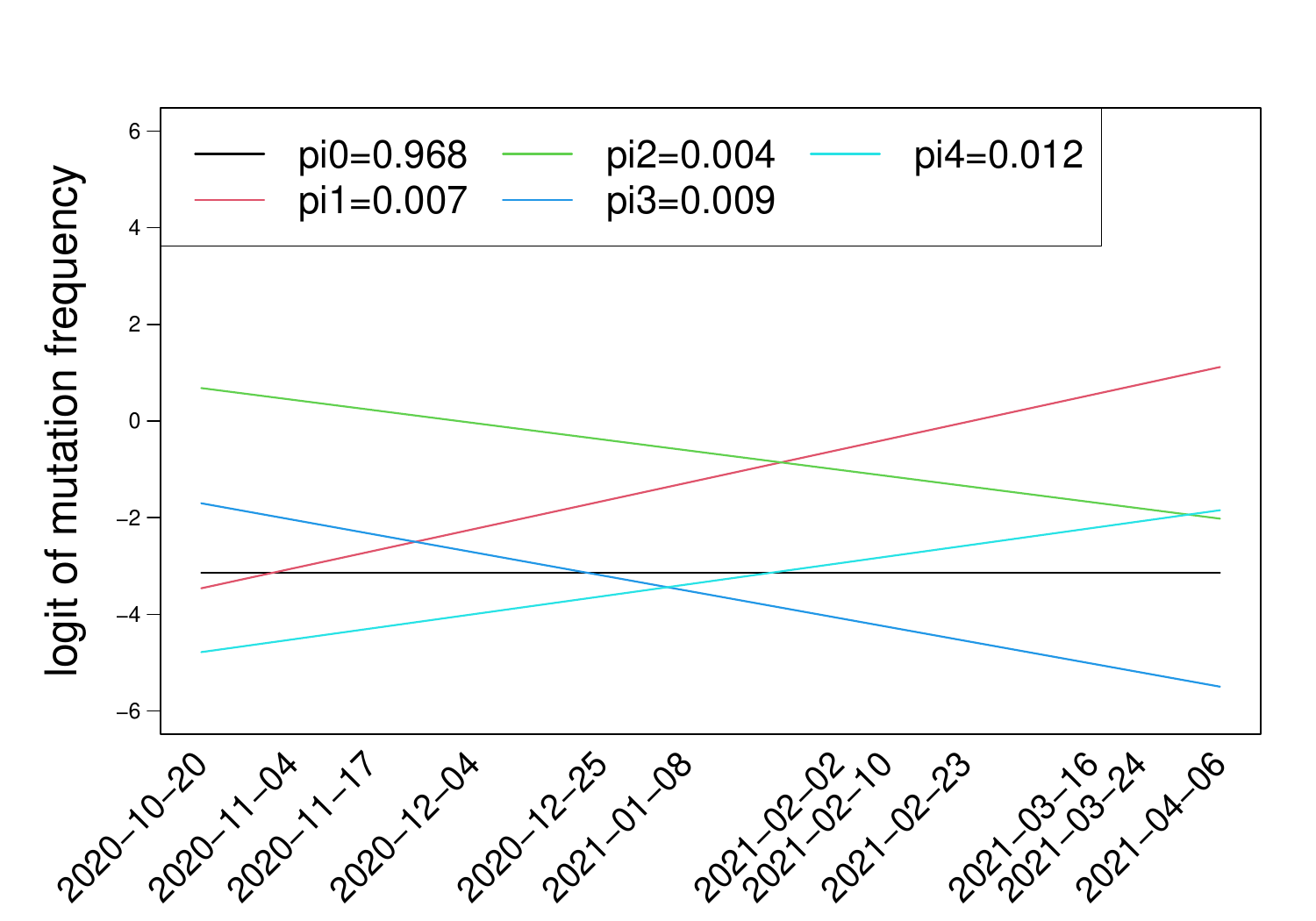}\par 
    \includegraphics[width=\linewidth]{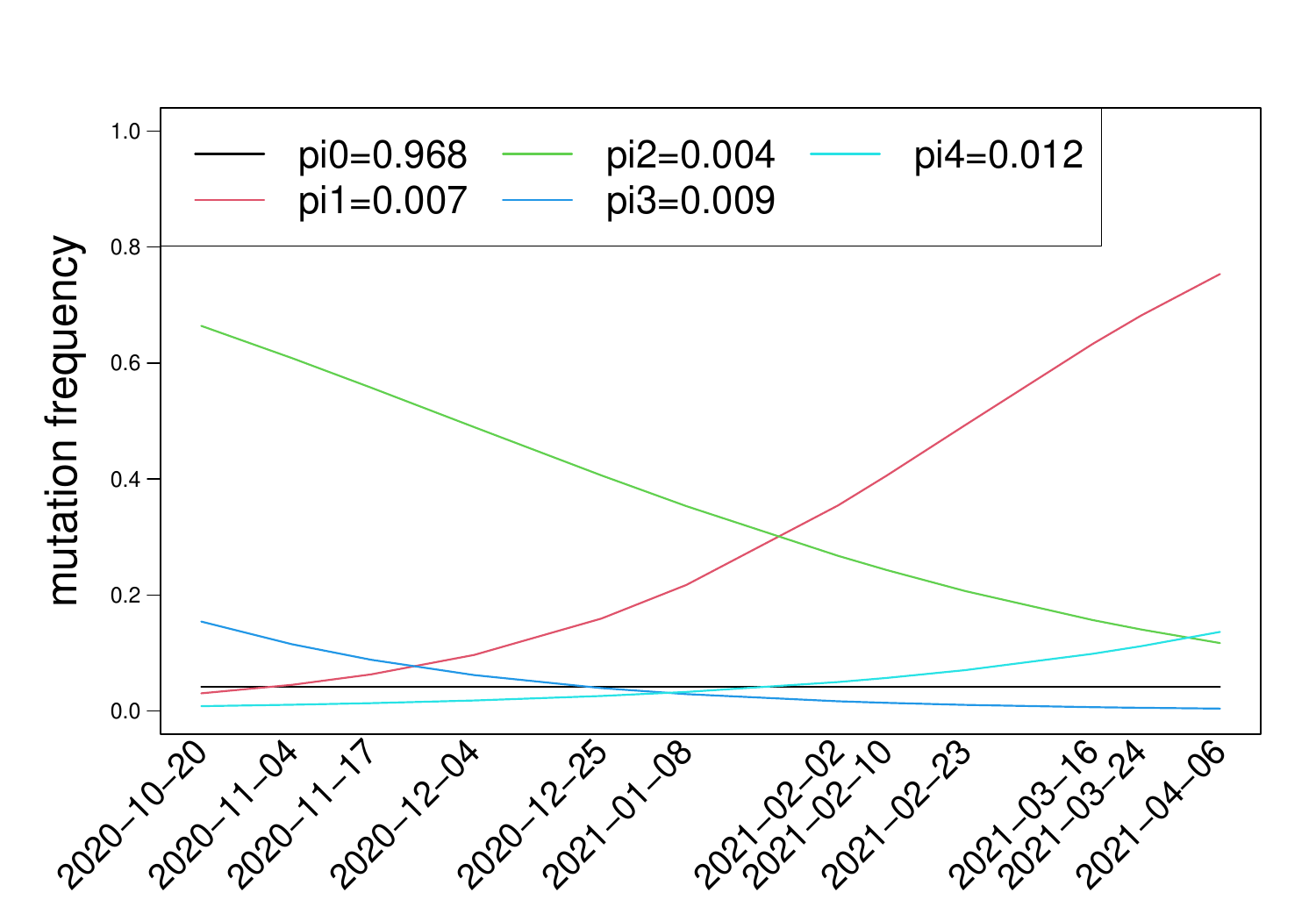}\par 
\end{multicols}
\caption{Estimated group log-frequency (left) and frequency (right) trajectories.}
\label{fig:traj.w1k4}
\end{subfigure}

\begin{subfigure}{1\textwidth}
\begin{multicols}{2}
    \includegraphics[width=\linewidth]{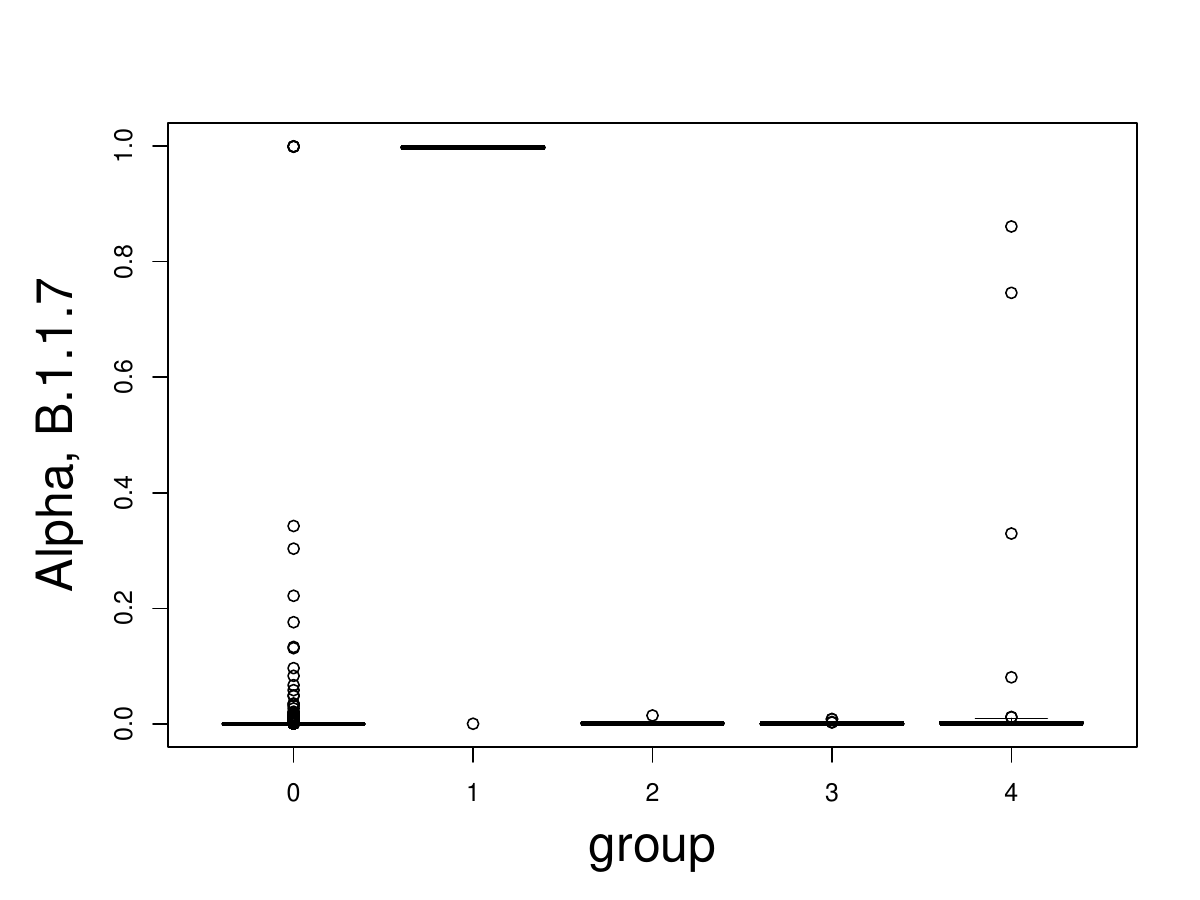}\par
    \includegraphics[width=\linewidth]{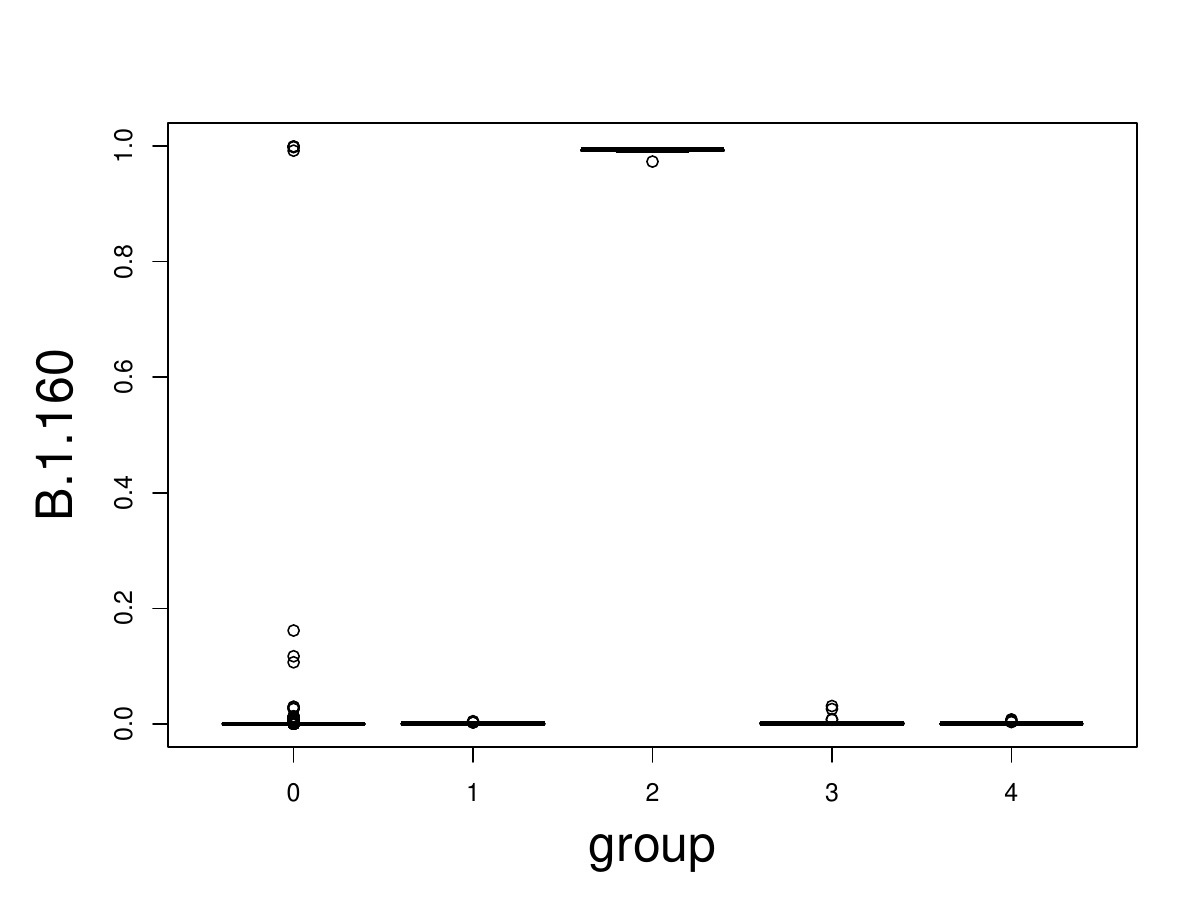}\par
\end{multicols}

\vspace{-3em}
\begin{multicols}{2}
    \includegraphics[width=\linewidth]{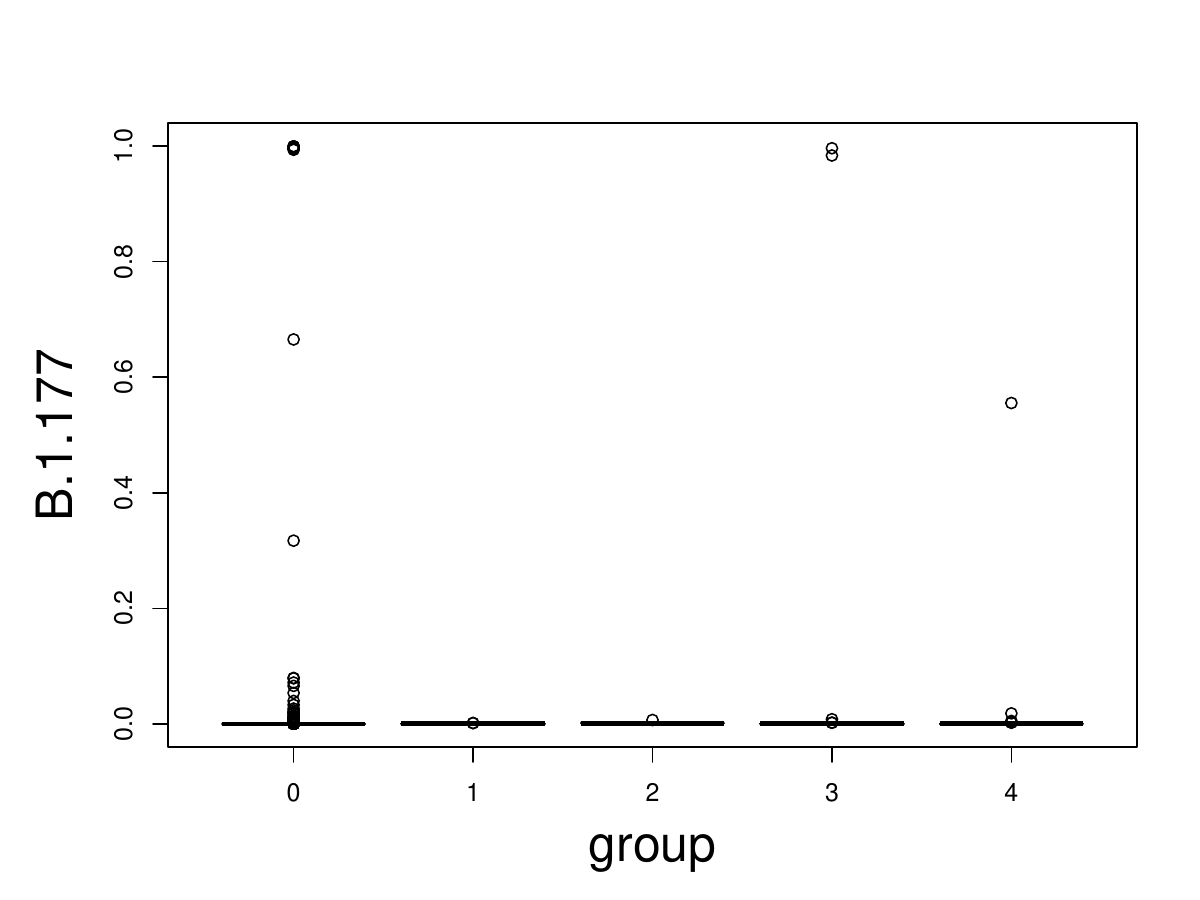}\par
    \includegraphics[width=\linewidth]{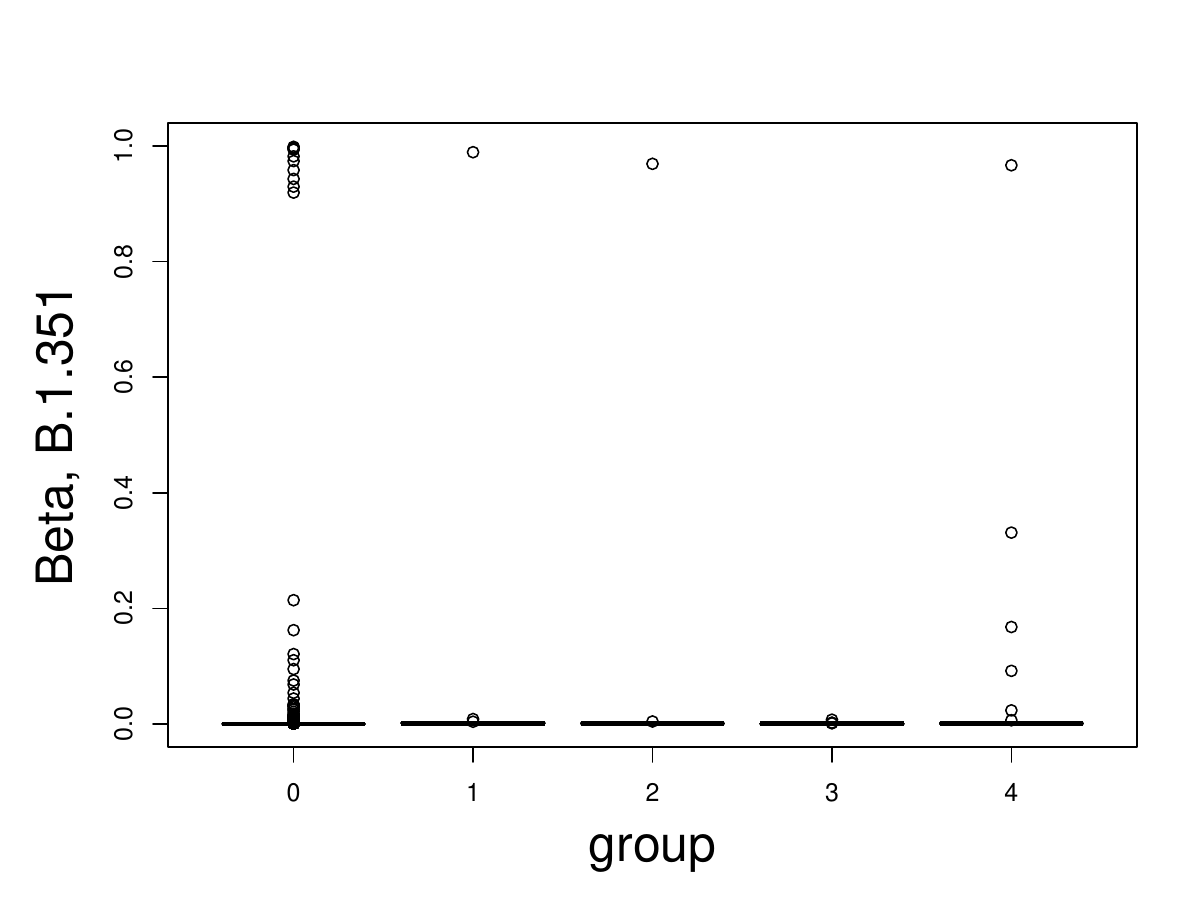}\par
\end{multicols}
\caption{Mutation signatures stratified on posterior group affectation.}
\label{fig:signature.w1k4}
\end{subfigure}

\caption{
Parameter estimates (Figure~\ref{fig:par.w1k4}), associated group log-frequency and frequency trajectories (Figure~\ref{fig:traj.w1k4}) and mutation profiles stratified on maximum a posteriori probability of group affectation (Figure~\ref{fig:signature.w1k4}) computed in Analysis A (conditional on WWTP1 dataset and $\hat K^\text{A}=4$ non-neutral groups).}
\label{fig:estim.w1k4}
\end{figure*}

%% file: fig_main_WWTP2_K3_estimation.tex
\begin{figure*}
\vspace{-6em}
\begin{subfigure}{1 \linewidth}
\centering
\begin{tabular}{|r|c|c|c|c|}
\cline{2-5}
\multicolumn{1}{c|}{} & \multicolumn{4}{c|}{Group} \\
\cline{2-5}
\multicolumn{1}{c|}{} & 0 & 1 & 2 & 3\\
\hline
$\hat\pi^\text{B}$ & 0.976 & 0.006 & 0.004 & 0.014\\
$\hat\mu^\text{B}$&-3.17 & -3.91 & 0.46 & -3.02 \\
$[95\% \text{CI}]$&-& [-3.93; -3.89] & [0.45; 0.48] &[-3.06; -2.98]\\
$\hat s^\text{B} \times 100$&-&3.42 & -2.05 & -1.30\\
$[95\% \text{CI}]$&- & [3.41; 3.44] &  [-2.07; -2.03] &[-1.37; -1.24]\\
\hline
\end{tabular}
\caption{Parameter estimates. Selection coefficients and their 95\%CI boundaries are multiplied by 100. The estimate $\hat\mu^\text{B}_0$ is computed from estimates $\hat\alpha^\text{B} = 1.88$ and $\hat\beta^\text{B} = 44.9$.}
\label{fig:par.w2k3}
\end{subfigure}

\begin{subfigure}[b]{1 \linewidth}
\begin{multicols}{2}
    \includegraphics[width=\linewidth]{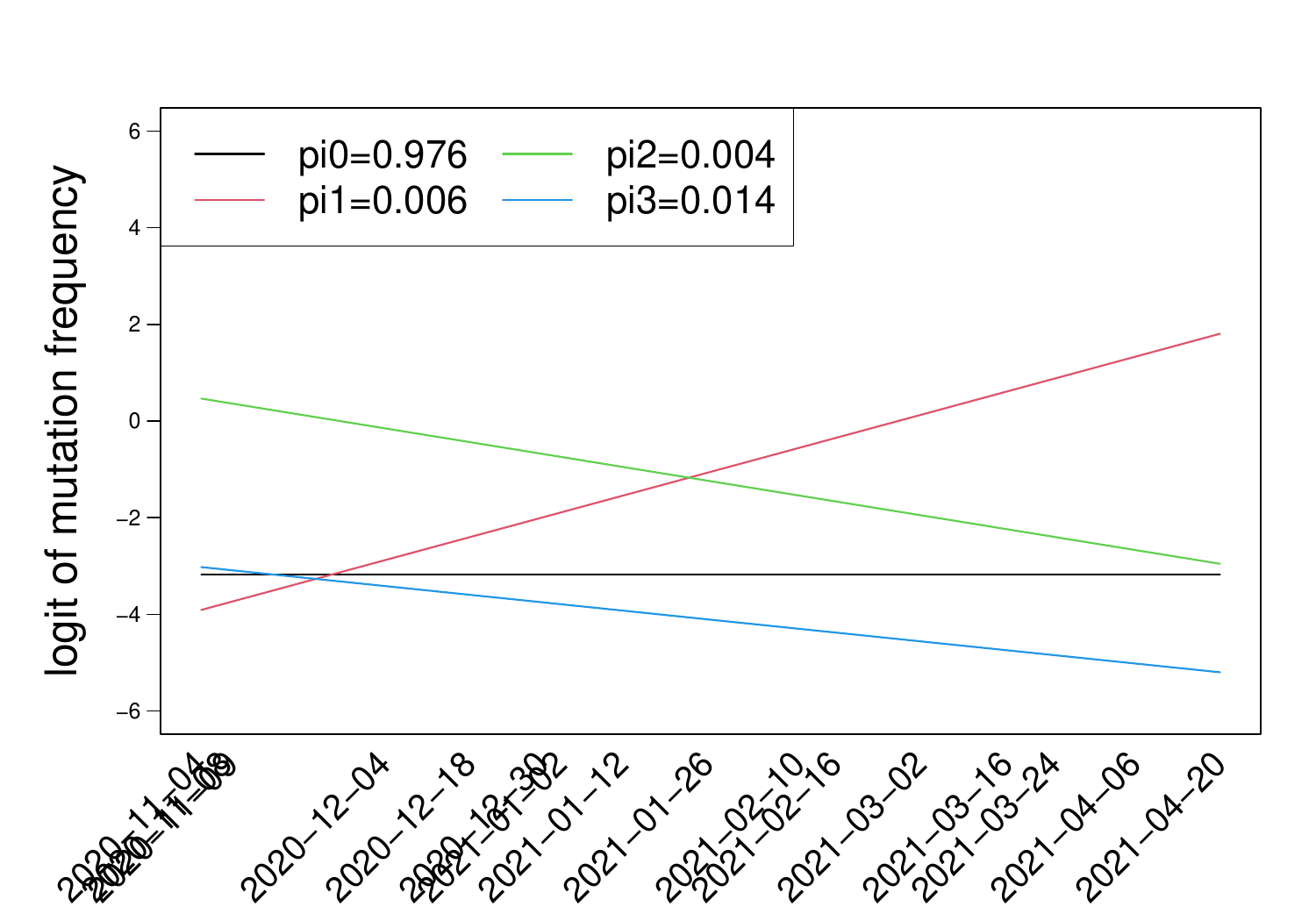}\par 
    \includegraphics[width=\linewidth]{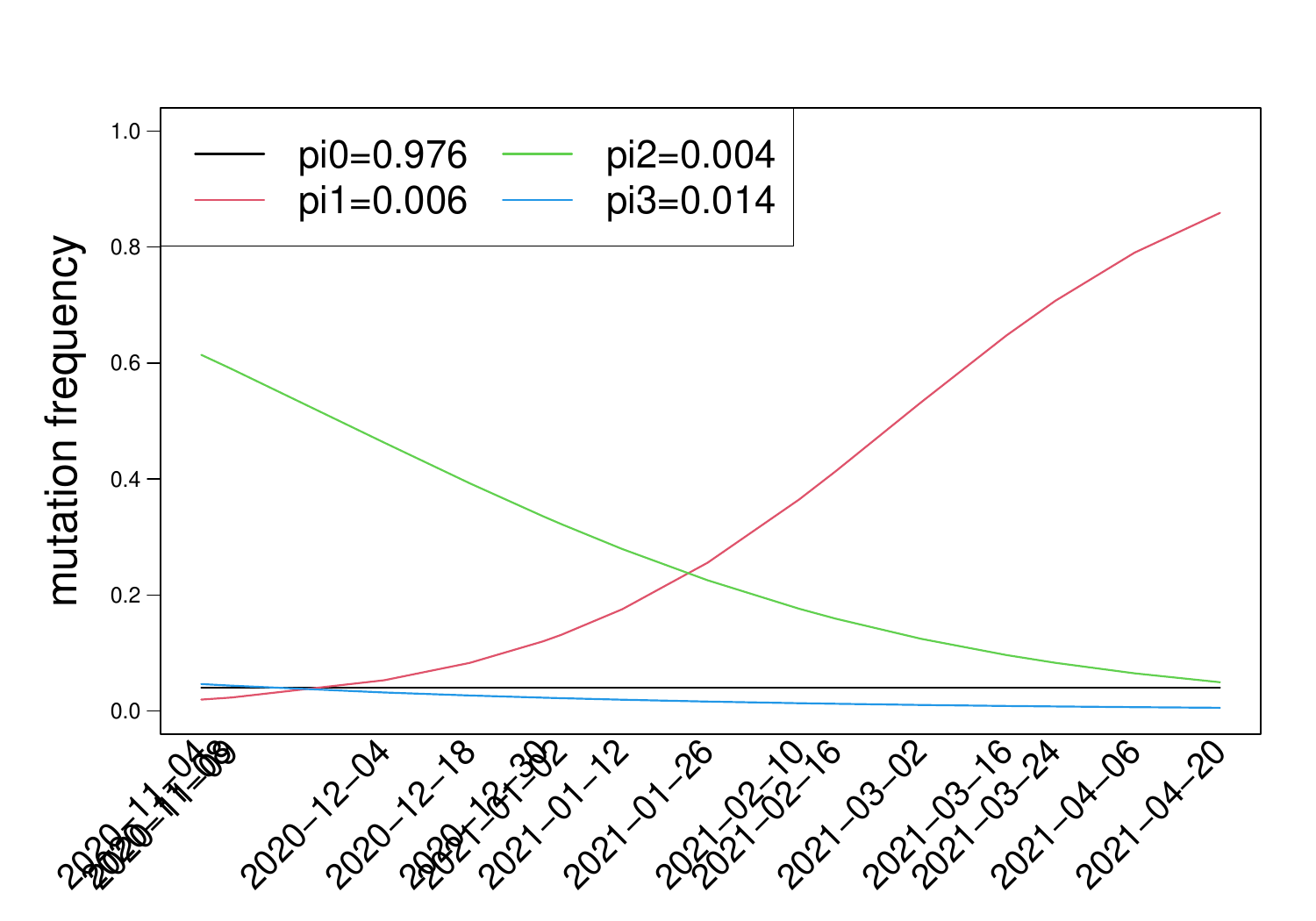}\par 
\end{multicols}
\caption{Estimated group logit frequency (left) and frequency (right) trajectories.}
\label{fig:traj.w2k3}
\end{subfigure}

\begin{subfigure}{1\linewidth}
\begin{multicols}{2}
    \includegraphics[width=\linewidth]{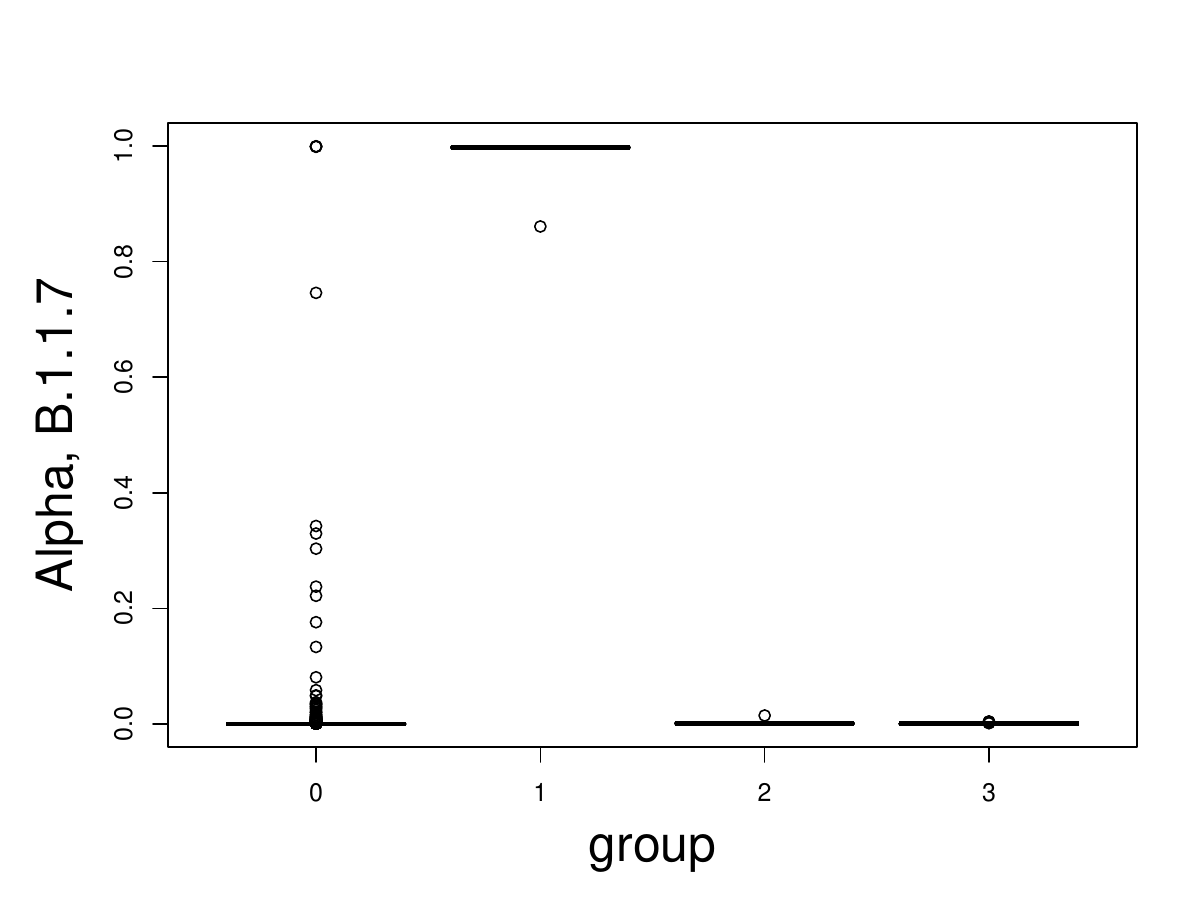}\par
    \includegraphics[width=\linewidth]{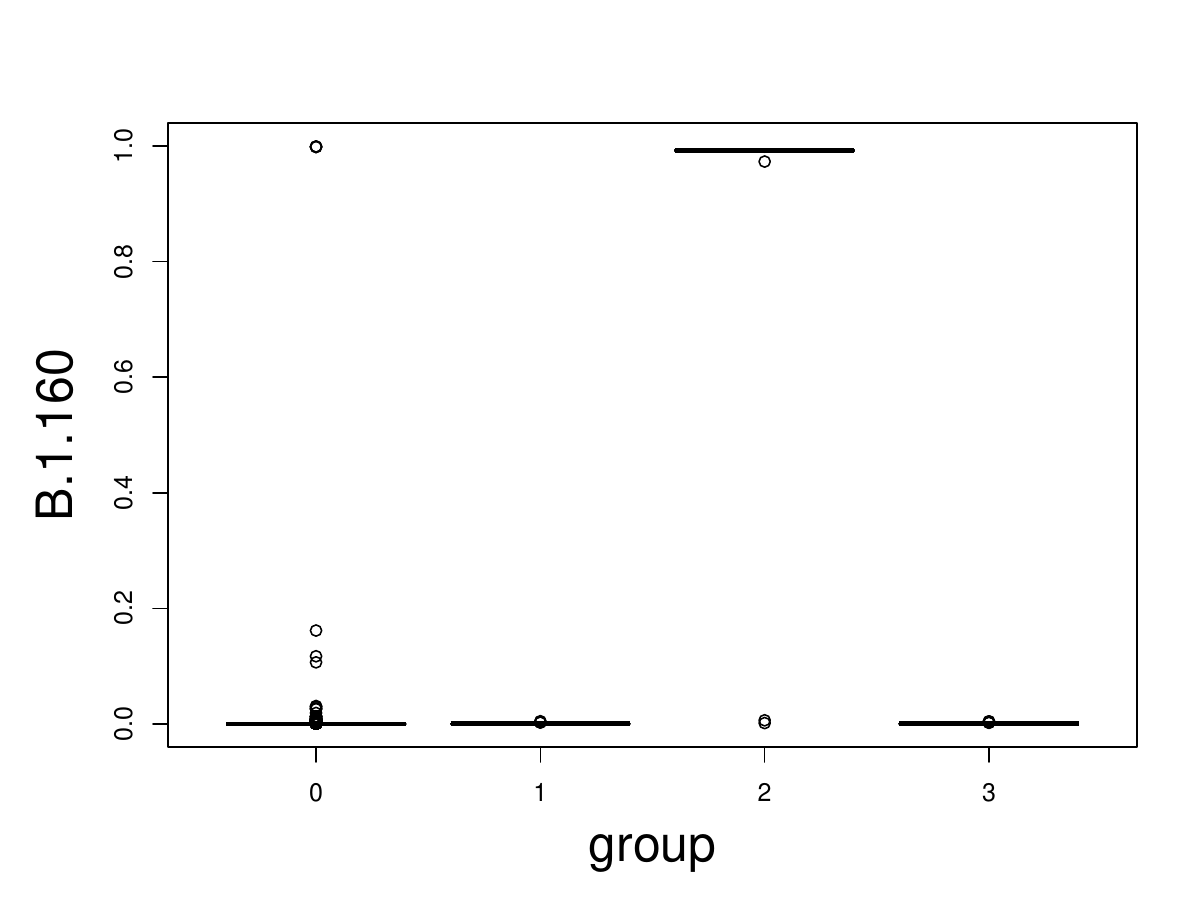}\par
\end{multicols}

\vspace{-4em}

\begin{multicols}{2}
    \includegraphics[width=\linewidth]{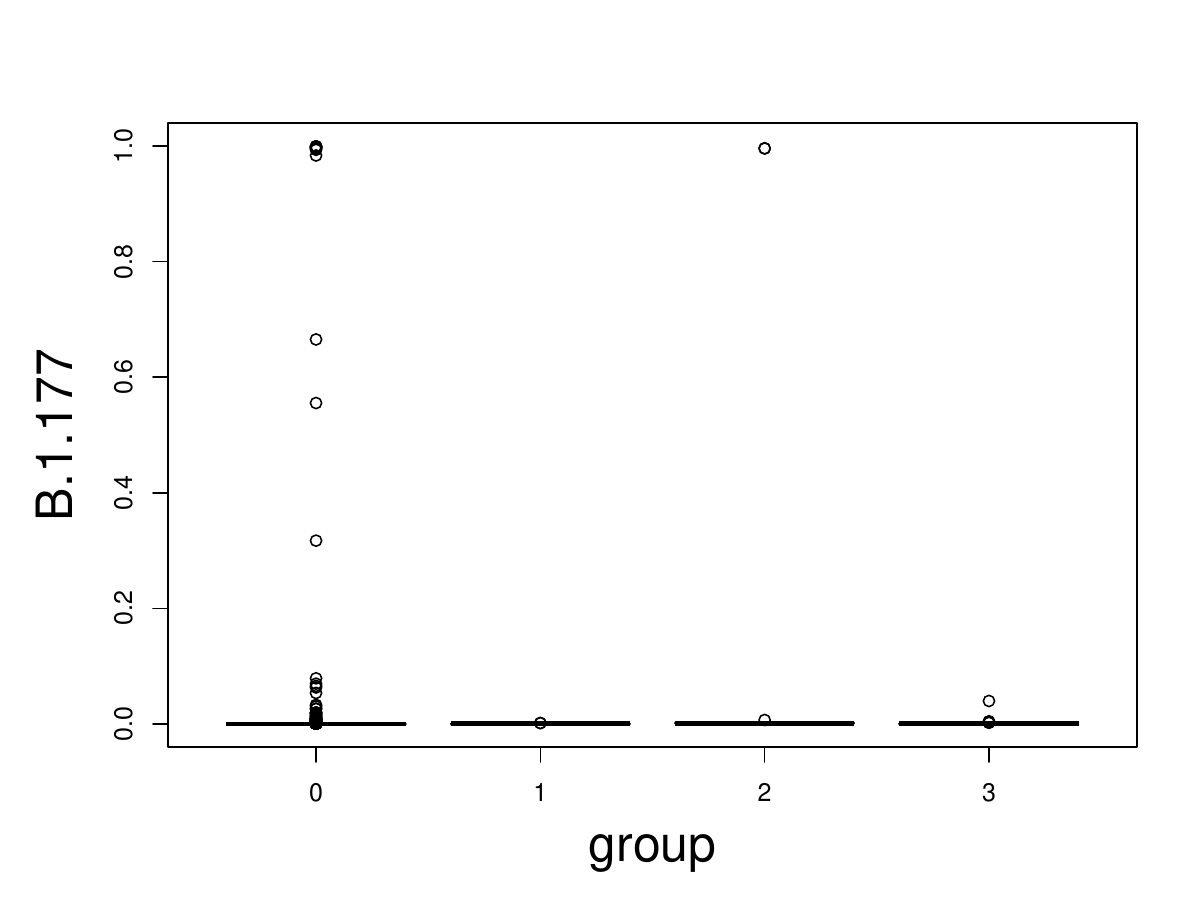}\par
    \includegraphics[width=\linewidth]{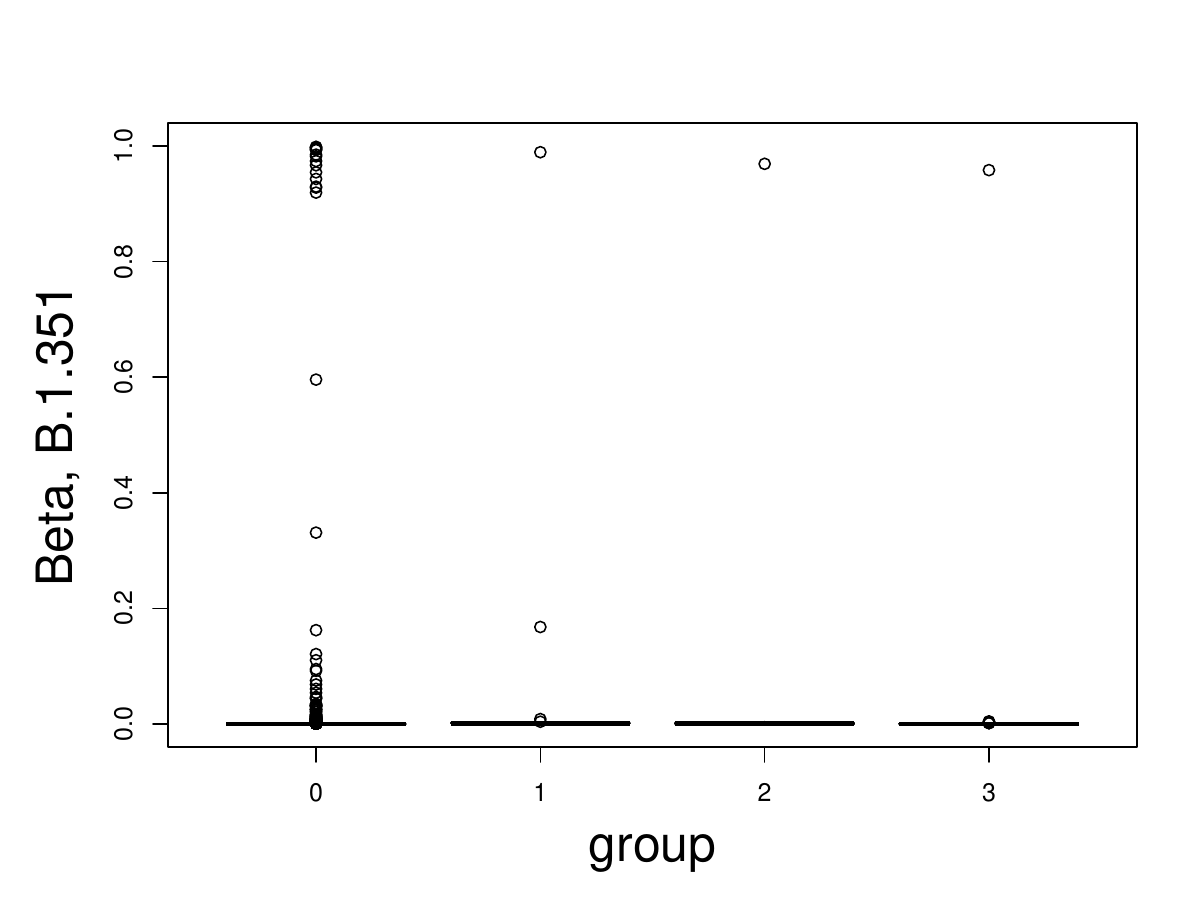}\par
\end{multicols}
\caption{Mutation signatures stratified on posterior cluster affectation.}
\label{fig:signature.w2k3}
\end{subfigure}

\caption{Parameter estimates (Figure~\ref{fig:par.w2k3}), associated group log-frequency and frequency trajectories (Figure~\ref{fig:traj.w2k3}) and mutation profiles stratified on maximum a posteriori probability of group affectation (Figure~\ref{fig:signature.w2k3}) computed in Analysis B (conditional on WWTP2 dataset and $\hat K^\text{B}=3$ non-neutral groups).}
\label{fig:estim.w2k3}
\end{figure*}

%% file: fig_main_WWTP2_K6_estimation.tex
\begin{figure*}
\vspace{-6em}
\begin{subfigure}{1 \linewidth}
\centering
\small{
\setlength{\tabcolsep}{4pt} 
\begin{tabular}{|r|c|c|c|c|c|c|c|}
\cline{2-8}
\multicolumn{1}{c|}{} & \multicolumn{7}{c|}{Group} \\
\cline{2-8}
\multicolumn{1}{c|}{} & 0 & 1 & 2 & 3 & 4 & 5 & 6\\
\hline
$\hat\pi^\text{C}$ & 0.954 & 0.006 & 0.004 & 0.004 & 0.010 & 0.011 & 0.011\\
$\hat\mu^\text{C}$ &-3.18 & -3.86 & 0.50 & -1.39 & -4.96 & -3.31 & -2.75\\
$[95\%\text{CI}]$ &-&[-3.88; -3.85]&[0.48; 0.52]&[-1.41; -1.36]&[-4.98; -4.93]&[-3.36; -3.26]&[-2.79; -2.72]\\
$\hat s^\text{C} \times 100$ & - & 3.40 & -2.06 & -1.30 & 1.92& -1.59  & -0.92\\
$[95\%\text{CI}]$ &-&[3.38; 3.41]&[-2.08; -2.04]&[-1.34; -1.27]&[1.90; 1.94]&[-1.68; -1.50]&[-0.96; -0.88]\\
\hline
\end{tabular}}
\caption{Parameter estimates. Selection coefficients and their 95\%CI boundaries are multiplied by 100. The estimate $\hat\mu^\text{C}_0$ is computed from estimates $\hat\alpha^\text{C} = 1.87$ and $\hat\beta^\text{C} = 45.0$.}
\label{fig:par.w2k6}
\end{subfigure}

\begin{subfigure}{1\textwidth}
\begin{multicols}{2}
    \includegraphics[width=\linewidth]{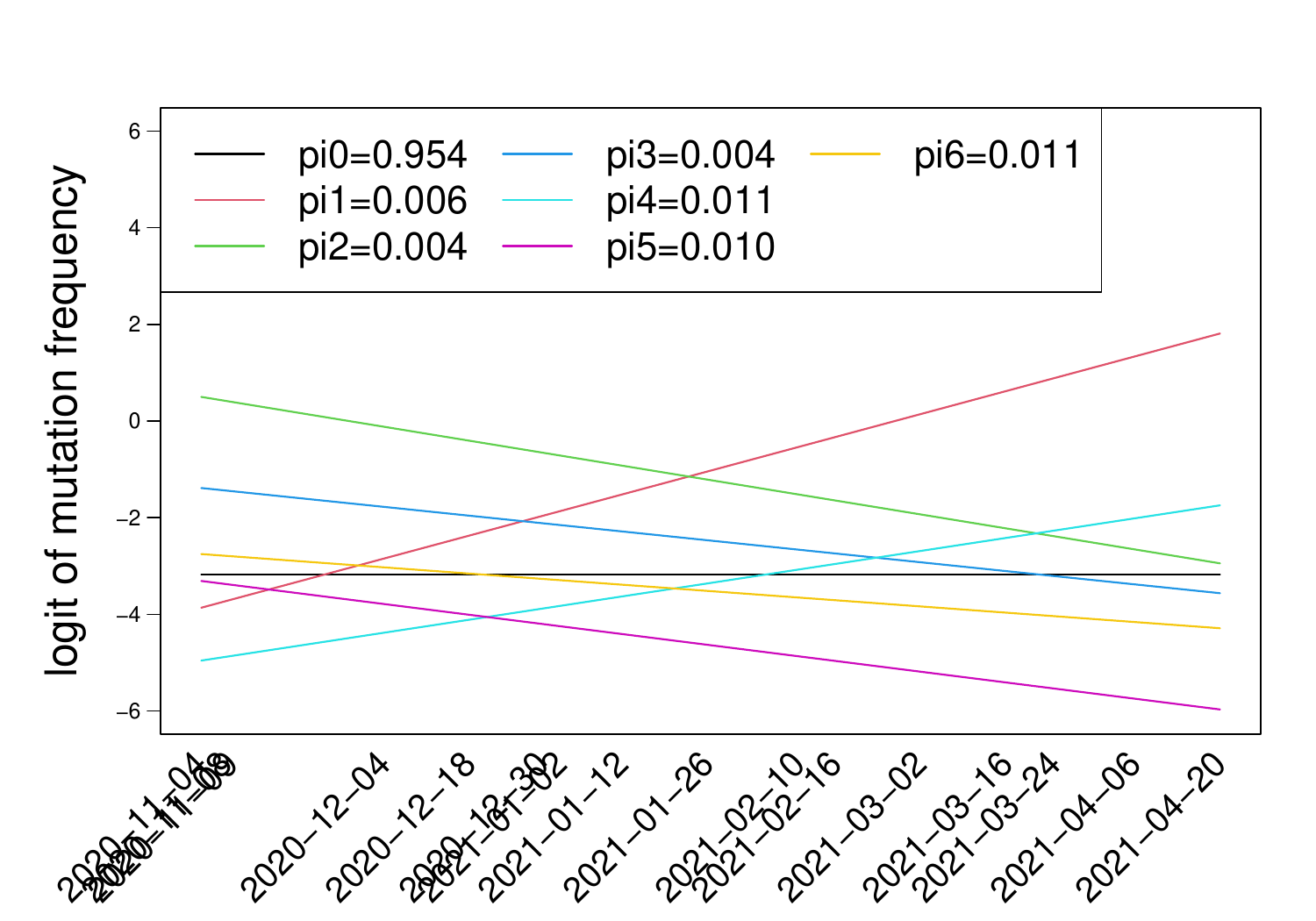}\par 
    \includegraphics[width=\linewidth]{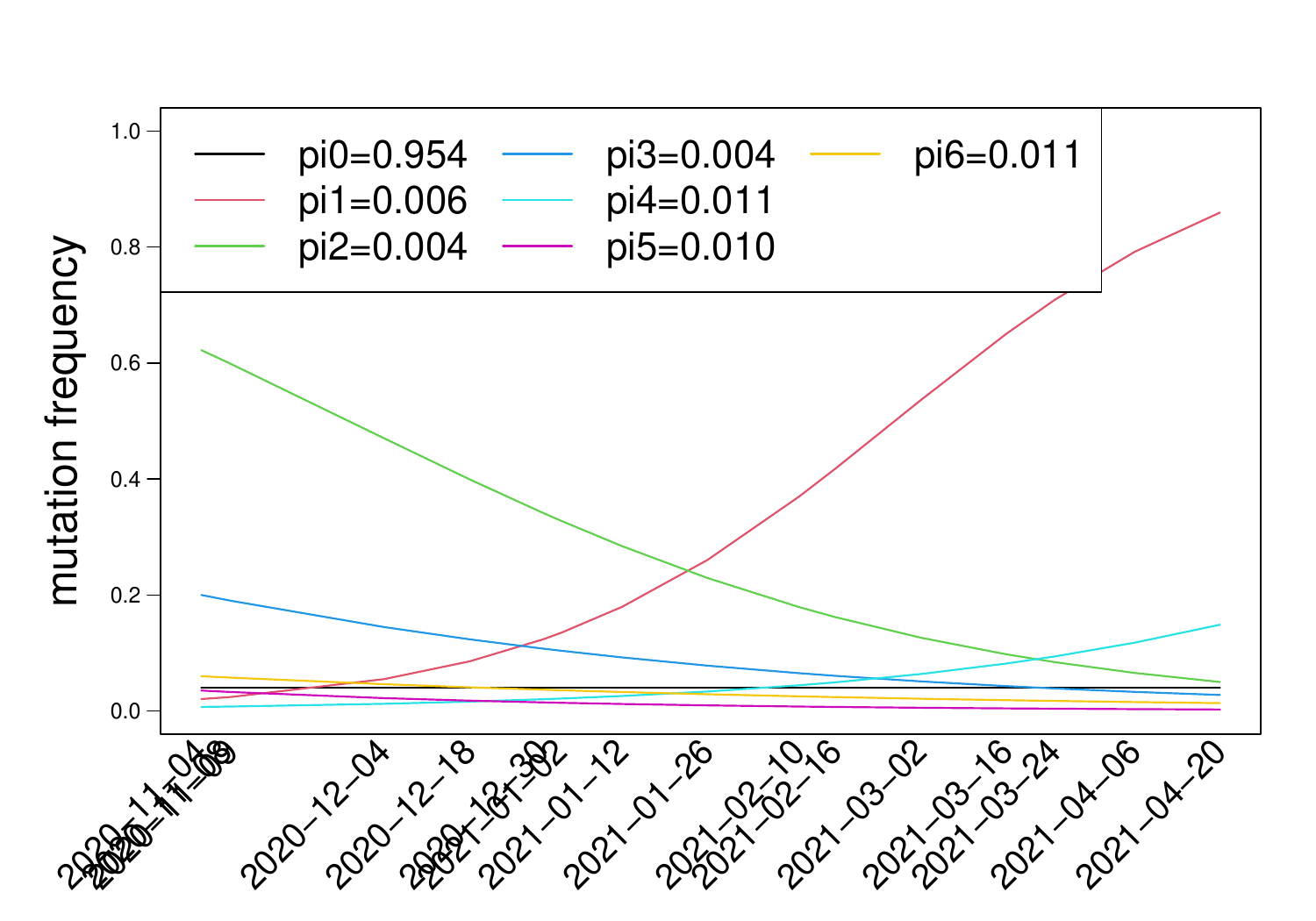}\par 
\end{multicols}
\caption{Estimated group log-frequency (left) and frequency (right) trajectories.}
\label{fig:traj.w2k6}
\end{subfigure}

\begin{subfigure}{1\textwidth}
\begin{multicols}{2}
    \includegraphics[width=\linewidth]{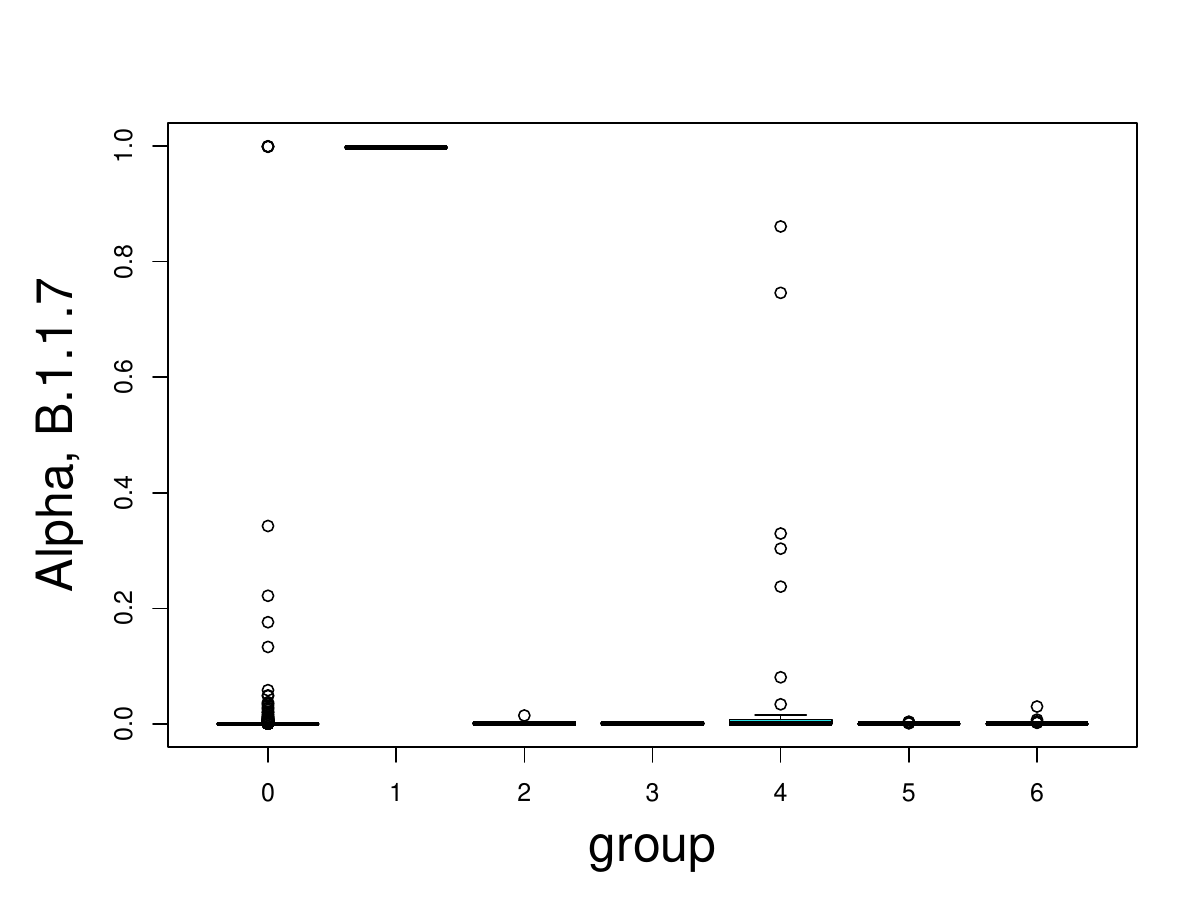}\par
    \includegraphics[width=\linewidth]{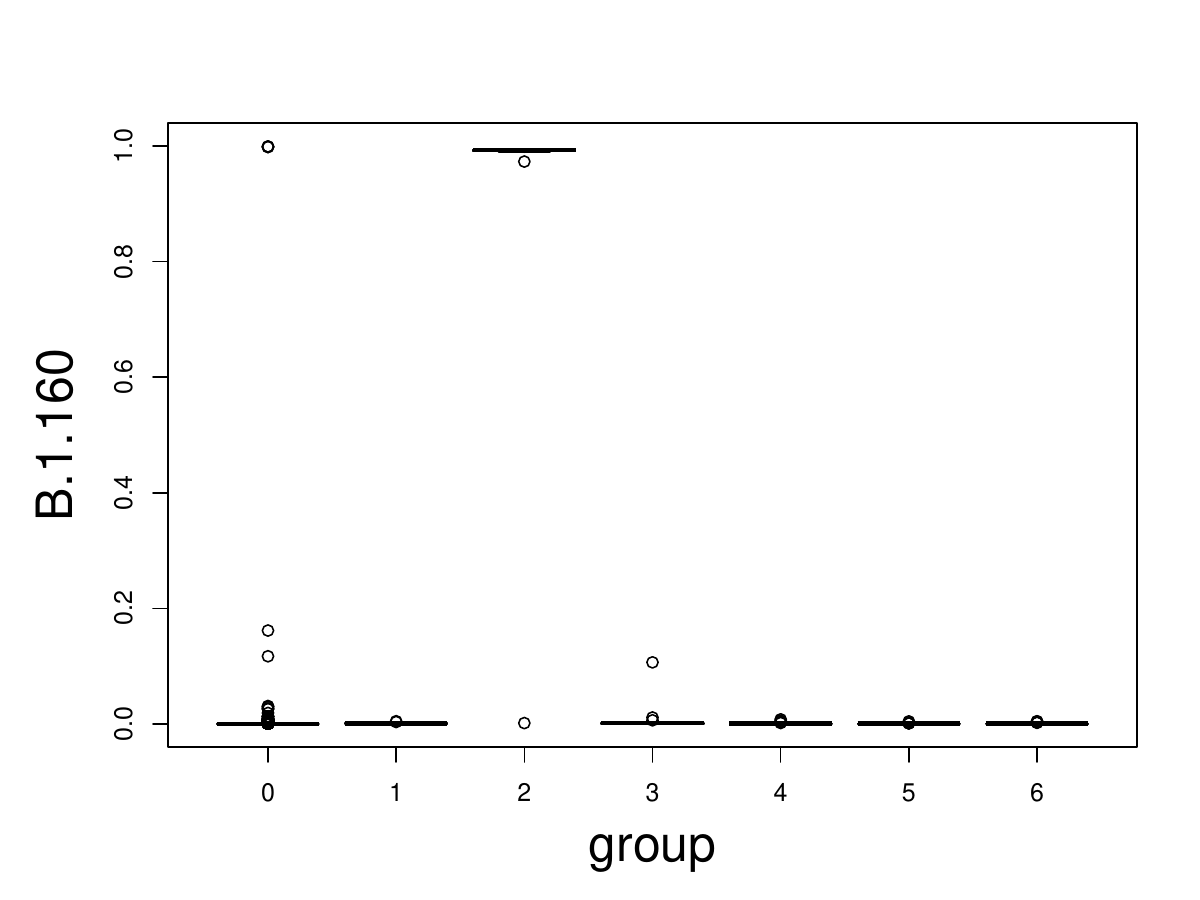}\par
\end{multicols}
\begin{multicols}{2}
    \includegraphics[width=\linewidth]{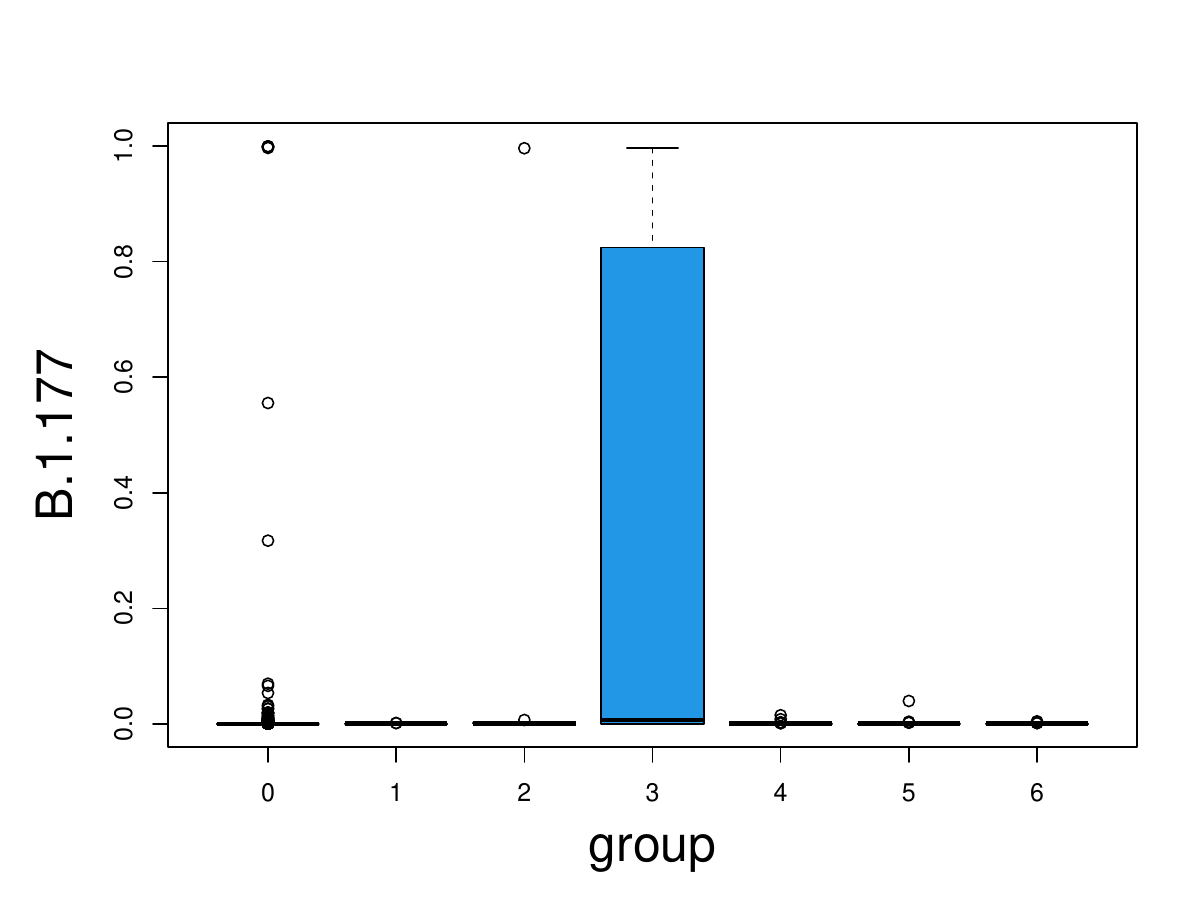}\par
    \includegraphics[width=\linewidth]{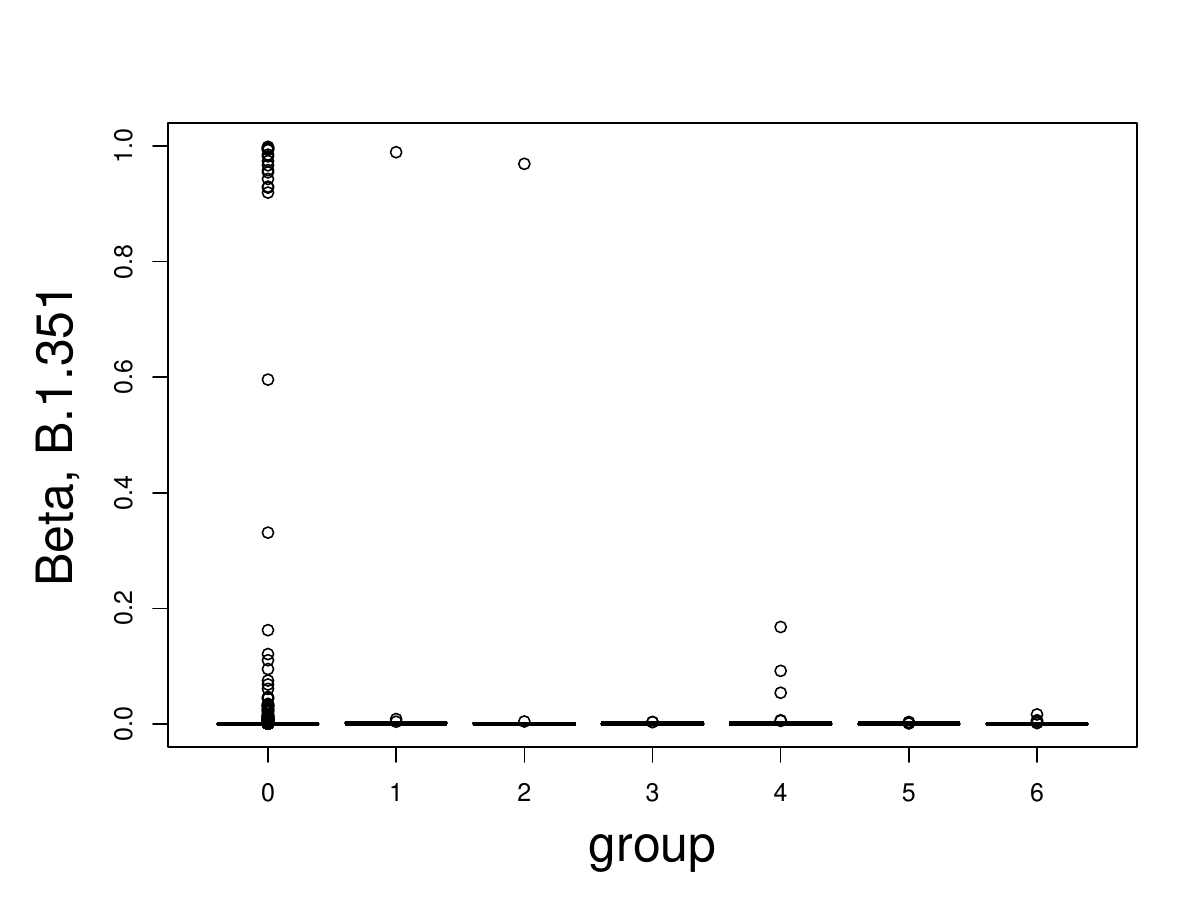}\par
\end{multicols}
\caption{Mutation signatures stratified on posterior cluster affectation.}
\label{fig:signature.w2k6}
\end{subfigure}

\caption{Parameter estimates (Figure~\ref{fig:par.w2k6}), associated group log-frequency and frequency trajectories (Figure~\ref{fig:traj.w2k6}) and mutation profiles stratified on maximum a posteriori probability of group affectation (Figure~\ref{fig:signature.w2k6}) computed in Analysis C (conditional on WWTP2 dataset and $\hat K^D=6$ non-neutral groups).}
\label{fig:estim.w2k6}
\end{figure*}

%% file: tab_freq.tex
\begin{table}
\centering
\begin{subtable}{1\textwidth}\centering
\begin{tabular}{|r|c|c|c|c|c|c|c|}
\cline{2-8}
\multicolumn{1}{c|}{} & 23/10 & 07/11 & 25/11 & 24/12 & 22/01 & 24/02 & 25/03\\
\multicolumn{1}{c|}{} & 2020 & 2020 & 2020 & 2020 & 2021 & 2021& 2021\\
\hline
Alpha (Ns.) & $<1$ & $<1$ & 11 & 43 & 39 & 47 & 59\\
\hline
$\hat f^\text{A}_1$ & 3 & 5 & 8 & 16 & 29 & 50 & 69\\
$\hat f^\text{B}_1$ & 1 & 2 & 4 & 10 & 23 & 48 & 71\\
$\hat f^\text{C}_1$ & 1 & 2 & 4 & 10 & 24 & 49 & 72\\
\hline
\end{tabular}
\caption{Alpha \& group $G_1$}
\label{tab:freq.alpha}
\end{subtable}

\vspace{1em}
\begin{subtable}{1\textwidth}\centering
\begin{tabular}{|r|c|c|c|c|c|c|c|}
\cline{2-8}
\multicolumn{1}{c|}{} & 23/10 & 07/11 & 25/11 & 24/12 & 22/01 & 24/02 & 25/03\\
\multicolumn{1}{c|}{} & 2020 & 2020 & 2020 & 2020 & 2021 & 2021& 2021\\
\hline
\;\; B.1 (Ns.) & 84 & 70 & 47 & 38 & 19 & 8 & 11\\
\hline
$\hat f^\text{A}_2$ & 65 & 60 & 53 & 41 & 30 & 20 & 14\\
$\hat f^\text{B}_2$ & 67 & 60 & 51 & 36 & 24 & 14 & 8\\
$\hat f^\text{C}_2$ & 68 & 61 & 52 & 37 & 24 & 14 & 8\\
\hline
\end{tabular}
\caption{B.1.160 \& group $G_2$}
\label{tab:freq.b1}
\end{subtable}

\caption{Alpha (Table~\ref{tab:freq.alpha}) and B.1.160 VOC (Table~\ref{tab:freq.b1}) frequency in France reported by Nextstrain (Ns. first line of each table) at various time points along with group $G_1$ (Table~\ref{tab:freq.alpha}) and group $G_2$  (Table~\ref{tab:freq.b1}) frequency estimates in Analysis A, B and C ($2^\text{nd}$, $3^\text{rd}$ and $4^\text{th}$ lines in each table). Frequencies are given in percent.}
\label{tab:freq}
\end{table}

%% file: fig_reduced_data_selection.tex
\begin{figure}
\centering 
\begin{subfigure}[b]{0.45\textwidth}
\includegraphics[width=\linewidth]{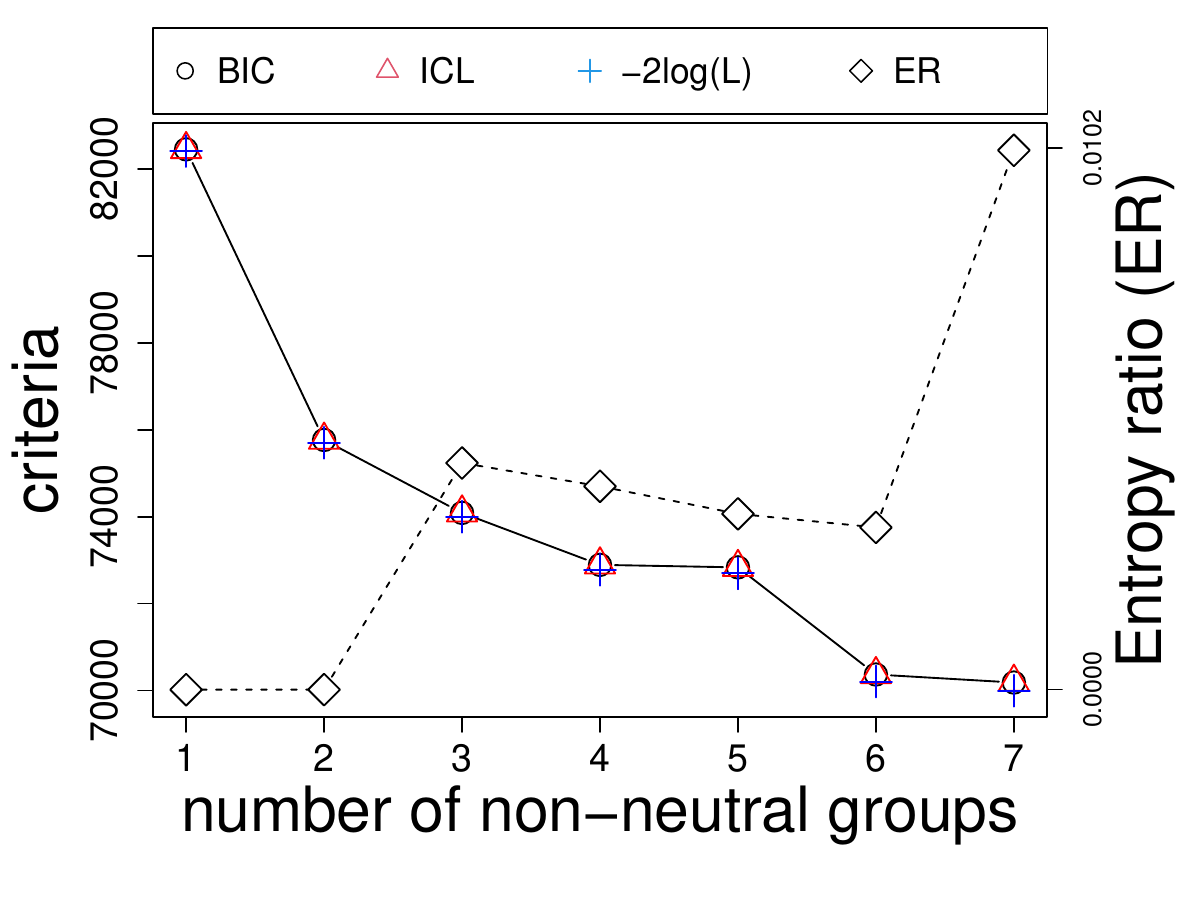}
\caption{Reduced WWTP1 dataset}
\end{subfigure}
\hfil
\hfil 
\begin{subfigure}[b]{0.45\textwidth}
\includegraphics[width=\linewidth]{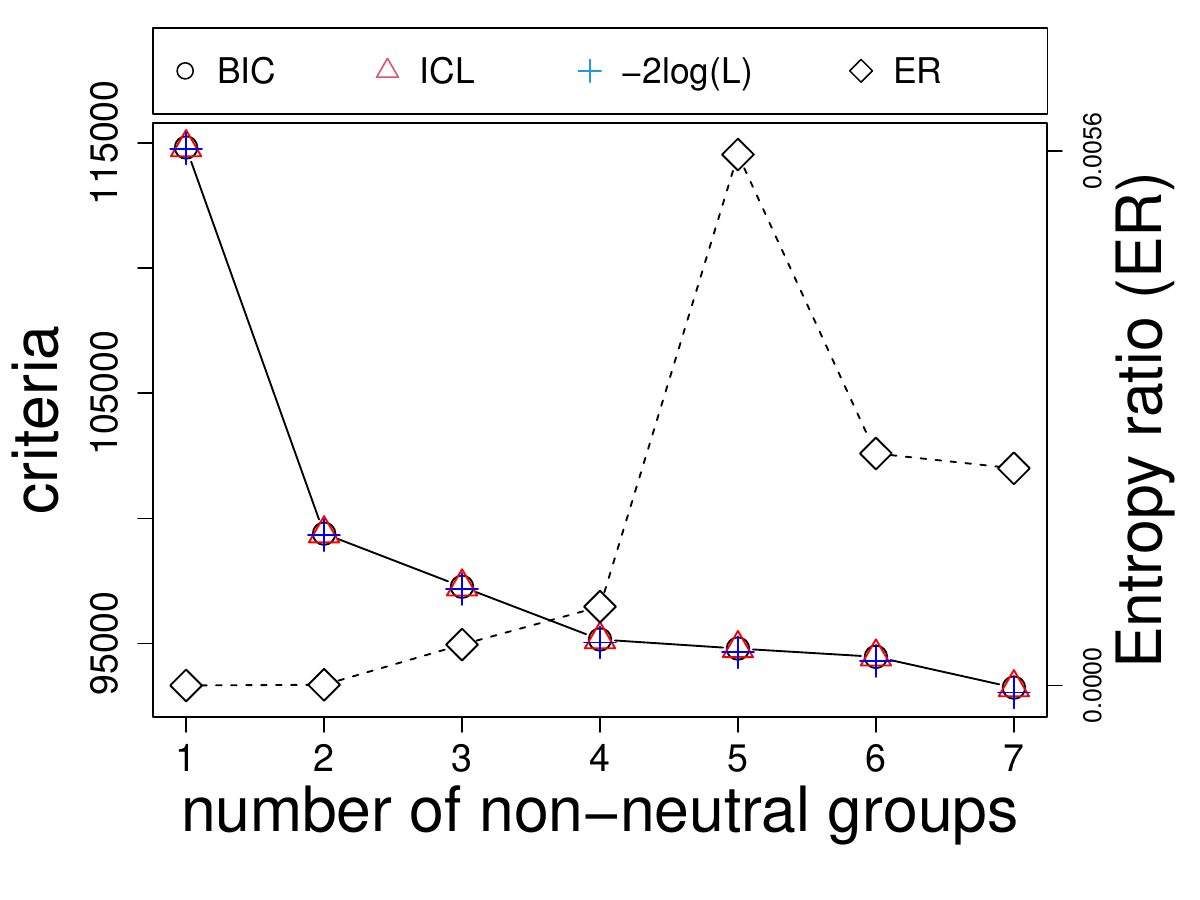}
\caption{Reduced WWTP2 dataset}
\end{subfigure}

\caption{BIC, ICL along with minus two times the log-likelihood ($-2\log L$) of models composed of 1 to 7 non-neutral groups with WWTP1 (left figure, left axis) or WWTP2 (right figure, left axis) datasets, both reduced to mutations associated with probabilities above 0.005 to belong to at least one lineage among B.1.1.7 (Alpha), B.1, B.1.1, B.1.160, B.1.177, B.1.351 (Beta), B.1.367. Entropy ratio (ER) added on both figures is associated with the right axis.}
\label{fig:reduced.sel}
\end{figure}

%% file: fig_reduced_data_WWTP_K2_estimation.tex
\begin{figure*}
\vspace{-6em}

\begin{subfigure}{1\textwidth}
\begin{multicols}{2}
    
    \begin{subfigure}{1\linewidth}
    \vspace{1em}
    \centering
    \small
    \begin{tabular}{|r|c|c|c|c|c|}
    \cline{2-4}
    \multicolumn{1}{c|}{} & \multicolumn{3}{c|}{Group} \\
    \cline{2-4}
    \multicolumn{1}{c|}{} & 0 & 1 & 2 \\
    \hline
    $\hat\pi^\text{D}$ &  0.865 & 0.085 & 0.050\\
    $\hat\mu^\text{D}$ & -2.46 & -3.43 & 0.68\\
    $[95\% \text{CI}]$& - & [-3.45; -3.41] & [0.66; 0.70]\\
    $\hat s^\text{D} \times 100$&-& 2.71 & -1.62\\
    $[95\% \text{CI}]$&- & [2.70; 2.73] & [-1.64; -1.60]\\
    \hline
    \end{tabular}
    \end{subfigure}
    
    \hspace{3em}
    \includegraphics[width=0.8\linewidth]{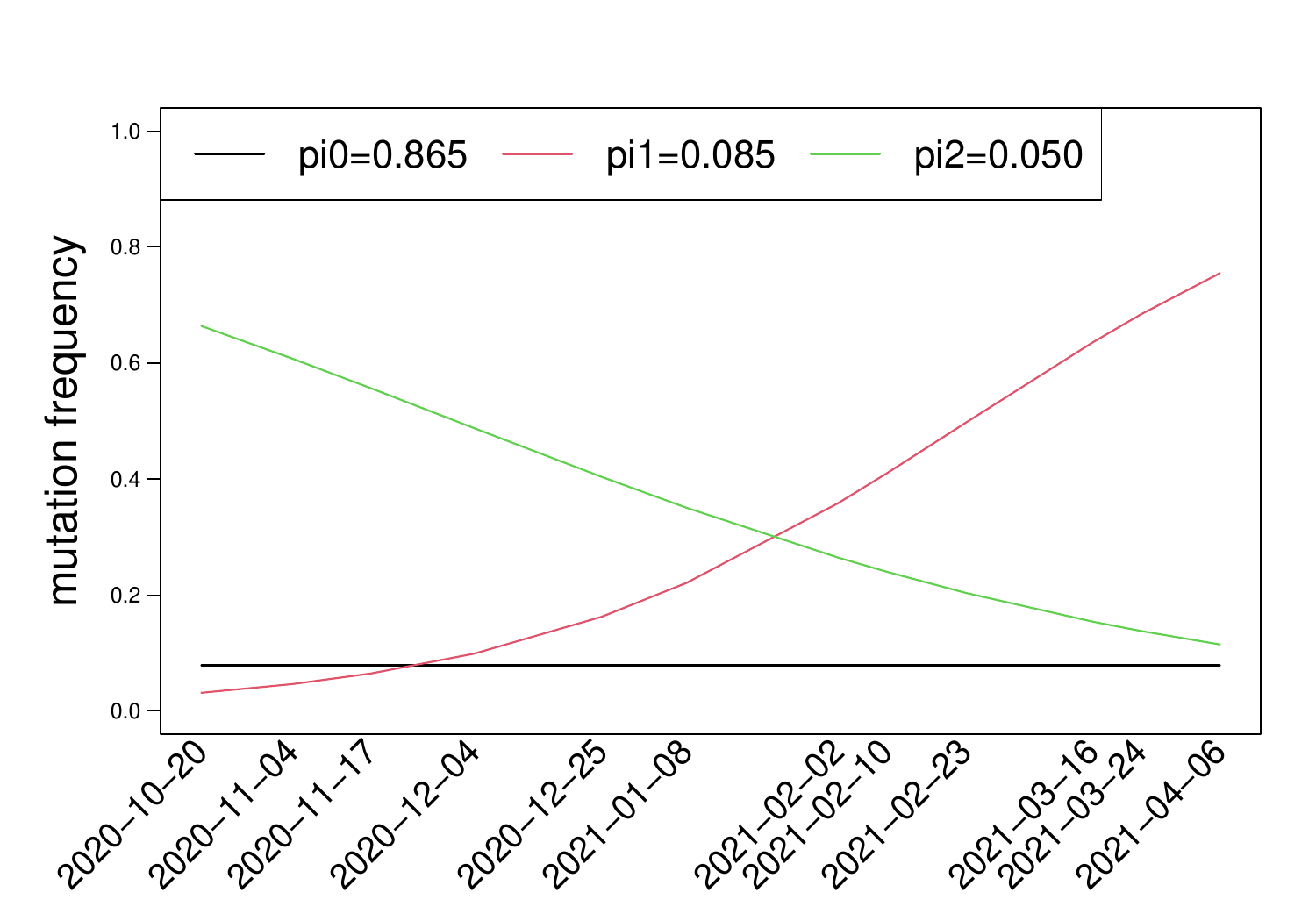}\par 
\end{multicols}
\vspace{-3em}
\begin{multicols}{2}
    \includegraphics[width=\linewidth]{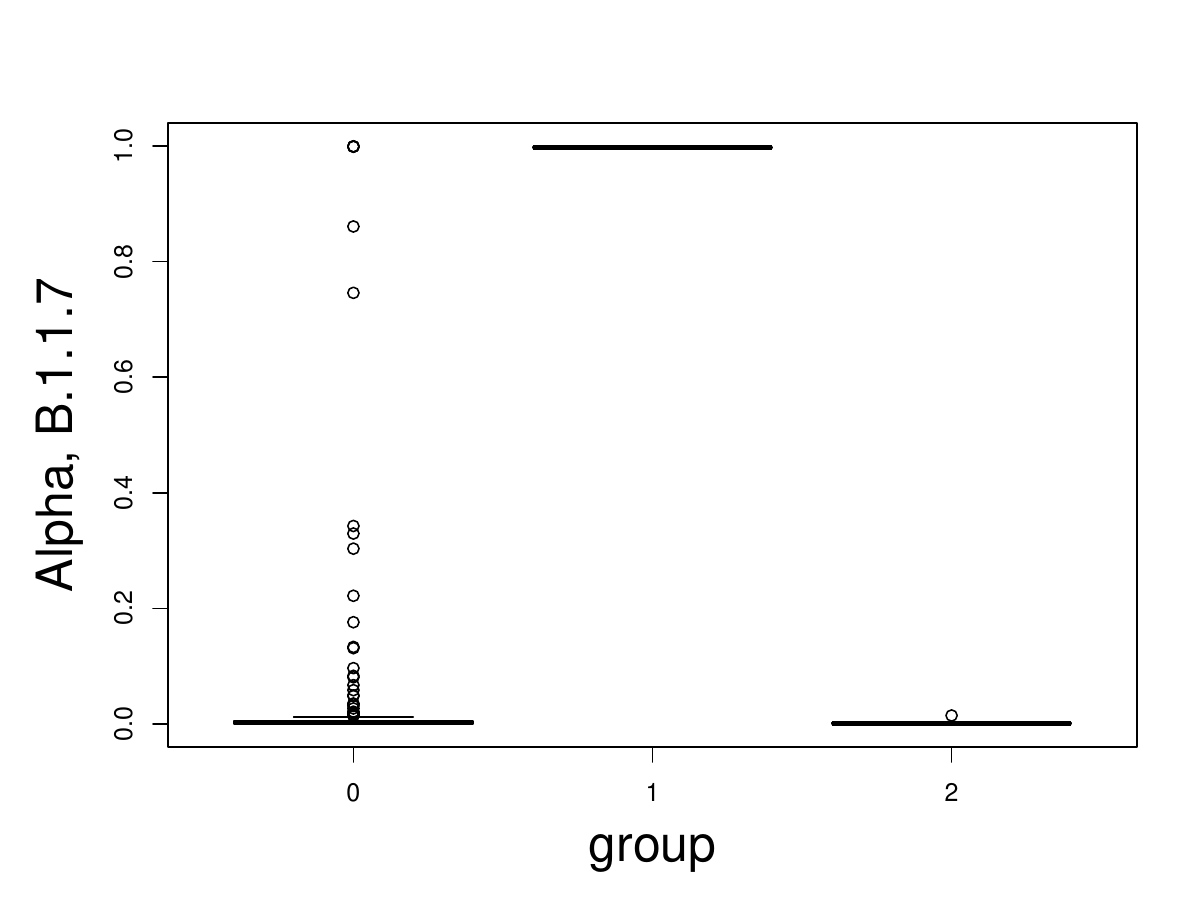}\par
    \includegraphics[width=\linewidth]{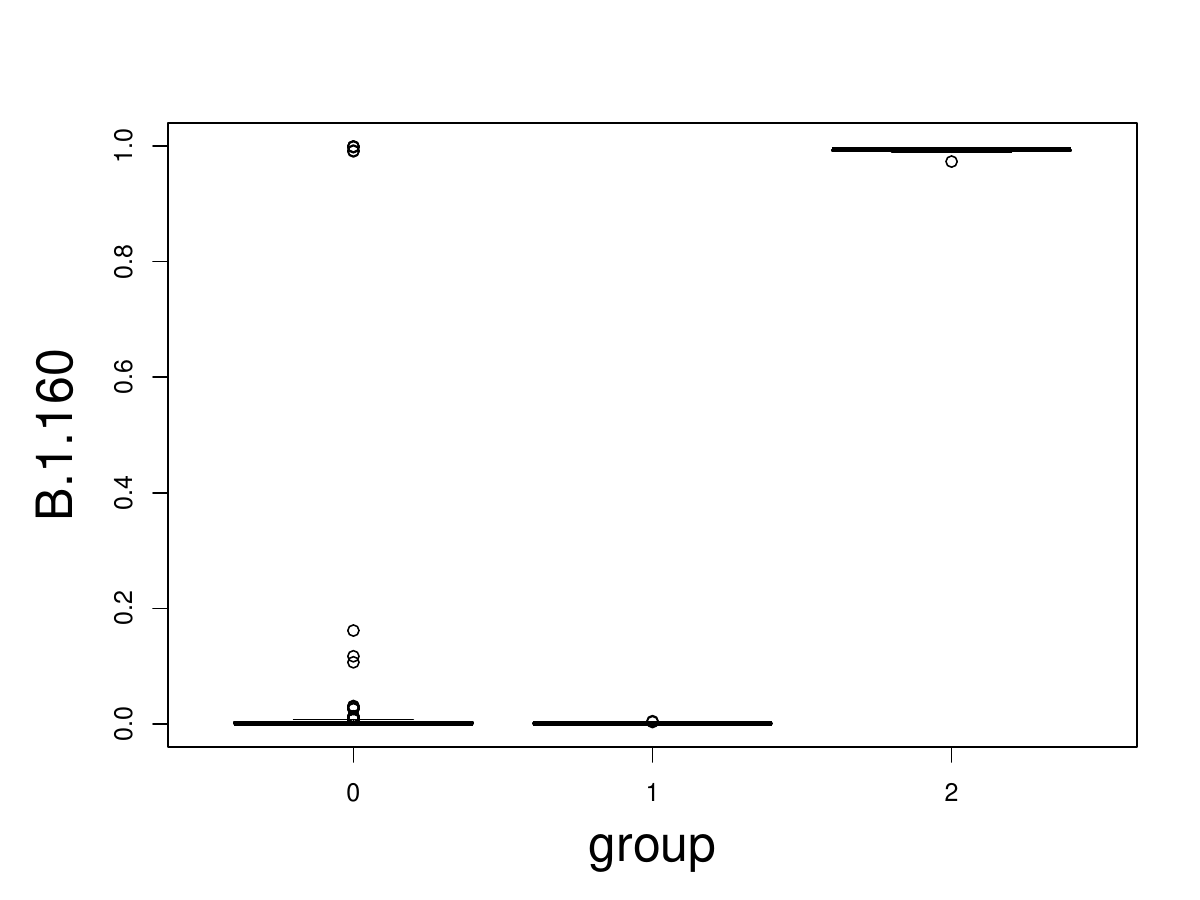}\par
\end{multicols}

\caption{Analysis D (reduced WWTP1 dataset and $\hat K^\text{D}=2$).}
\label{fig:reduced.estim.w1k2}
\end{subfigure}


\begin{subfigure}{1\textwidth}
\begin{multicols}{2}
    
    \begin{subfigure}{1 \linewidth}
    \vspace{1em}
    \centering
    \small
    \begin{tabular}{|r|c|c|c|c|c|}
    \cline{2-4}
    \multicolumn{1}{c|}{} & \multicolumn{3}{c|}{Group} \\
    \cline{2-4}
    \multicolumn{1}{c|}{} & 0 & 1 & 2 \\
    \hline
    $\hat\pi^\text{E}$ &  0.861 & 0.081 & 0.058\\
    $\hat\mu^\text{E}$ &-2.57 & -3.91 & 0.46\\
    $[95\% \text{CI}]$& - & [-3.93; -3.89] & [0.45; 0.48]\\
    $\hat s^\text{E} \times 100$&-&3.42 & -2.05\\
    $[95\% \text{CI}]$&- & [3.41; 3.44] & [-2.07; -2.03]\\
    \hline
    \end{tabular}
    \end{subfigure}
    
    \hspace{3em}
    \includegraphics[width=0.8\linewidth]{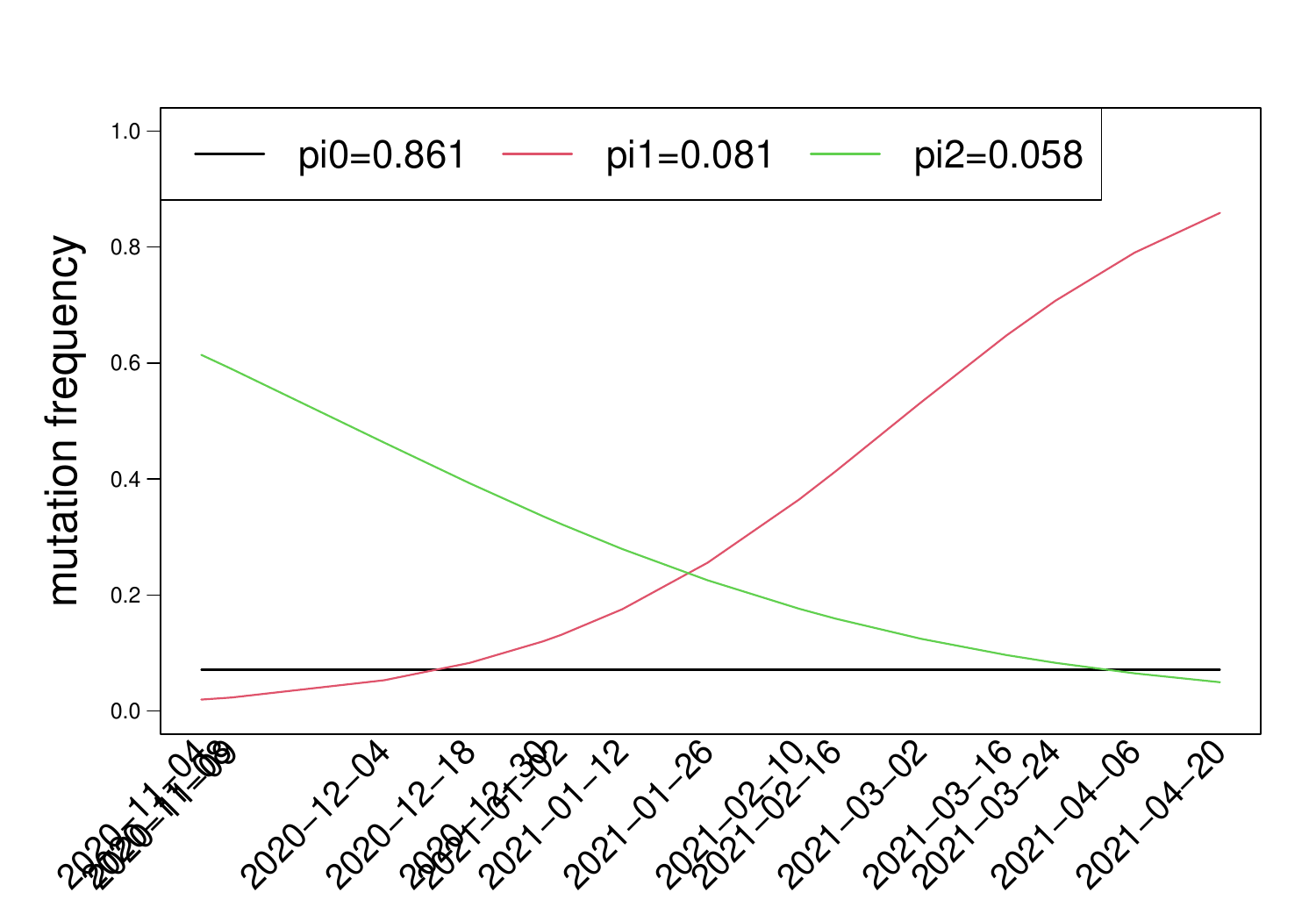}\par 
\end{multicols}
\vspace{-3em}
\begin{multicols}{2}
    \includegraphics[width=\linewidth]{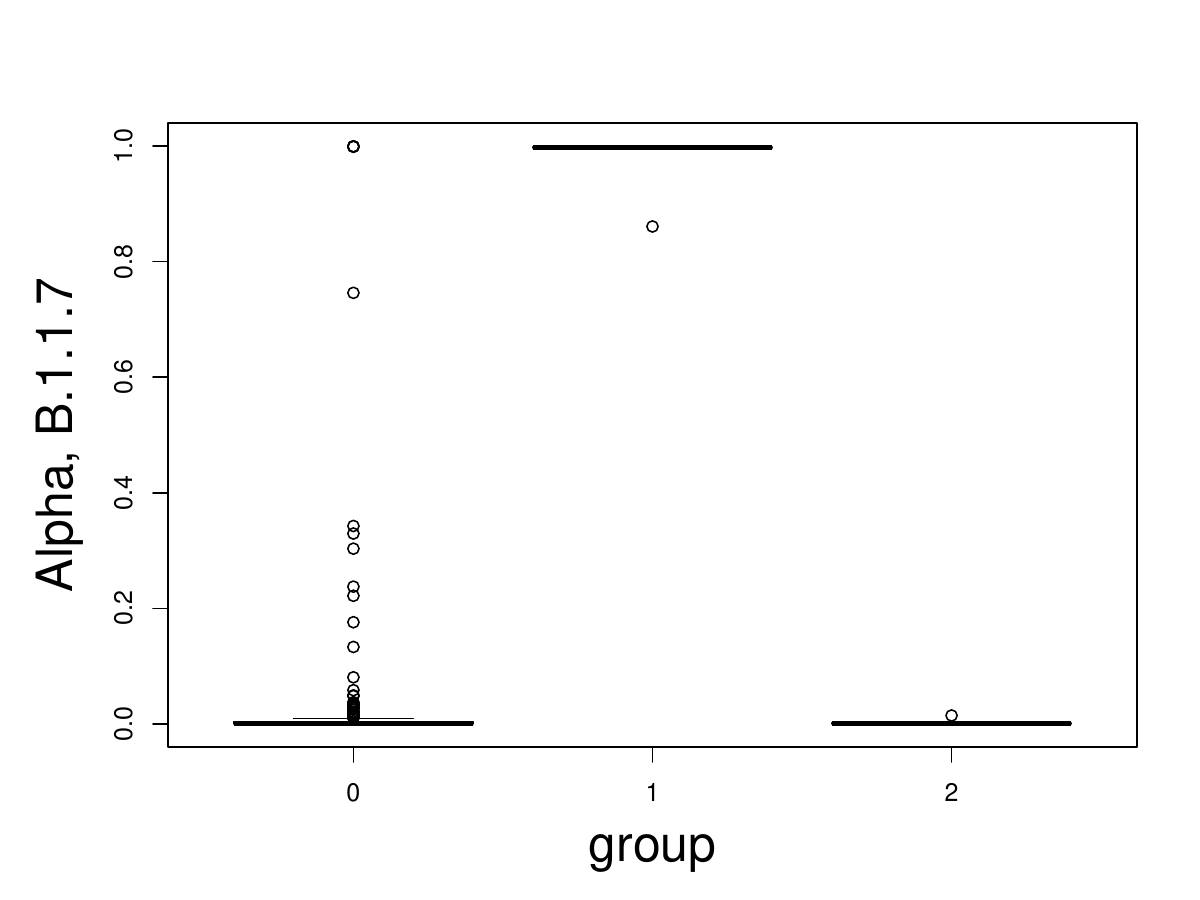}\par
    \includegraphics[width=\linewidth]{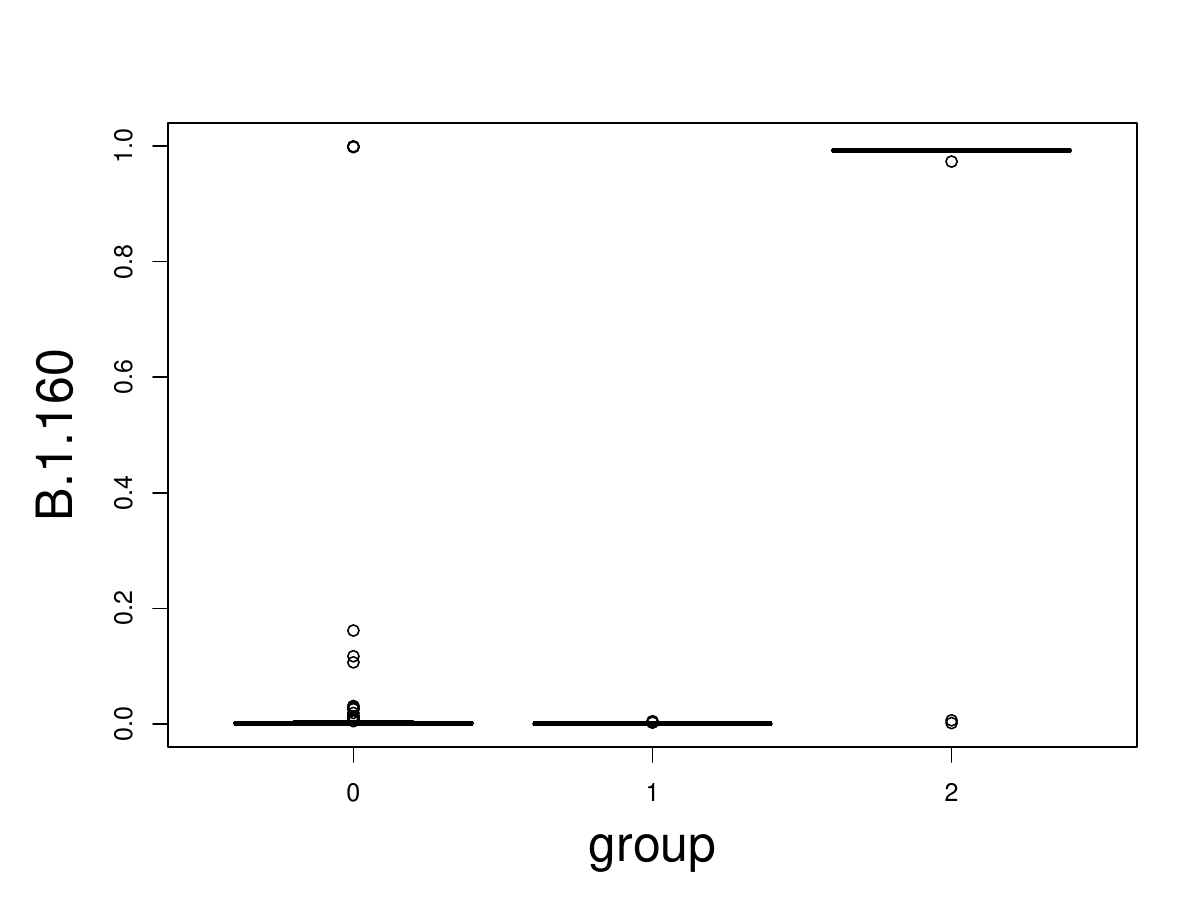}\par
\end{multicols}

\caption{Analysis E (reduced WWTP2 dataset and $\hat K^\text{E}=2$).}
\label{fig:reduced.estim.w2k2}
\end{subfigure}

\caption{Parameter estimates, associated group frequency trajectories and mutation profiles stratified on maximum a posteriori probability of group affectation computed in Analysis D and Analysis E performed respectively over WWTP1 and WWTP2 datasets both reduced to mutations associated with a probability above 0.005 to belong to at least one main circulating lineage at the time of the analysis.}
\label{fig:reduced.estim.w.k2}
\end{figure*}

%% file: fig_reduced_data_WWTP1_K6_estimation.tex
\begin{figure*}
\vspace{-6em}
\begin{subfigure}{1 \linewidth}
\centering
\small
\setlength{\tabcolsep}{3.5pt}
\begin{tabular}{|r|c|c|c|c|c|c|c|}
\cline{2-8}
\multicolumn{1}{c|}{} & \multicolumn{7}{c|}{Group} \\
\cline{2-8}
\multicolumn{1}{c|}{} & 0 & 1 & 2 & 3 & 4 & 5 & 6\\
\hline
$\hat\pi^\text{F}$ & 0.764 & 0.050 & 0.050 & 0.036 & 0.055 & 0.036 & 0.011\\
$\hat\mu^\text{F}$ & -2.89 & -2.91 & 0.68 & -1.48 & -4.82 & -5.65 & 2.53\\
$[95\% \text{CI}]$& - & [-2.94; -2.89] & [0.66; 0.70]& [-1.52; -1.44] & [-4.87; -4.78] & [-5.68; -5.61] & [2.45; 2.60]\\
$\hat s^\text{F} \times 100$&-& 2.45 & -1.62 & -1.52 & 1.80 & 4.08 & -0.11\\
$[95\% \text{CI}]$&- & [2.43; 2.47] & [-1.64; -1.60] & [-1.58; -1.46] & [1.77; 1.84] & [4.05; 4.11] & [-0.19; -0.04]\\
\hline
\end{tabular}
\caption{Parameter estimates. Selection coefficients and their 95\%CI boundaries are multiplied by 100. The estimate $\hat\mu^\text{F}_0$ is computed from estimates $\hat\alpha^\text{F} = 1.16$ and $\hat\beta^\text{F} = 21.0$.}
\label{fig:par.w1k6.reduced}
\end{subfigure}

\begin{subfigure}{1\textwidth}
\begin{multicols}{2}
    \includegraphics[width=1\linewidth]{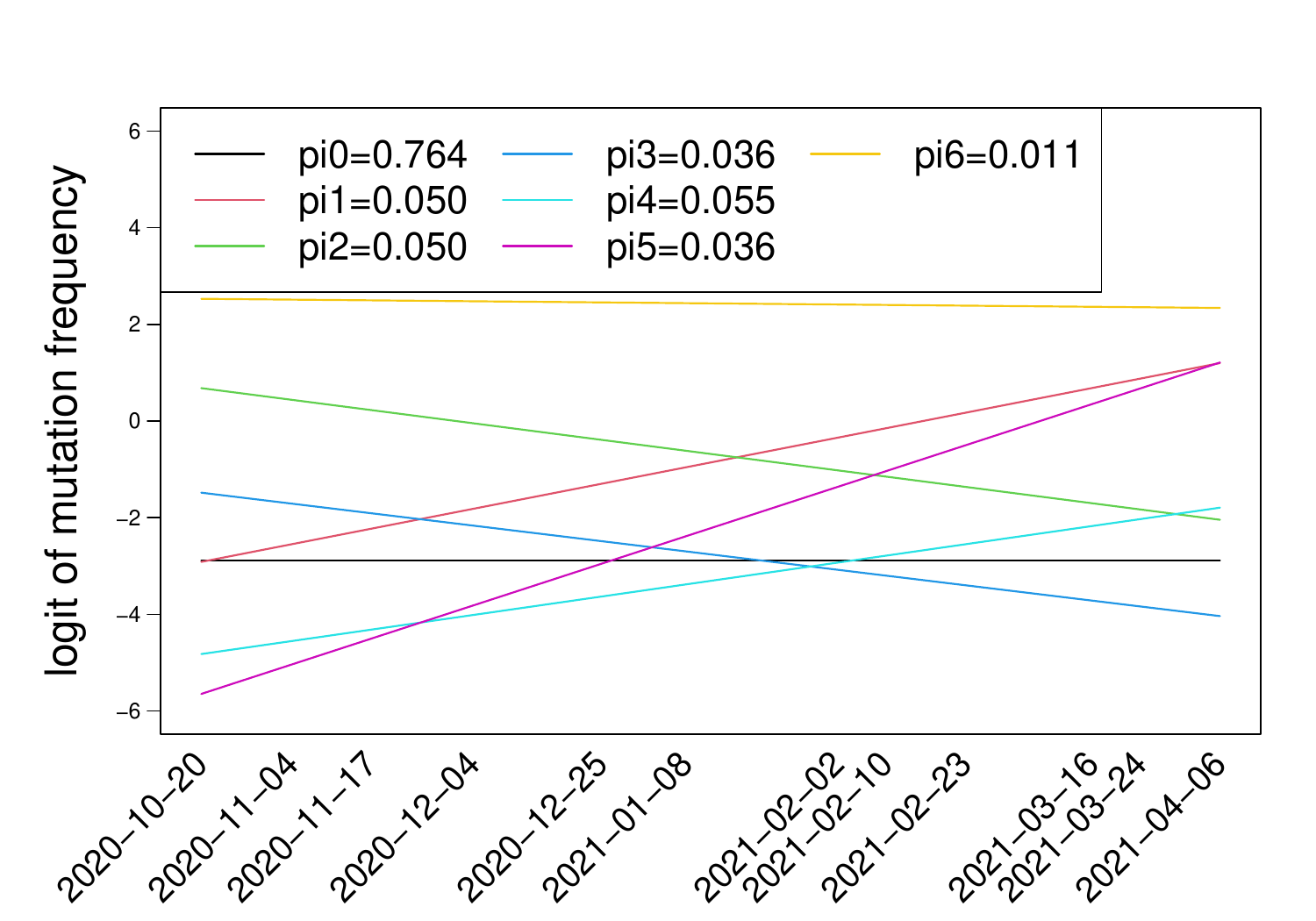}\par 
    \includegraphics[width=\linewidth]{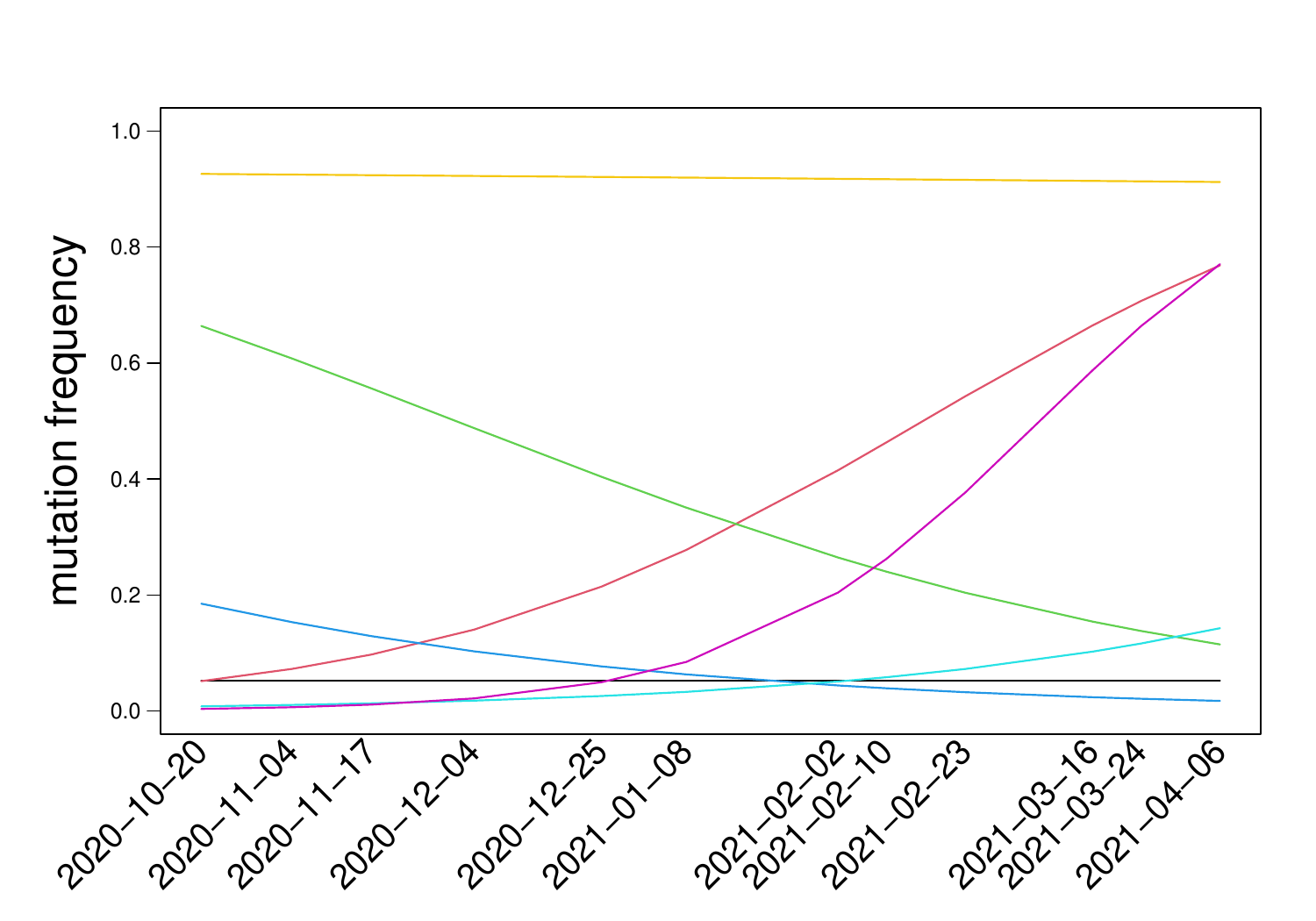}\par 
\end{multicols}
\caption{Estimated group log-frequency (left) and frequency (right) trajectories.}
\label{fig:traj.w1k6.reduced}
\end{subfigure}

\begin{subfigure}{1\textwidth}
\begin{multicols}{2}
    \includegraphics[width=\linewidth]{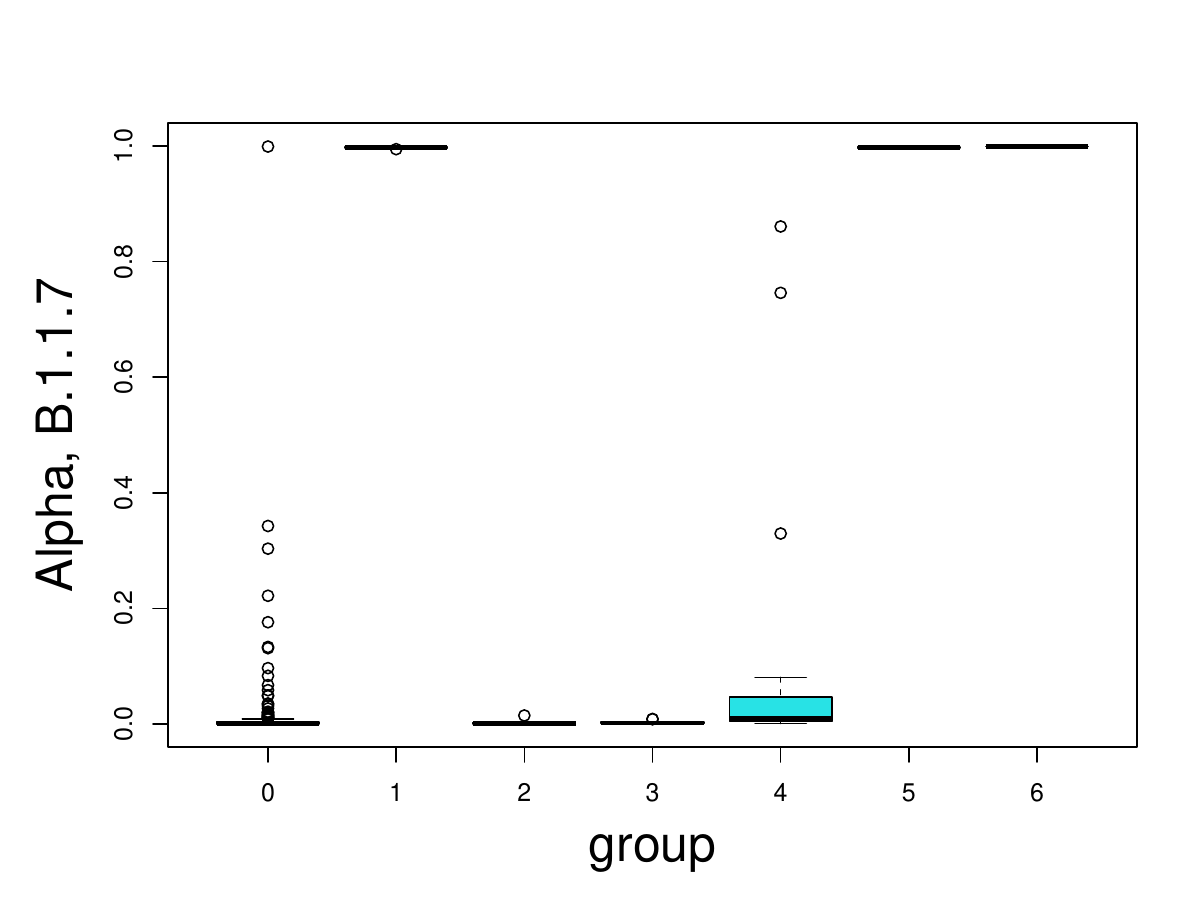}\par
    \includegraphics[width=\linewidth]{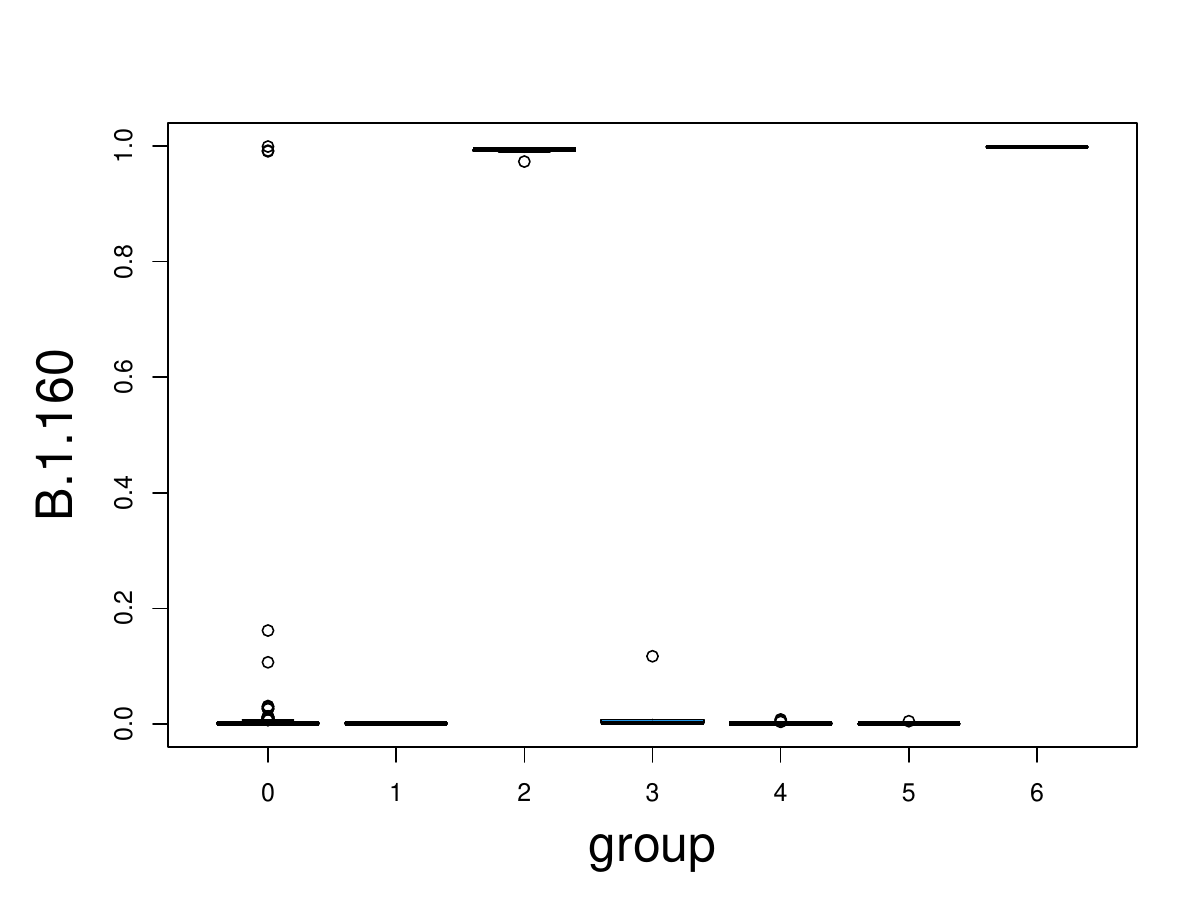}\par
\end{multicols}

\vspace{-3em}
\begin{multicols}{2}
    \includegraphics[width=\linewidth]{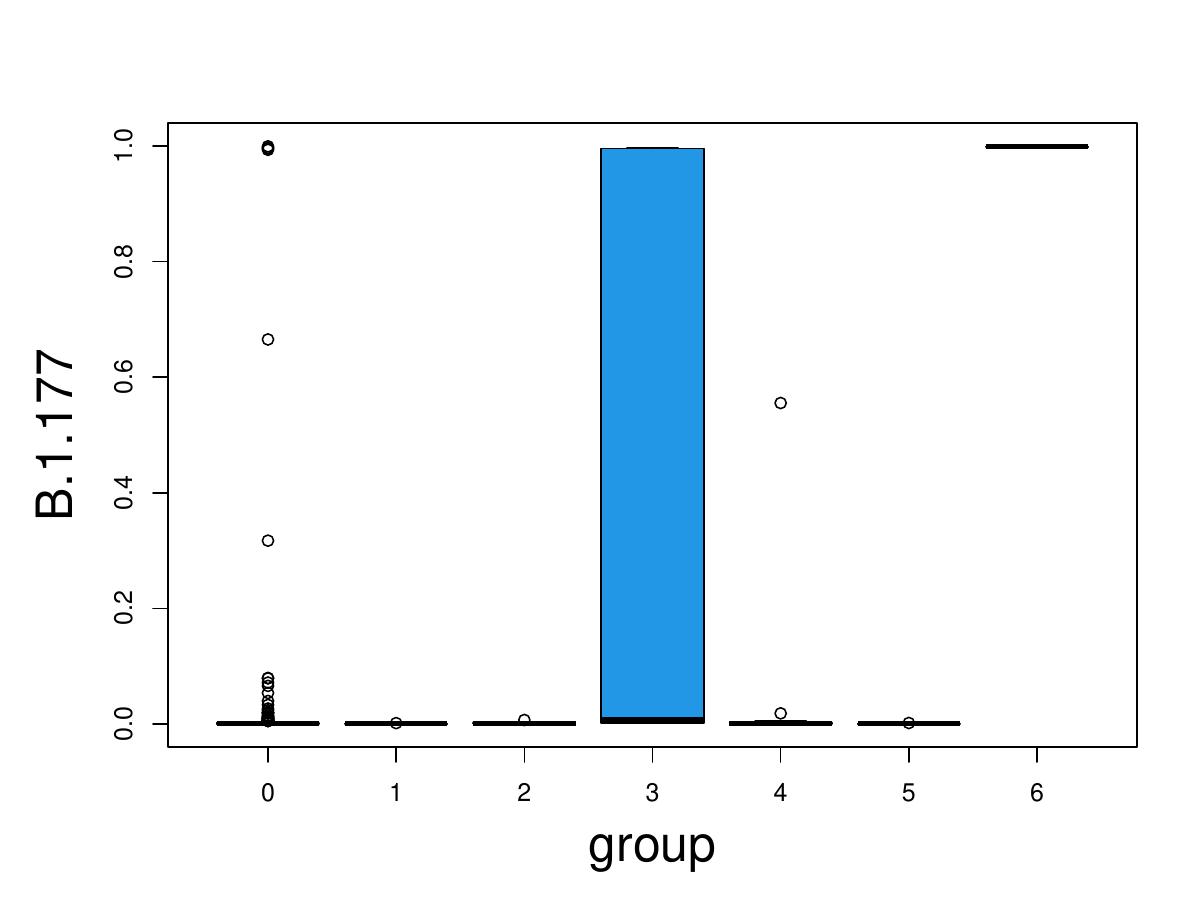}\par
    \includegraphics[width=\linewidth]{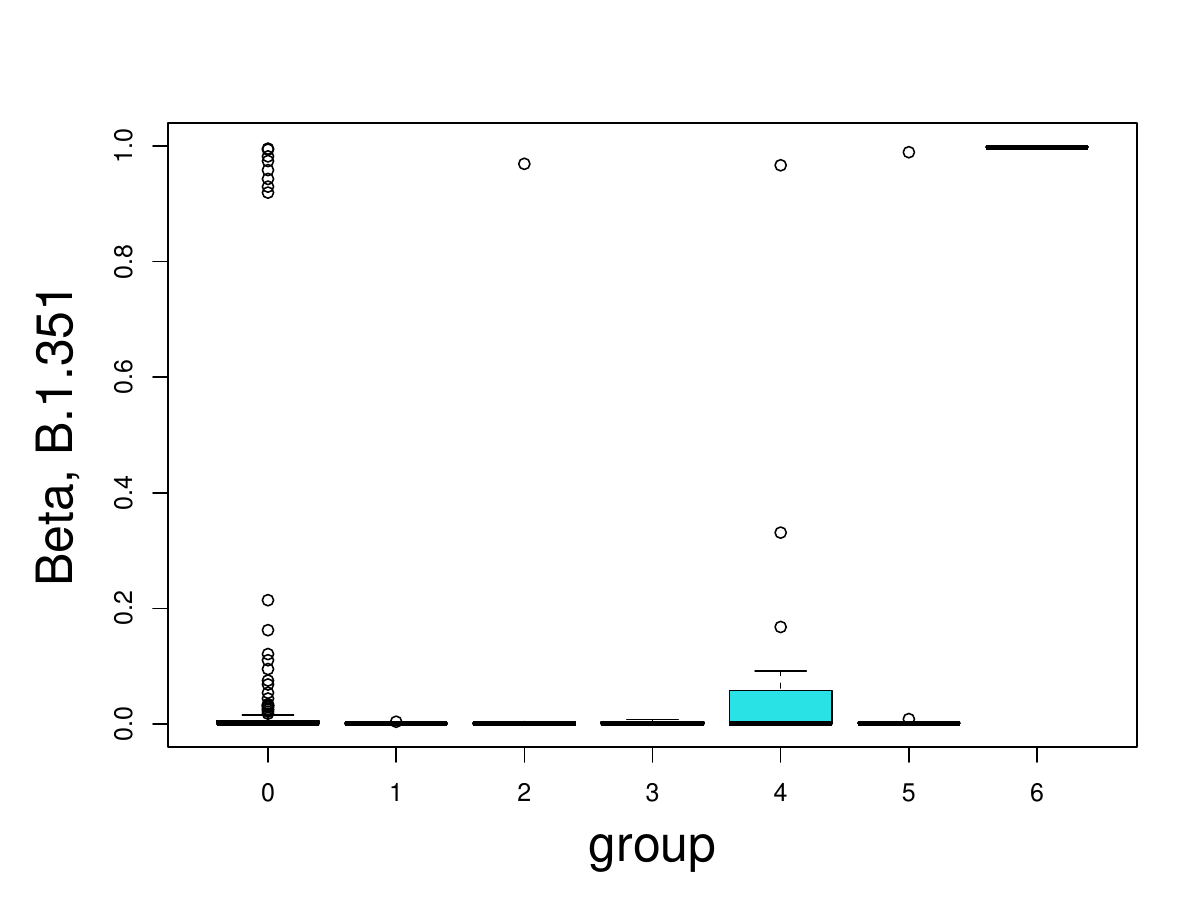}\par
\end{multicols}
\caption{Mutation signatures stratified on posterior group affectation.}
\label{fig:signature.w1k6.reduced}
\end{subfigure}

\caption{Parameter estimates (Figure~\ref{fig:par.w1k6.reduced}), associated group log-frequency and frequency trajectories (Figure~\ref{fig:traj.w1k6.reduced}) and mutation profiles stratified on maximum a posteriori probability of group affectation (Figure~\ref{fig:signature.w1k6.reduced}) computed in Analysis F (conditional on $\hat K^\text{F}=6$ non-neutral groups and WWTP1 dataset reduced to mutations associated with probability above 0.005 to belong to at least one main circulating lineage at the time of the analysis).}
\label{fig:reduced.estim.w1k6}
\end{figure*}

%% file: tab_signature_WWTP1_rising.tex
\begin{table}
\vspace{-8em}
\centering
\begin{subtable}{1\textwidth}
\begin{tabular}{r||c|c|c|c|c|}
\multicolumn{1}{c||}{Mutations} & \multicolumn{5}{c}{Lineages} \\ 
\hline 
\multicolumn{1}{l||}{\textcolor{red}{Group $G_1$}}  & B.1.1.7 (Alpha) & B.1.1 & B.1.160 & B.1.177 & B.1.351\\ 
\hline
\hline
C913T & \textbf{99.9} & 0.6 & 0.1 & - & 0.1\\
C3267T & \textbf{99.8} & 0.7 & 0.1 & 0.1 & -\\
C14676T & \textbf{99.8} & 0.8 & 0.2 & 0.1 & 0.2\\
T16176C & \textbf{99.9} & 0.4 & - & - & 0.4\\
C23271A & \textbf{99.8} & 0.5 & - & - & -\\
C23604A & \textbf{99.8} & 1.7 & - & - & 0.1\\
C23709T & \textbf{99.8} & 0.4 & 0.1 & 0.1 & 0.1\\
A28111G & \textbf{99.8} & 0.6 & - & - & -\\
G28280C & \textbf{99.5} & 0.5 & - & - & -\\
A28281T & \textbf{99.5} & 0.4 & - & - & -\\
T28282A & \textbf{99.6} & 0.5 & - & - & -\\
G28881A & \textbf{99.8} & \textbf{99.9} & 0.2 & 0.1 & 0.1\\
G28882A & \textbf{99.7} & \textbf{99.9} & 0.3 & 0.1 & 0.1\\
G28883C & \textbf{99.7} & \textbf{99.8} & 0.2 & - & 0.1\\
\hline
\end{tabular}
\caption{Group $G_1$ (red in Figure~\ref{fig:reduced.estim.w1k6})}
\label{tab:signature.red}
\end{subtable}

\vspace{1em}

\begin{subtable}{1\textwidth}
\begin{tabular}{r||c|c|c|c|c|}
\multicolumn{1}{l||}{\textcolor{magenta}{Group $G_5$}}  & B.1.1.7 (Alpha) & B.1.1 & B.1.160 & B.1.177 & B.1.351\\ 
\hline
\hline
C5388A & \textbf{99.8} & 0.4 & 0.2 & - & 0.1\\
C5986T & \textbf{99.8} & 0.4 & 0.1 & 0.2 & 0.8\\
T6954C & \textbf{99.7} & 1.1 & - & 0.1 & 0.1\\
C15279T & \textbf{99.8} & 0.7 & 0.1 & - & 0.1\\
A23063T & \textbf{99.8} & 0.7 & - & - & \textbf{98.9}\\
T24506G & \textbf{99.8} & 0.8 & - & - & -\\
G24914C & \textbf{99.6} & 0.4 & - & - & 0.1\\
C27972T & \textbf{99.9} & 0.5 & 0.5 & 0.1 & 0.1\\
G28048T & \textbf{99.7} & 0.4 & 0.1 & - & 0.1\\
C28977T & \textbf{99.8} & 1.3 & 0.1 & - & 0.1\\
\hline
\end{tabular}
\caption{Group $G_5$ (magenta in Figure~\ref{fig:reduced.estim.w1k6})}
\label{tab:signature.magenta}
\end{subtable}

\caption{Mutations assigned to group $G_1$ (top) and to group $G_5$ (bottom) by maximum a posteriori probability in Analysis F (Figure~\ref{fig:reduced.estim.w1k6}) along with mutation signature for some of the main circulating variants. Probabilities above 90\% are in bold. Zeros are replaced by sign `-' for readability.}
\label{tab:signature.w1}
\end{table}

%% file: fig_detection_selection.tex
\begin{figure}

\centering

\begin{subfigure}{0.45\textwidth}
\includegraphics[width=\linewidth]{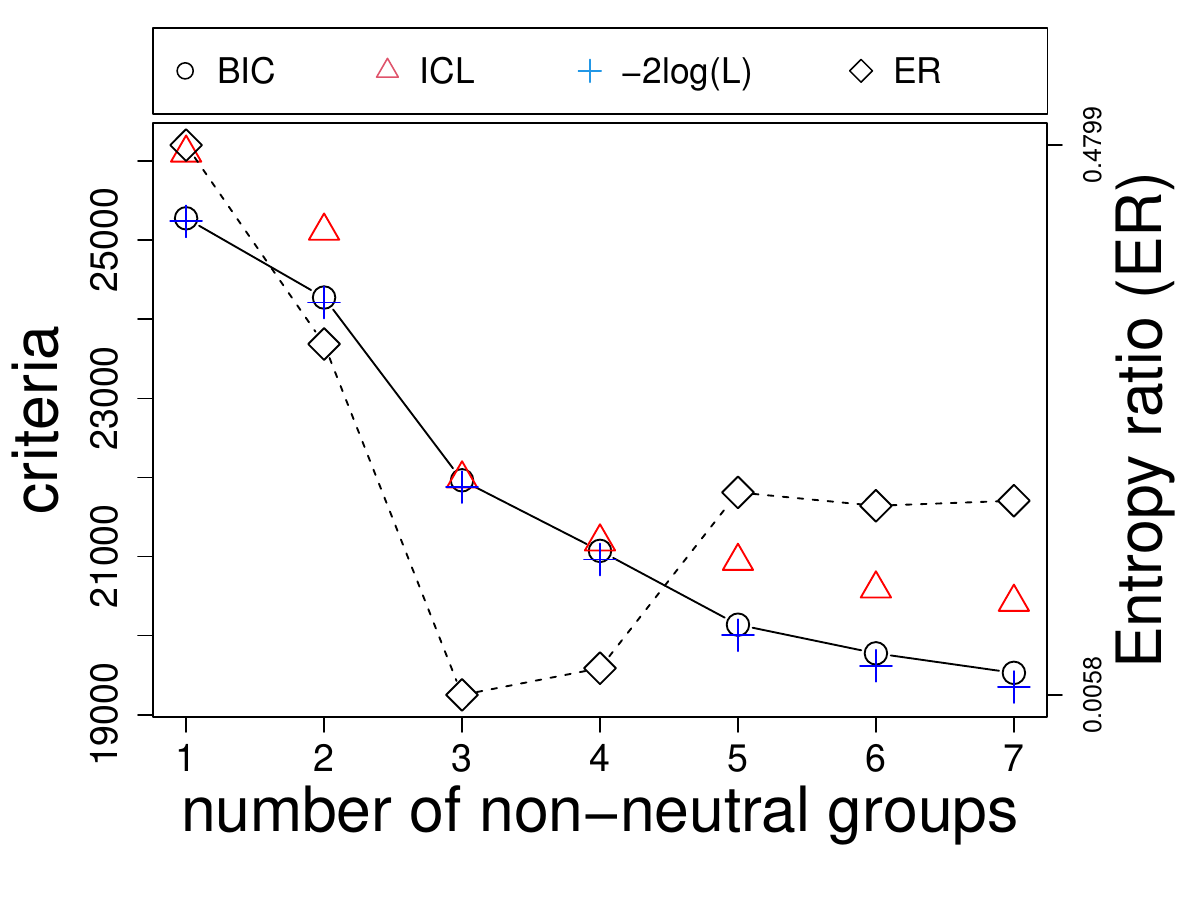}
\caption*{Analysis G}
\end{subfigure}
\hfil\hfil\hfil
\begin{subfigure}{0.45\textwidth}
\includegraphics[width=\linewidth]{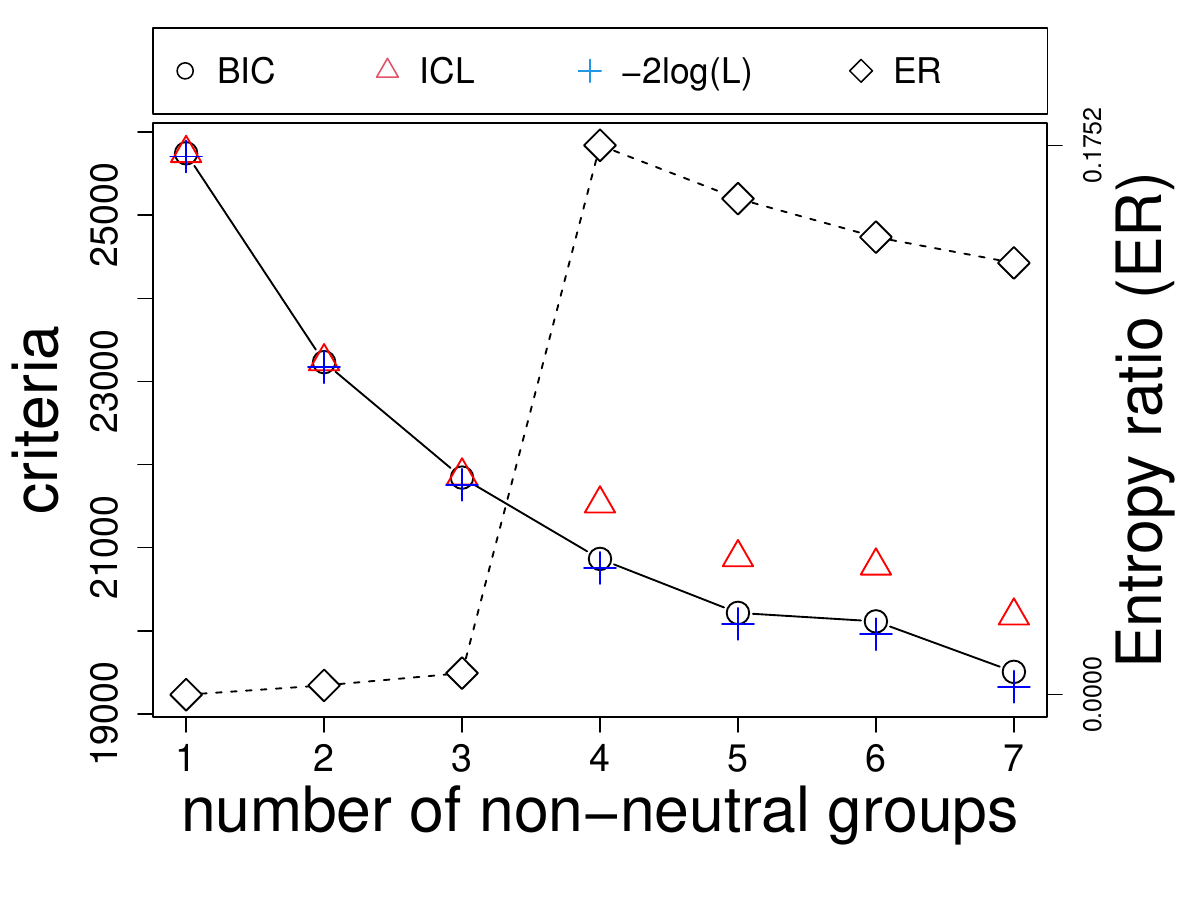}
\caption*{Analysis H}
\end{subfigure}

\vspace{1.5em}
\begin{subfigure}{0.45\textwidth}
\includegraphics[width=\linewidth]{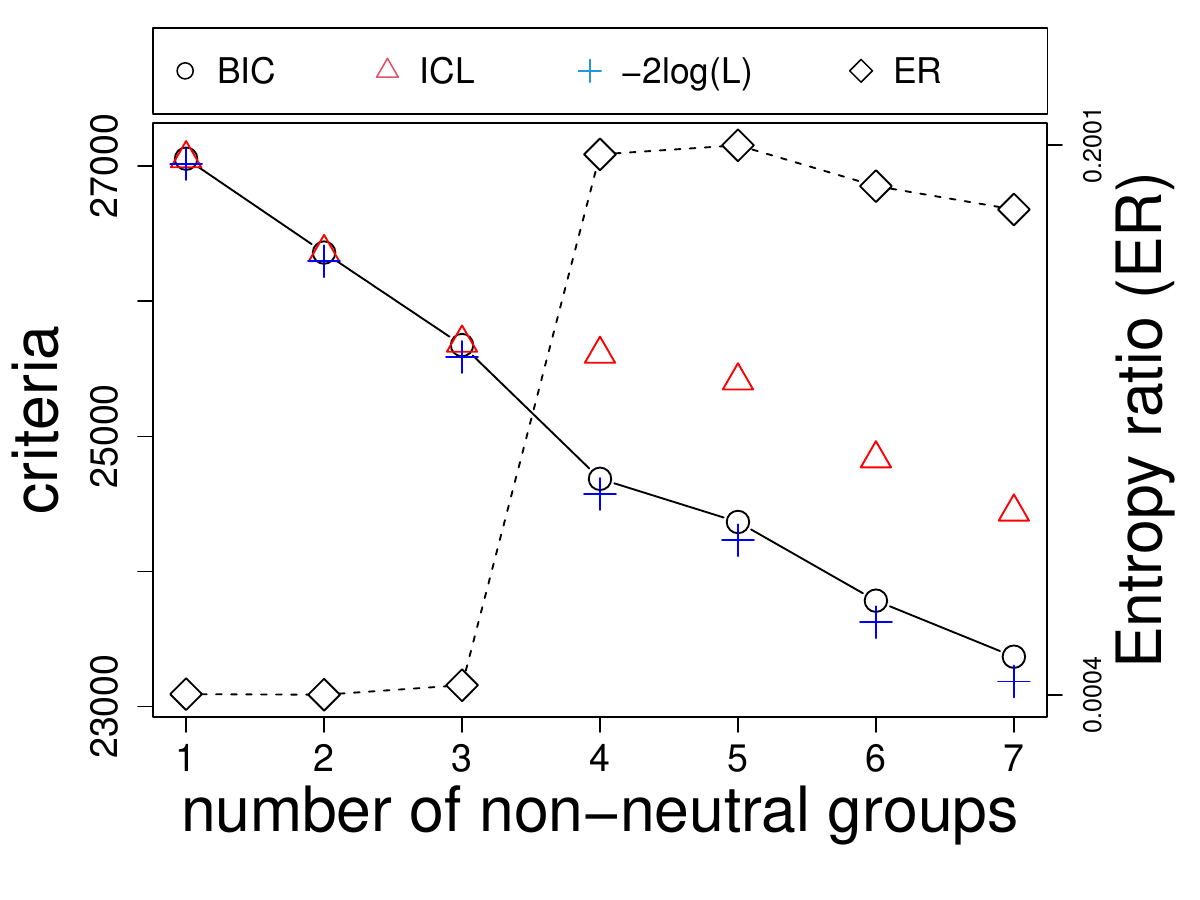}
\caption*{Analysis I}
\end{subfigure}
\hfil\hfil\hfil
\begin{subfigure}{0.45\textwidth}
\includegraphics[width=\linewidth]{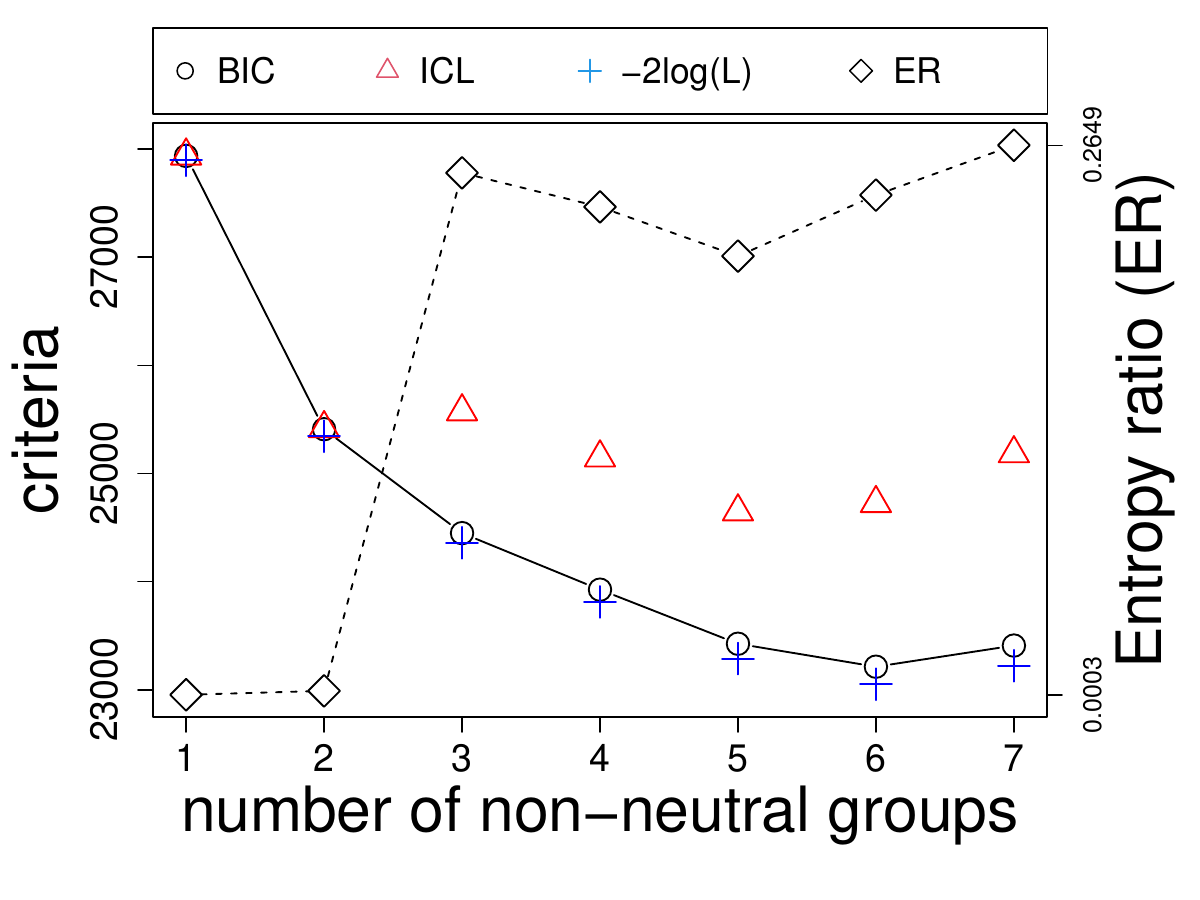}
\caption*{Analysis J}
\end{subfigure}

\caption{BIC, ICL along with minus two times the log-likelihood ($-2\log L$) of models composed of 1 to 7 non-neutral groups with WWTP1 dataset restricted to time points 2020-10-20 and 2020-11-04 (Analysis G) or 2020-11-04 and 2020-11-17 (Analysis H) and WWTP2 dataset restricted to time points 2020-11-09 and 2020-12-04 (Analysis I) or 2020-12-04 and 2020-12-18 (Analysis J). Entropy ratio (ER) added on each figure is associated with the right axis.}
\label{fig:detect.sel}
\end{figure}

%% file: fig_detection_estimation.tex
\begin{figure*}
\vspace{-4em}

\begin{subfigure}{1\textwidth}
\caption*{Analysis G}

\vspace{-2em}
\begin{multicols}{3}
    \includegraphics[width=\linewidth]{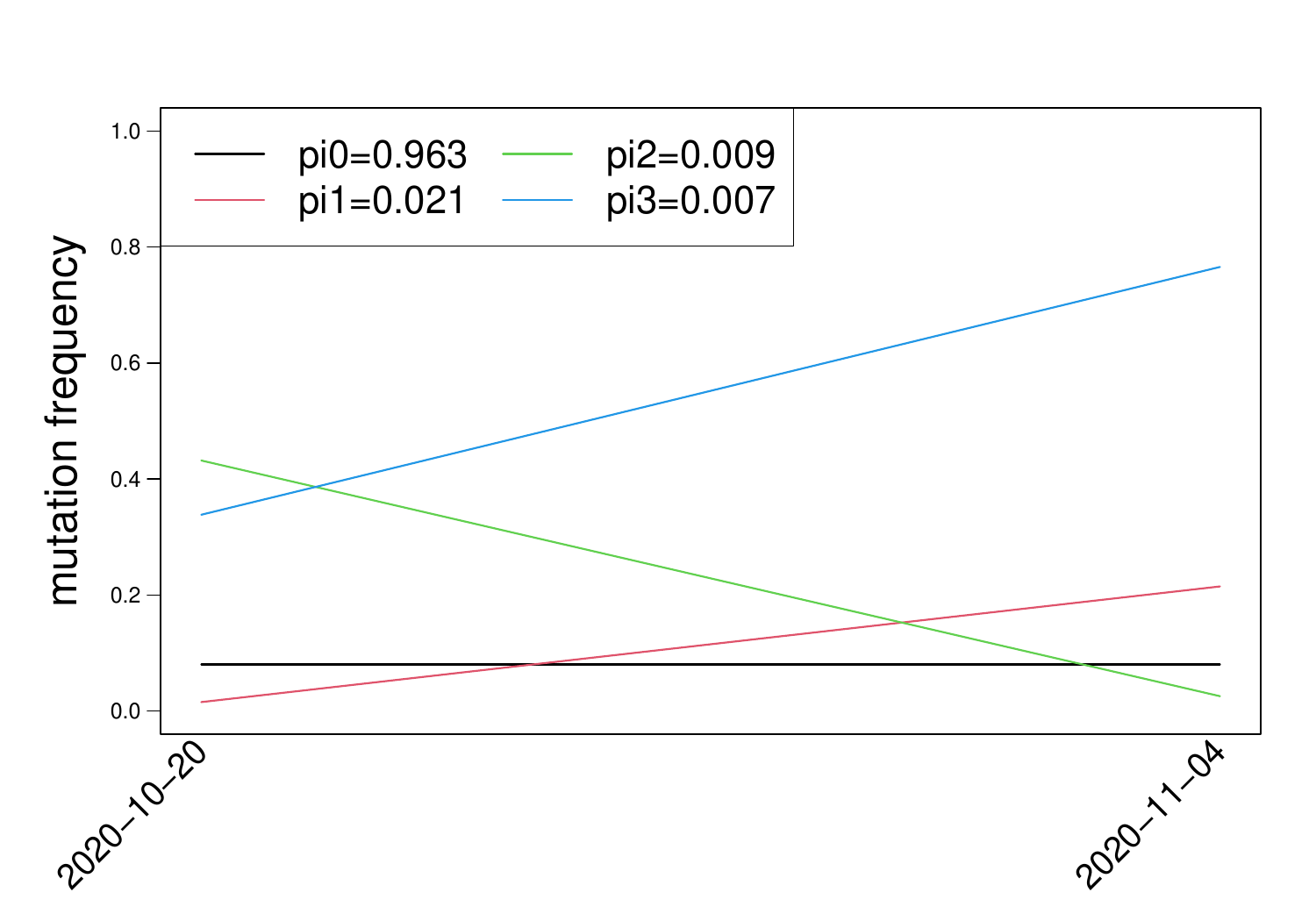}\par
    \includegraphics[width=\linewidth]{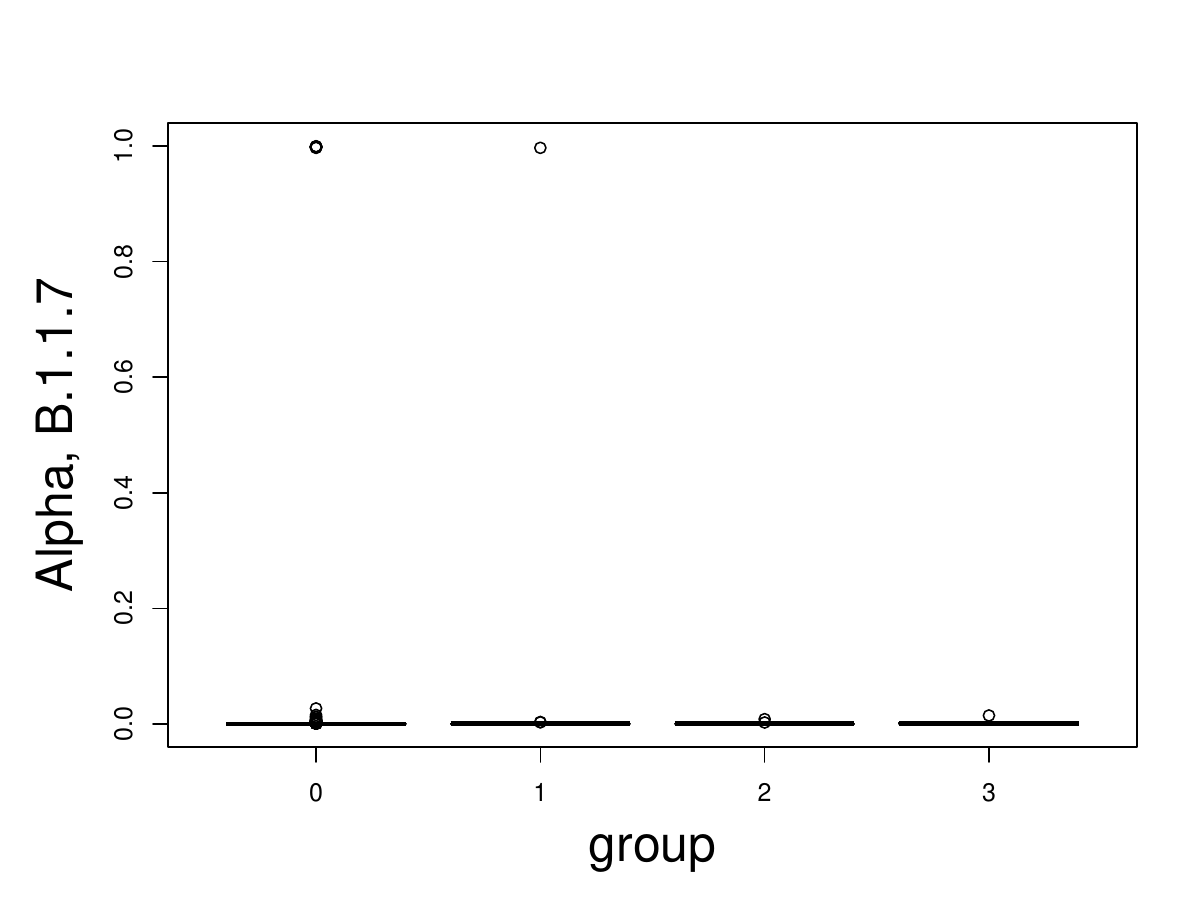}\par 
    \includegraphics[width=\linewidth]{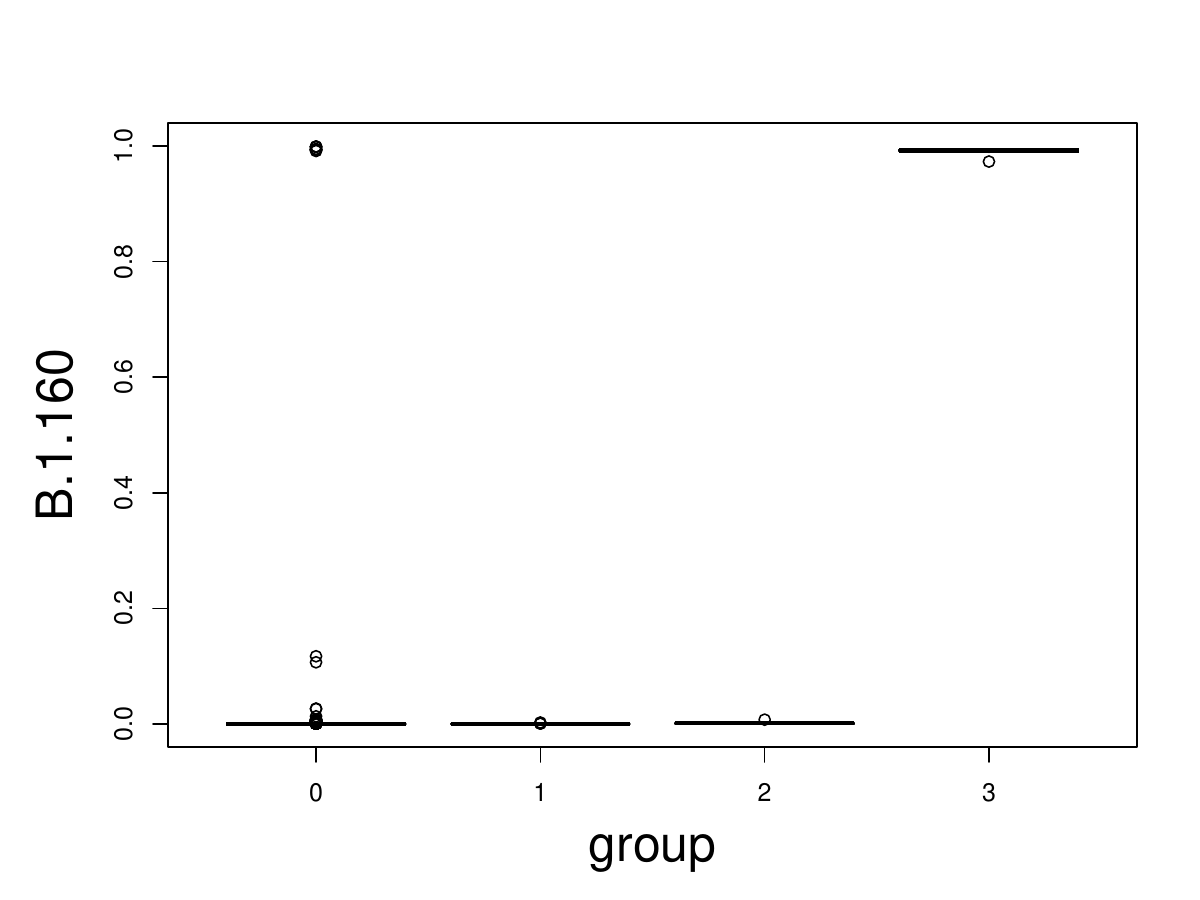}\par     
\end{multicols}


\caption*{Analysis H}

\vspace{-2em}

\begin{multicols}{3}
    \includegraphics[width=\linewidth]{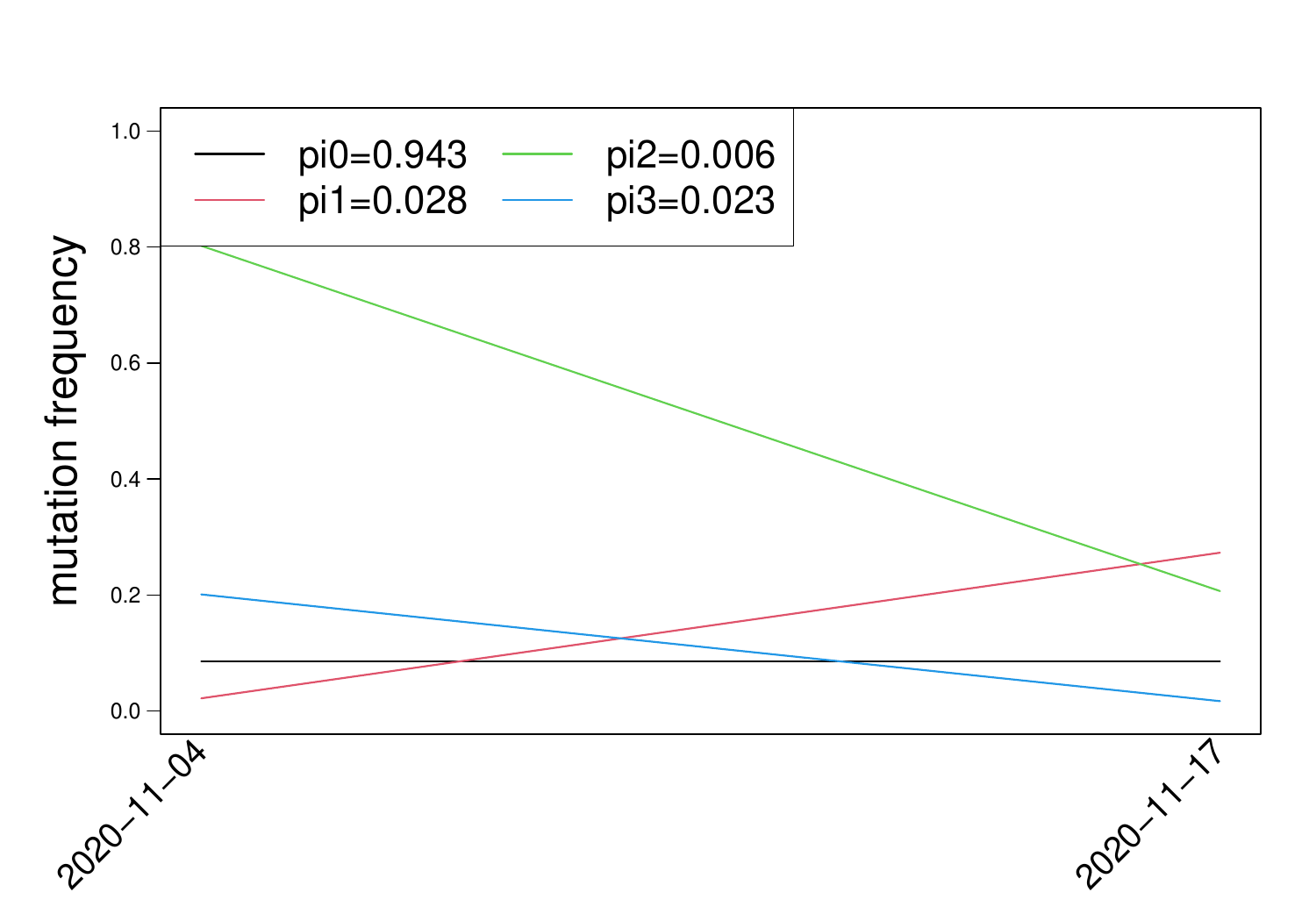}\par
    \includegraphics[width=\linewidth]{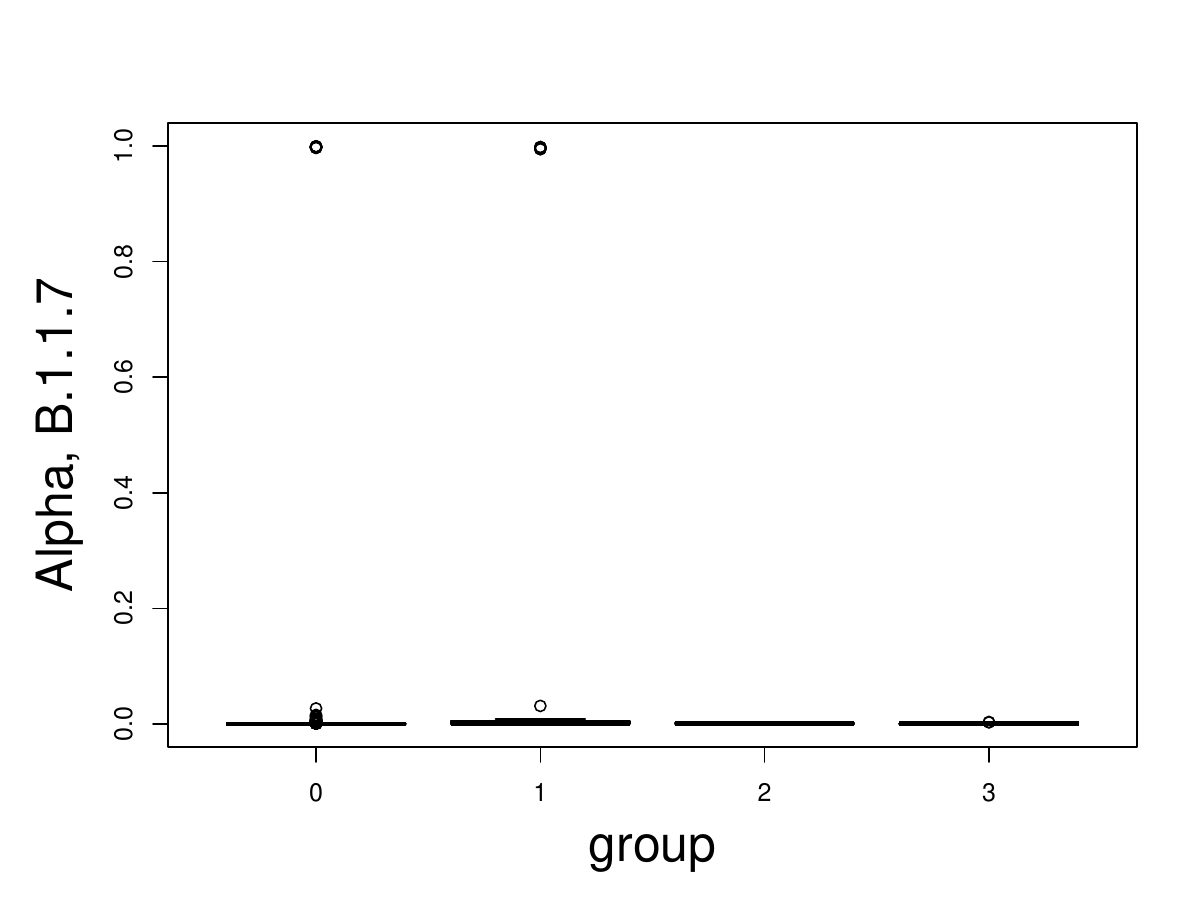}\par 
    \includegraphics[width=\linewidth]{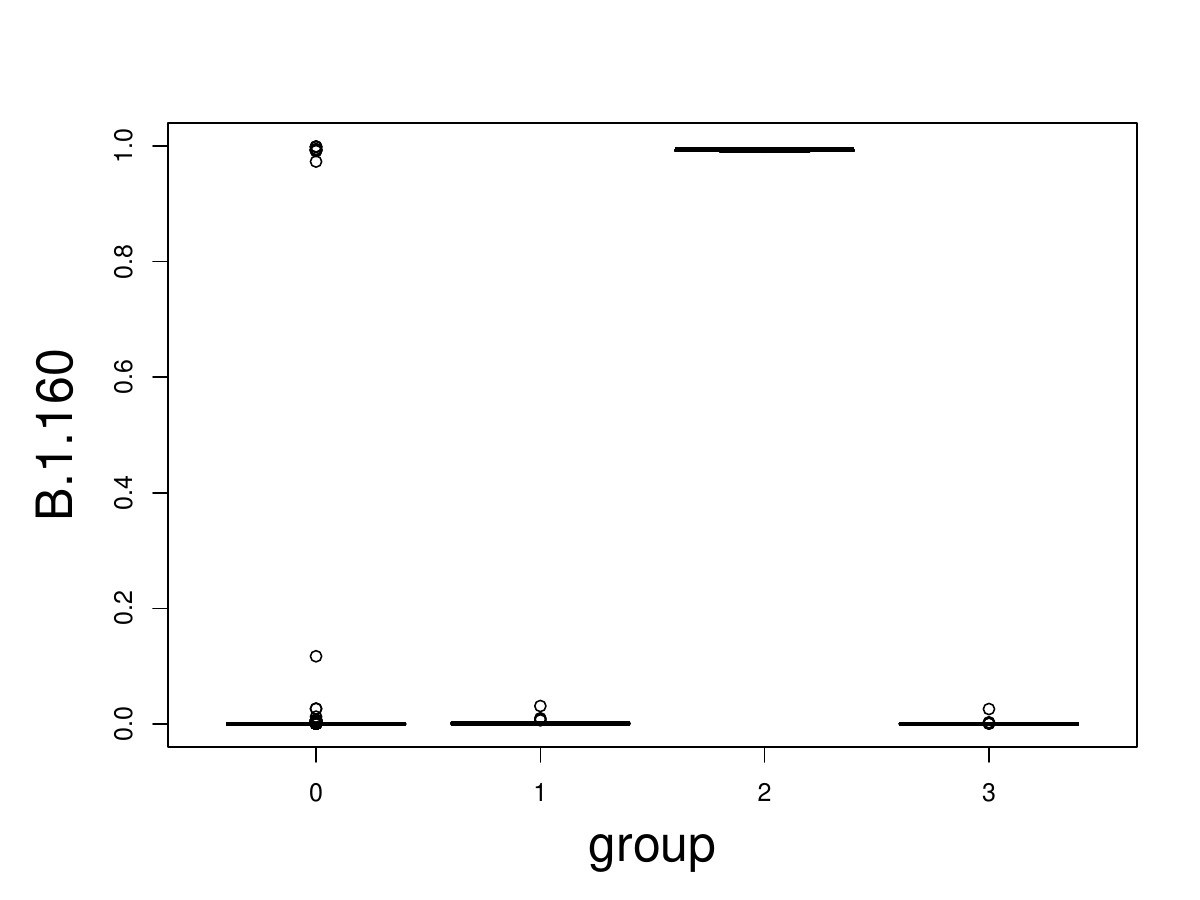}\par        
\end{multicols}

\vspace{-1.5em}
\caption{WWTP1}
\label{fig:detect.estim1}
\end{subfigure}

\vspace{1.5em}
\begin{subfigure}{1\textwidth}

\caption*{Analysis I}
\vspace{-2em}\begin{multicols}{3}
    \includegraphics[width=\linewidth]{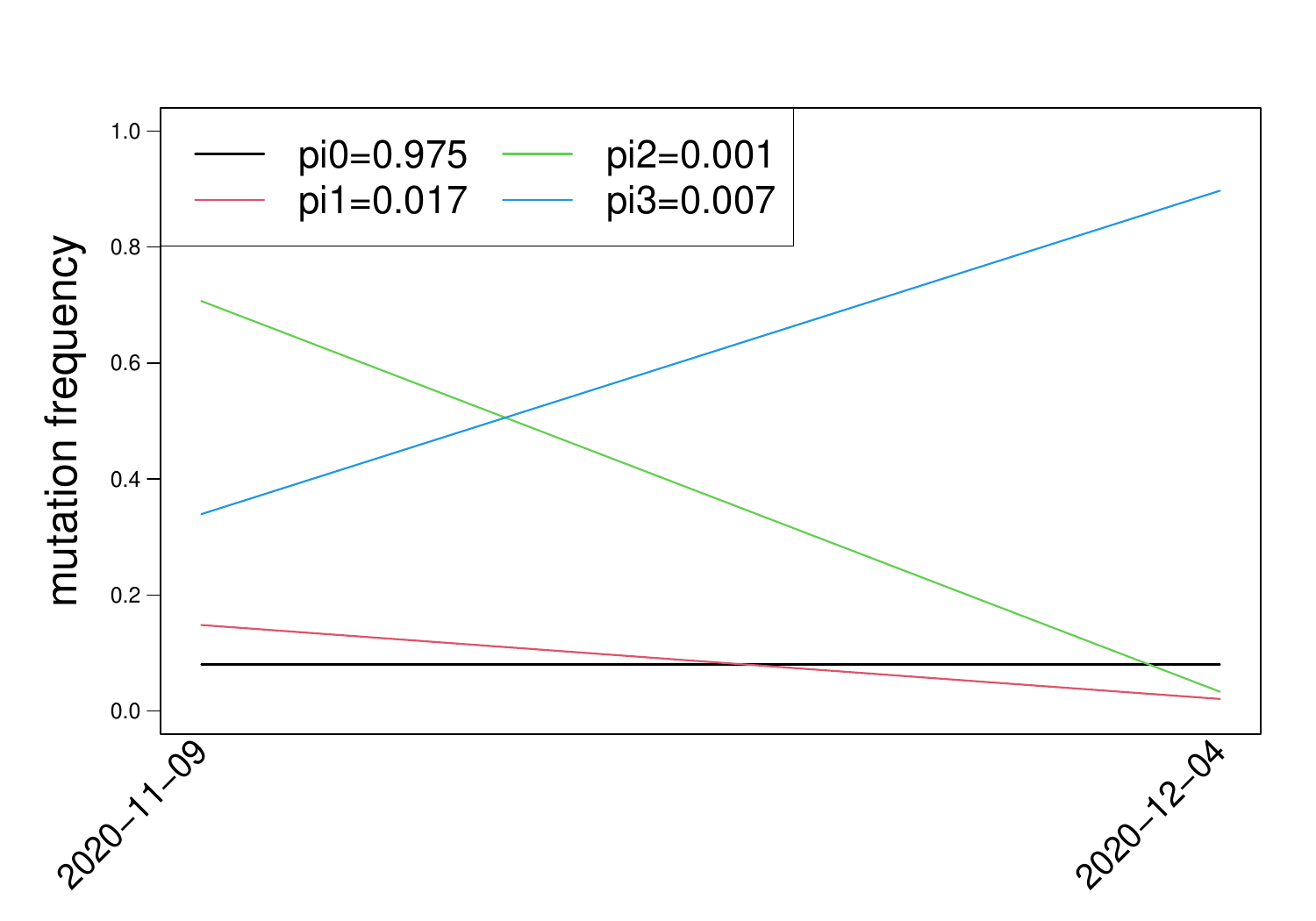}\par
    \includegraphics[width=\linewidth]{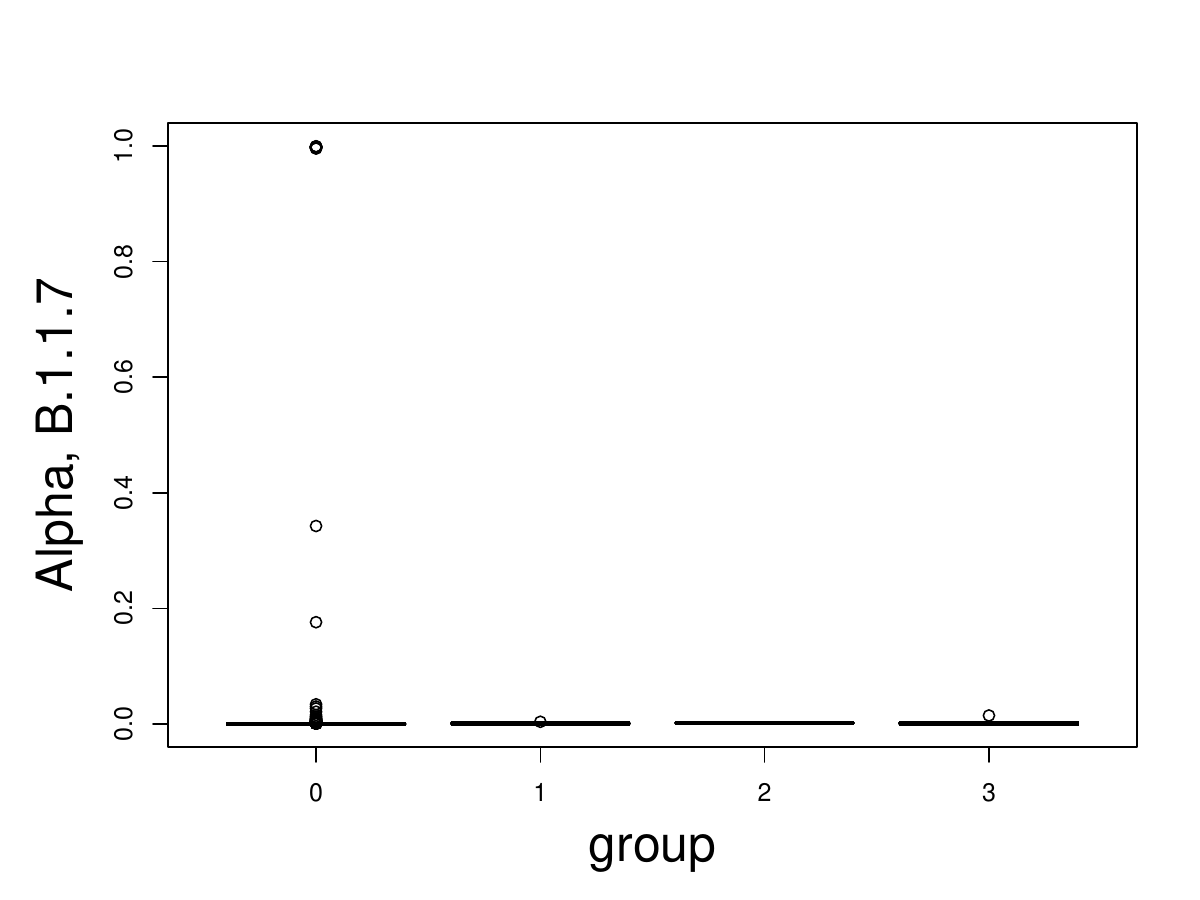}\par 
    \includegraphics[width=\linewidth]{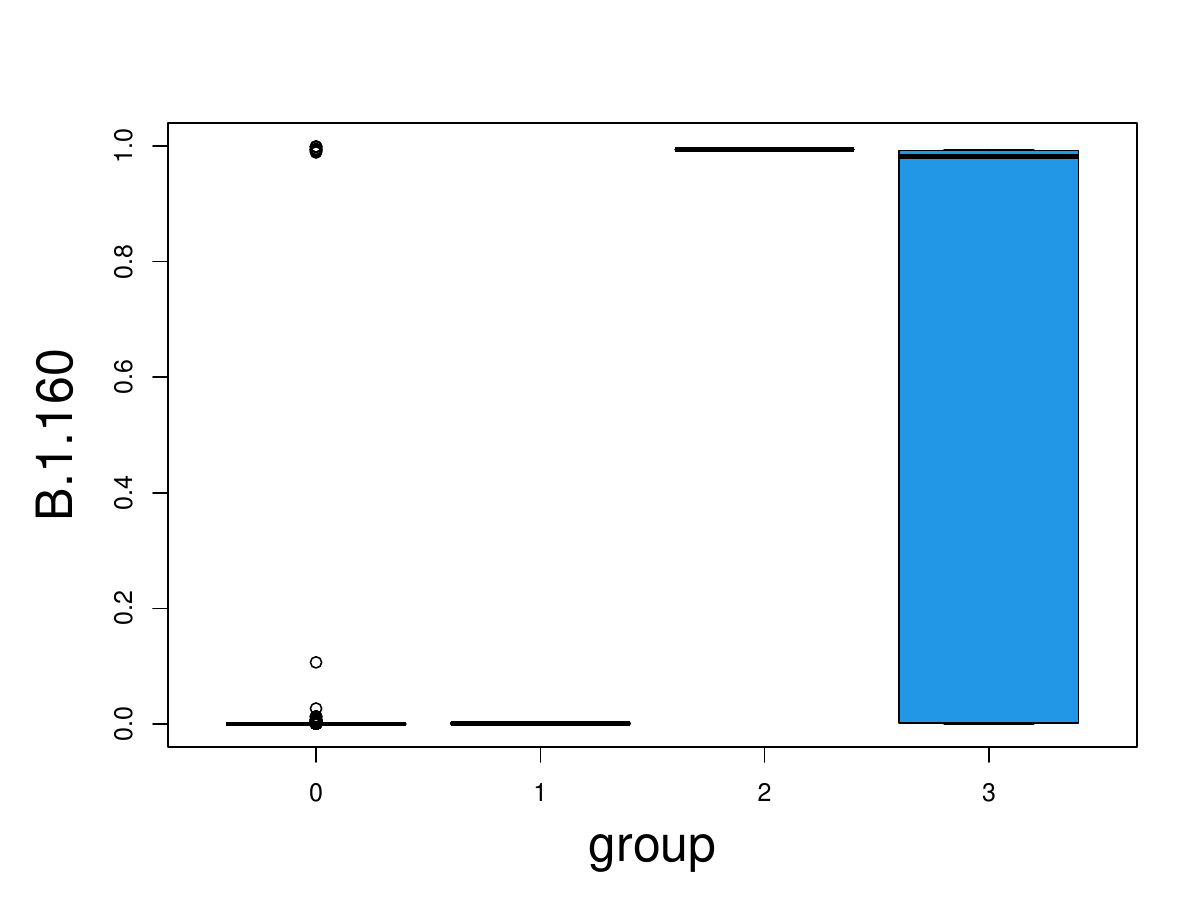}\par     
\end{multicols}

\vspace{-1em}
\caption*{Analysis J}
\vspace{-2em}
\begin{multicols}{3}
    \includegraphics[width=\linewidth]{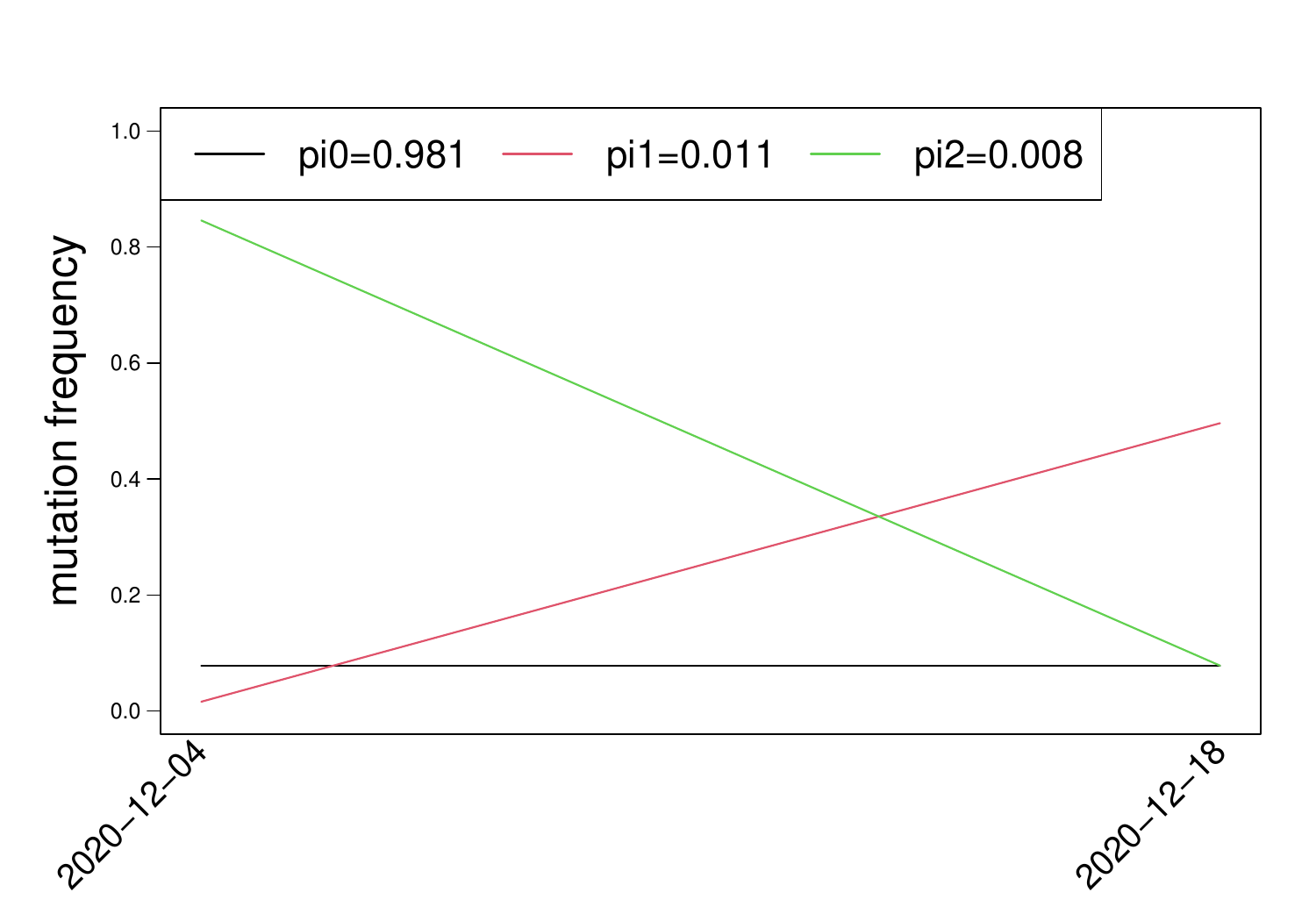}\par
    \includegraphics[width=\linewidth]{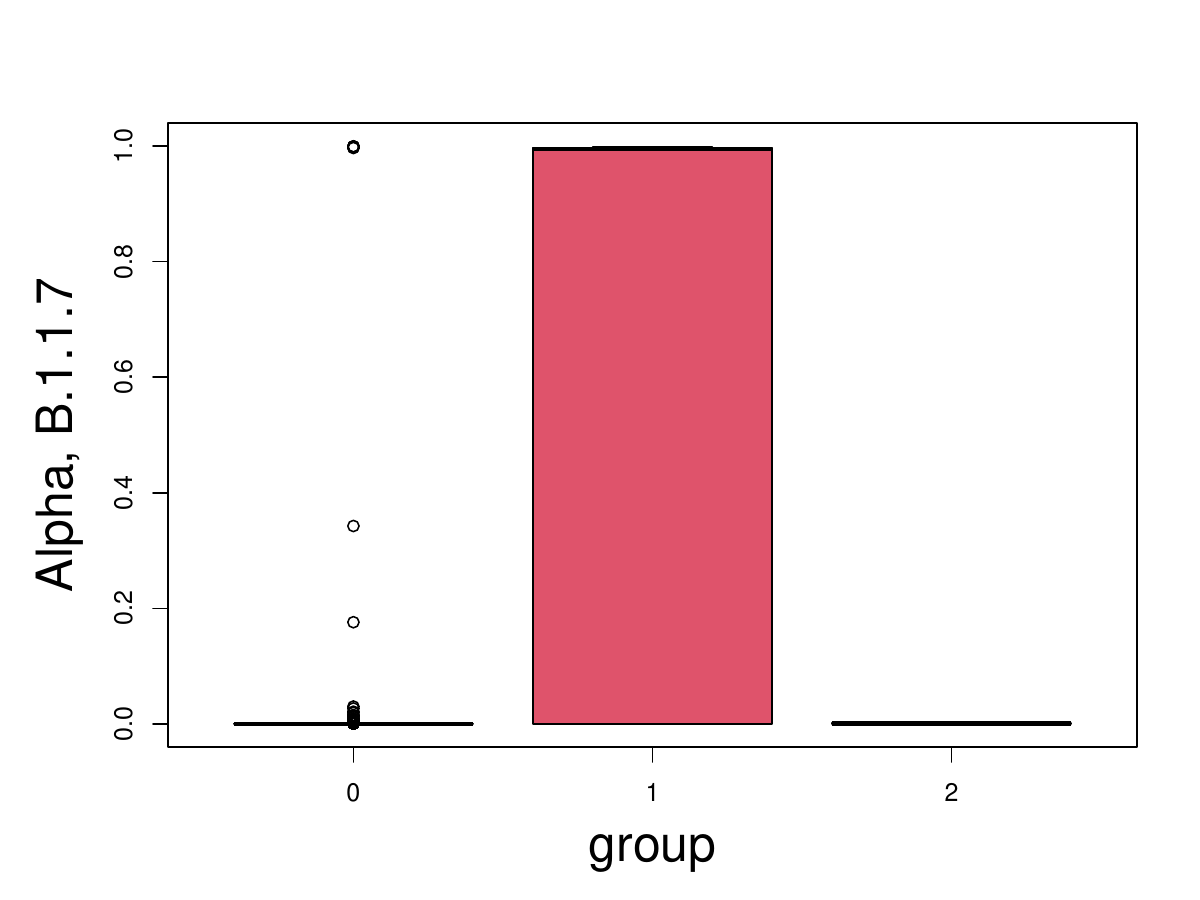}\par 
    \includegraphics[width=\linewidth]{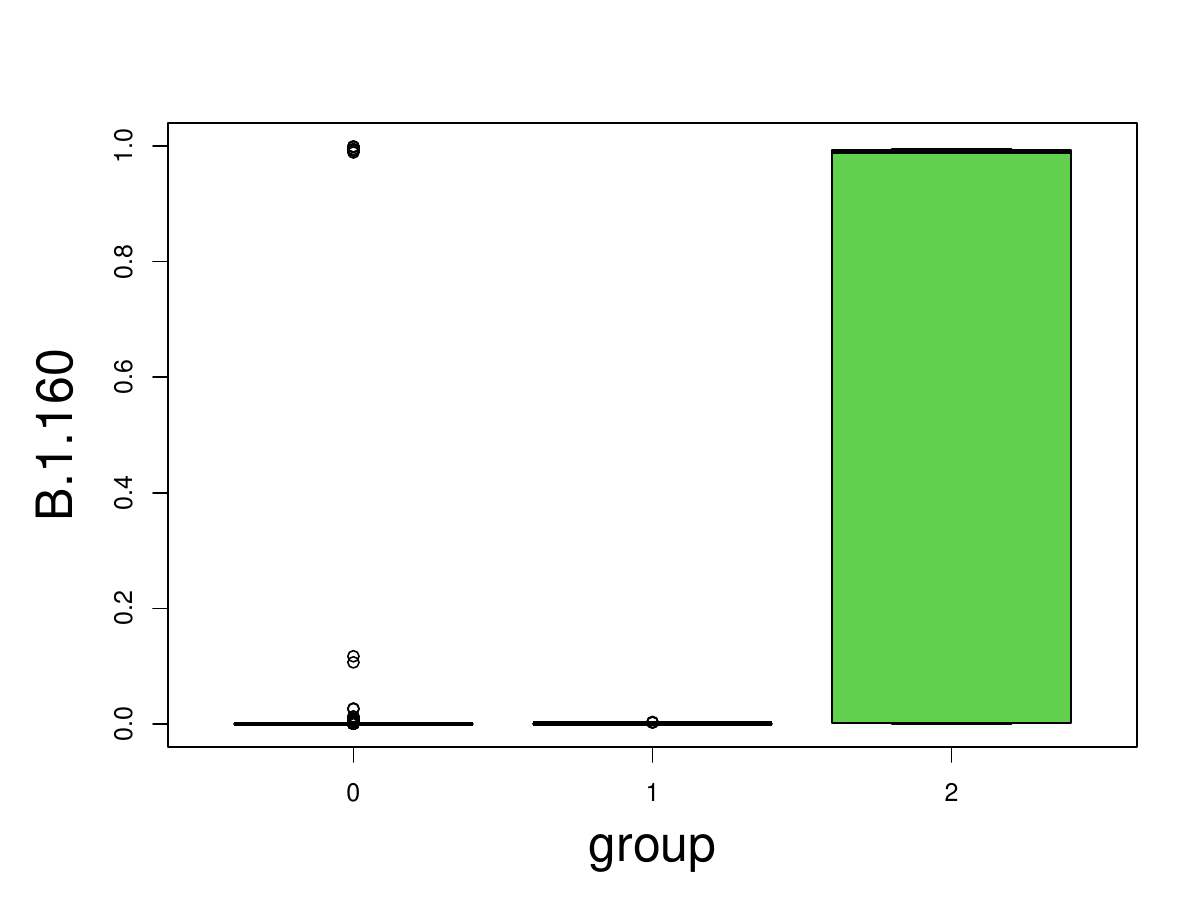}\par     
\end{multicols}
\vspace{-1.5em}
\caption{WWTP2}
\label{fig:detect.estim2}
\end{subfigure}

\caption{Mutation frequency trajectories and mutation signature of B.1.1.7 and B.1.160 VOC stratified on posterior group assignment estimated over dataset WWTP1 (Figure~\ref{fig:detect.estim1}) and WWTP2 (Figure~\ref{fig:detect.estim2}) restricted to different time periods over Octobre - December 2020.}
\label{fig:detect.estim}
\end{figure*}

%% file: tab_signature_detection.tex
\begin{table}
\vspace{-6em}
\centering
\begin{tabular}{r||c|c|c|c|c|}
\multicolumn{1}{c}{} & \multicolumn{5}{c}{Main circulating lineages} \\ 
\hline
& B.1.1.7 (Alpha) &  B.1.1 & B.1.160 & B.1.177 & B.1.367\\
\hline
T445C & - & 0.1 & 0.2 & \textbf{99.4} & -\\
G571A & - & - & 3.1 & - & -\\
C3177T & 3.1 & 0.1 & - & 0.1 & -\\
C3267T & \textbf{99.8} & 0.7 & 0.1 & 0.1 & \textbf{1.00}\\
C3924T & 0.1 & 0.1 & 0.1 & 0.1 & -\\
G4006T & - & - & - & 4 & -\\
T5071C & - & - & 1 & - & -\\
C10582T & 0.2 & 0.1 & - & 0.1 & \textbf{99.4}\\
G11132T & - & - & - & 7.9 & -\\
A11217G & - & - & - & - & -\\
C11916T & - & 0.1 & - & - & -\\
T12015G & - & - & - & 0.1 & -\\
G13201A & 0.1 & - & - & - & -\\
C13694T & 0.1 & 0.1 & 0.1 & 0.1 & -\\
G13723T & - & - & - & - & -\\
T15009C & - & - & - & - & -\\
A15267G & - & - & - & - & -\\
G16288A & - & - & - & - & -\\
G17302T & - & 0.1 & - & - & -\\
C22227T & 0.1 & 0.2 & 0.6 & \textbf{99.6} & -\\
C23185T & 0.1 & 0.2 & 0.2 & - & -\\
C23271A & \textbf{99.8} & 0.5 & - & - & -\\
G23948C & - & 0.1 & - & - & -\\
C25571T & 0.1 & - & 0.2 & 0.1 & -\\
G26314T & - & - & - & - & -\\
T27384C & 0.8 & 0.4 & 0.8 & 0.8 & -\\
A28111G & \textbf{99.8} & 0.6 & - & - & -\\
G28141T & - & - & - & - & -\\
G28280C & \textbf{99.5} & 0.5 & - & - & -\\
A28281T & \textbf{99.5} & 0.4 & - & - & -\\
T28282A & \textbf{99.6} & 0.5 & - & - & -\\
C28830A & - & - & - & - & \textbf{99.5}\\
\hline
\end{tabular}
\caption{Mutations assigned group $G_1$ of rising frequency trajectory in Analysis H (Figure~\Ref{fig:detect.estim1}, bottom) by maximum a posteriori probability along with mutation signature on main circulating variants 2 months later. Probabilities above 90\% are in bold. Zeros are replaced by sign `-' for readability.} 
\label{tab:signature.detection}
\end{table}